\pdfoutput=1
\documentclass{aa}
\usepackage[varg]{txfonts}
\usepackage{graphicx}
\usepackage{lscape}
\usepackage{multicol}
\usepackage{xcolor}

\begin{document}
\title{Spectral type, temperature and evolutionary stage in cool supergiants}
\author{Ricardo Dorda\inst{\ref{inst1}}
\and Ignacio Negueruela\inst{\ref{inst1}}
\and Carlos Gonz\'alez-Fern\'andez\inst{\ref{inst2}}
\and Hugo M. Tabernero\inst{\ref{inst1},\ref{inst3}}}
\institute{Departamento de F\'{\i}sica, Ingenier\'{\i}a de Sistemas y Teor\'{\i}a de la Se\~nal, Universidad de Alicante, Carretera de San Vicente s/n, San Vicente del Raspeig E03690, Alicante, Spain\label{inst1}
\and Institute of Astronomy, University of Cambridge, Madingley Road, Cambridge CB3 0HA, United Kingdom\label{inst2}
\and Departamento de Astrof\'{\i}sica, Universidad Complutense de Madrid, Facultad de CC F\'{\i}sicas, Avenida Complutense s/n, 28040 Madrid \label{inst3}
}

\abstract{In recent years, our understanding of red supergiants has been questioned by strong disagreements between stellar atmospheric parameters derived with different techniques. Temperature scales have been disputed, and the possibility that spectral types do not depend primarily on temperature has been raised.}{We explore the relations between different observed parameters and the capability of deriving accurate intrinsic stellar parameters from them through the analysis of the largest spectroscopic sample of red supergiants to date.}{We have obtained intermediate-resolution spectra of a sample of about 500 red supergiants in the Large and the Small Magellanic Cloud. From them, we derive spectral types and measure a large set of photospheric atomic lines. We explore possible correlations between different observational parameters, also making use of near- and mid-infrared colours and literature on photometric variability.}{Direct comparison between the behaviour of atomic lines (\ion{Fe}{i}, \ion{Ti}{i}, and \ion{Ca}{ii}) in the observed spectra and a comprehensive set of synthetic atmospheric models provides compelling evidence that effective temperature is the prime underlying variable driving the spectral-type sequence between early G and M2 for supergiants. In spite of this, there is a clear correlation between spectral type and luminosity, with later spectral types tending to correspond to more luminous stars with heavier mass loss. This trend is much more marked in the LMC than in the SMC. The population of red supergiants in the SMC is characterised by a higher degree of spectral variability, early spectral types (centred on type K1) and low mass-loss rates (at least as measured by dust-sensitive mid-infrared colours). The population in the LMC displays less spectroscopic variability and later spectral types. The distribution of spectral types is not single-peaked. Instead, the brightest supergiants have a significantly different distribution from less luminous objects, presenting mostly M subtypes (centred on M2), and increasing mass-loss rates for later types. In this, the behaviour of red supergiants in the LMC is not very different from that of Milky Way objects.}{The observed properties of red supergiants in the SMC and the LMC cannot be described correctly by standard evolutionary models. The very strong correlation between spectral type and bolometric luminosity, supported by all data from the Milky Way, cannot be reproduced at all by current evolutionary tracks.}


\maketitle

\section{Introduction}
\label{intro}

According to evolutionary models \citep[e.g.][]{bro2011,eks2012,eks2013}, when stars with initial masses between $\sim10$ and $\sim40\:M_{\sun}$ deplete the H in their cores, they evolve quickly from the hot to the cool side of the Hertzsprung-Russel (HR) diagram, at approximately constant luminosity. This decrease in temperature has to be compensated by a huge increase in radius, and these stars become red supergiants (RSGs).

RSGs have late spectral types (SpTs) and low effective temperatures. Traditionally, very luminous stars of spectral types K and M have been known as RSGs, but -- as will be discussed later -- the separation from G supergiants may be artificial, at least at metallicities much lower than that of the Sun. It is thus also frequent to use the term cool supergiants (CSGs) to refer to the range from early G to M. Different studies of galaxy-wide RSG populations have suggested that the spectral subtypes of RSGs adopt a distribution around a typical SpT. \cite{hum1979a} found that this distribution may span a different range of SpTs depending on the galaxy hosting the population: the lower its metallicity, the earlier its RSGs. This behaviour has been repeatedly confirmed since then \citep{eli1985,mas2003b,lev2012}.

\cite{eli1985} put forward two possible causes for this dependence of the typical SpT on metallicity. The first one is the effect of the metallicity on the Hayashi limit (i.e.\ the lowest temperature that RSGs can reach in their evolution), as this limit is expected to appear at higher temperatures for lower metallicities. RSGs with higher metallicity would thus evolve down to lower temperatures, and therefore reach later SpTs. The second one is the effect of metallicity on the TiO abundance, because the strength of its molecular bandheads is the main criterion for spectral classification in the K to M range, with later SpTs defined by deeper bandheads. For the same temperature, a RSG with a lower metal content should have a lower TiO abundance, and thus weaker bandheads, leading to a classification as an earlier-type star.

The effective temperature ($T_{\textrm{eff}}$) scale, i.e.\ the relation between SpT and temperature, for M supergiants was initially estimated to span from $3600\:$K at M0 to $2800\:$K at M5 \citep{lee1970}. \cite{hum1984} confirmed these values and extended the scale to earlier SpTs, with temperatures stretching the range from $4300$\:K at K0 to $2800$\:K at M5. Over the following two decades, this relation was not revisited, until \cite{mas2003b} calculated a slightly different scale, with temperatures a bit cooler for the K subtypes and a bit warmer for the M ones. In any case, all these works agreed on RSGs being cooler than the lowest temperature predicted by their contemporary evolutionary models.

Some years later, \cite{lev2005,lev2006} derived a new effective temperature scale by fitting synthetic spectra generated using MARCS atmospheric models to their spectrophotometric observations, in the range from $4\,000$ to $9\,500\:$\AA{}. Their results brought the galactic RSGs into agreement with the temperatures predicted by the evolutionary models of \cite{mey2000}. They also placed RSGs from the Small Magellanic Cloud (SMC) and Large Magellanic Cloud (LMC) closer to the theoretical predictions, without achieving a very good agreement, especially in the case of the SMC.

The effective temperature scale obtained by \citet{lev2005} for galactic RSGs has a flatter slope and is warmer than previous ones, going from $4\,100$\:K at K1 to $3\,450$\:K at M5. The LMC and SMC RSGs span almost the same temperature range \citep{lev2006}, from $\sim4\,200$\:K at K1 to $3\,475$\:K at M2 for the SMC, and from $\sim4\,300$\:K at K1 to $3\,450$\:K at M4 for the LMC. This implies that, at a given temperature and for M subtypes, stars from the SMC appear earlier than those from the LMC, which in turn are earlier than Milky Way (MW) objects.

\citet{lev2006} explore the arguments presented by \cite{eli1985} under the light of their results, finding that the difference in SpT between RSGs from the SMC and the LMC (or the Galaxy) at a given temperature (for M~RSGs) is significantly smaller than the difference between the mean SpTs of both galaxies. In view of this, they conclude that the effect of metallicity on the TiO bandheads is not enough to explain the shift in SpT from one galaxy to another, and, in consequence, a varying Hayashi limit must be the main reason behind the shift in the typical SpTs of RSGs between galaxies. Moreover, using the same method, \cite{dro2012} find that RSGs in M33 have different typical temperatures at different galactocentric distances, presumably because of the radial metallicity gradient present in this galaxy. With this result, they confirm the behaviour observed for different galaxies within a given galaxy.

There are, however, a number of unresolved issues in these works that must be noted before accepting these effective temperature scales. Firstly, in addition to synthetic spectra fitting, \cite{lev2006} also used synthetic colours to calculate alternative temperatures, finding that the results from fits to $(V-K)_{0}$ are systematically warmer than those calculated from MARCS stellar atmospheric models. They argue that this discrepancy is caused by the MARCS synthetic spectra not reproducing correctly the near infrared (NIR) fluxes. It must be noted, however, that the temperatures obtained from $(V-K)_{0}$ do agree with the predictions of their contemporary Geneva evolutionary models. Secondly, \cite{lev2007} find a number of RSGs in the Magellanic Clouds (MCs) with extremely late SpTs, which lay clearly under the coolest temperature predicted by evolutionary models for their respective metallicities. Thirdly, they find that for M-type supergiants, at a given SpT, RSGs in the SMC are cooler than those in the LMC or the MW. They explain this difference as a consequence of the effect of metallicity variations on TiO-band strengths. On the other hand, for K-type supergiants they find that stars from all three galaxies have roughly the same temperature at a given SpT.

\cite{dav2013} obtained spectrophotometry for a small number of targets from both MCs. They used three different methods to calculate their temperatures. On one hand, they used the strengths of the TiO bands, fitting their spectra with synthetic spectra generated using MARCS stellar atmospheric models, as \cite{lev2006} did. On the other hand, they performed fits to the optical and infrared spectral energy distribution (SED). Finally, they also used the flux integration method (FIM), though leaving $A_{V}$ as a free parameter and utilizing these results only as a constraint for the SED and TiO scales. They find that the TiO scale is significantly cooler than the SED temperatures. They present three strong arguments against the TiO scale and in support of temperatures derived from the SED: the TiO scale is cooler than the lowest temperatures derived from the FIM at the lowest reddening; the TiO temperatures overpredict the IR flux; and there is a lack of correlation between the reddening derived from the TiO methods and the diffuse interstellar bands measured. The SED temperatures are in agreement with the Geneva evolutionary models, but all the RSGs have temperatures inside a narrow range ($4150\pm150$\:K), regardless of which galaxy they belong to and thus which SpT they have. In consequence, if SpTs, which are determined from the strength of the TiO bands, do not depend mainly on temperature, then they have to depend on luminosity, which seems related to the evolutionary stage. From this, they suggest that RSGs with early SpTs have arrived from the main sequence recently, while those with late SpTs have moved steadily in the HR diagram towards higher luminosities, increasing at the same time their mass loss and their circumstellar envelopes.

Since then, a flurry of works \citep{gaz2014,dav2015,pat2015} have obtained similar temperatures for RSGs from different environments. Ordered by increasing metallicity, these are NGC~6822, the SMC, the LMC and Perseus OB1. They used the spectral synthesis method on $J$-band spectra, as initially proposed by \cite{dav2010}. \cite{gaz2014} found temperatures for all but one of their RSGs in the Galactic Perseus OB1 association to lie in the range from $3800$\:K to $4100$\:K. \cite{pat2015} obtained temperatures between $3790$\:K and $4000$\:K for all their RSGs in NGC~6882. \cite{dav2015} re-analysed the same data from \cite{dav2013} with this method, finding all the stars from both MCs within the range $3800$\:K to $4200$\:K, without any differences between RSGs from each galaxy. They also find that $J$-band temperatures agree well with those calculated through the SED method for the SMC RSGs, but there is a significant offset ($160\pm110$\:K) for RSGs in the LMC. Finally, \citet{gaz2015} studied RSGs in NGC~300, paying attention to their galactocentric distances, because of the expected abundance gradient. They found that all but one were inside the range between $3800$ and $4300$\:K, but also that there is no dependence between the metallicity and temperature, even though their range of metallicities spans from solar  down to $-0.6$~[dex]. Although none of these works provides SpTs for their RSGs, all of them subscribe the conclusion of \cite{dav2013}, because they are finding a single temperature range for all the RSGs, even when shifts in the mean SpT of RSGs are expected between these different environments. 

The situation described leaves many open issues that we explore in the present work. The main one is the relation of SpT with luminosity, evolutionary stage (mass-loss), and temperature. The works discussed in the two preceding paragraphs suggest a relation between SpT, luminosity and evolutionary stage, but there is no statistically significant analysis that proves this relation or describes it. Similarly, all these papers ignore SpT, as they do not expect it to be related to temperature. In addition, the differences in mean SpT between different galaxies have been considered implicitly as a consequence of the effect of metallicity on the TiO bands, which define the M sequence. Unfortunately, this argument ignores that most of the SMC stars are early-K stars whose SpT is not determined mainly from the strength of TiO bands, but from atomic line ratios and the shy rise of TiO bands. Thus, the hypothesis of increasing luminosity along the evolution as an interpretation for the M sequence does not give a satisfactory explanation for the low-metallicity populations with early mean SpTs: do these stars evolve without a SpT progression? Or instead, are they also changing along their own "early" sequence? Finally, if the relation between luminosity and SpT is eventually confirmed, what does this imply for the observed changes in SpT demonstrated by many CSGs?

\section{Data and measurements}

\subsection{The sample}
\label{sample}

The sample of CSGs used in this work has already been published in \citet[from now on, Paper~I]{gon2015}. Detailed explanations about data selection, observation, and reduction are presented there, while in this paper we give only a short overview of the main points.

We used the fibre-fed dual-beam AAOmega spectrograph on the 3.9~m Anglo-Australian Telescope (AAT) to observe one of the largest samples of CSGs from both MCs to date. These observations were done along four different epochs (three for the SMC and two for the LMC), including $\sim$100 previously known CSGs from each cloud plus a large number of candidates to CSGs, among which we identified a large fraction as supergiants (SGs), most of them previously unknown. All these new candidates in both galaxies were observed only on one epoch, while most RSGs already listed in the literature were observed at least twice. The properties of the sample are summarized in Table~\ref{observations}.

Spectra in the optical range were used to classify the SpT and luminosity class (LC) for all the targets observed (see Paper~I for a detailed description of the criteria utilised). We used classical criteria based on atomic line ratios along with TiO band depths, when these features were present. Radial velocities (RVs) were obtained, and we used this information to complement and confirm our LC classification. As many of the targets were observed more than once on the same epoch, the differences between their assigned classifications and RVs were used to quantify the typical dispersion in RV ($\sim1.0$~km~s$^{-1}$) and the typical errors in SpT and LC, which is about one subtype for SpT and half subclass for LC. The final SpT and LC for each one these CSGs at each epoch are the mean values weighted by the signal-to-noise of each spectrum.

Thanks to the dual-beam of AAOmega all the targets were observed simultaneously in the region of the infrared Calcium Triplet (CaT). We used the 1700D grating, which provides a 500~\AA{}-wide range with nominal resolving power ($\lambda/\delta\lambda$) of $11\,000$ at the wavelengths considered. The blaze was centred on $8600$~\AA{} in the 2010 observations, but on $8700$~\AA{} at all other epochs.

\begin{table*}
\caption{Summary of CSGs observed along our four epochs.}
\label{observations}
\centering
\begin{tabular}{c | c | c | c c | c}
\hline\hline
\noalign{\smallskip}
Galaxy&Epoch&Selected CSGs&\multicolumn{2}{| c |}{CSGs from candidate list}&Total\\
&&&Already known&Previously unknown&\\
\noalign{\smallskip}
\hline
\noalign{\smallskip}
&2010&107&0&1&108\\
SMC&2011&104&0&0&104\\
&2012&146&40&117&303\\
&All\tablefootmark{a}&158&40&117&315\\
\noalign{\smallskip}
\hline
\noalign{\smallskip}
&2010&84&0&0&84\\
LMC&2013&97&37&90&224\\
&All\tablefootmark{a}&102&37&90&229\\
\noalign{\smallskip}
\hline
\end{tabular}
\tablefoot{
\tablefoottext{a}{Unique targets observed in any of the the epochs.}
}
\end{table*}

\subsection{Equivalent widths}
\label{EW}

\begin{figure}[h!]
   \centering
   \includegraphics[trim=1cm 0.5cm 2cm 1.2cm,clip,width=9cm]{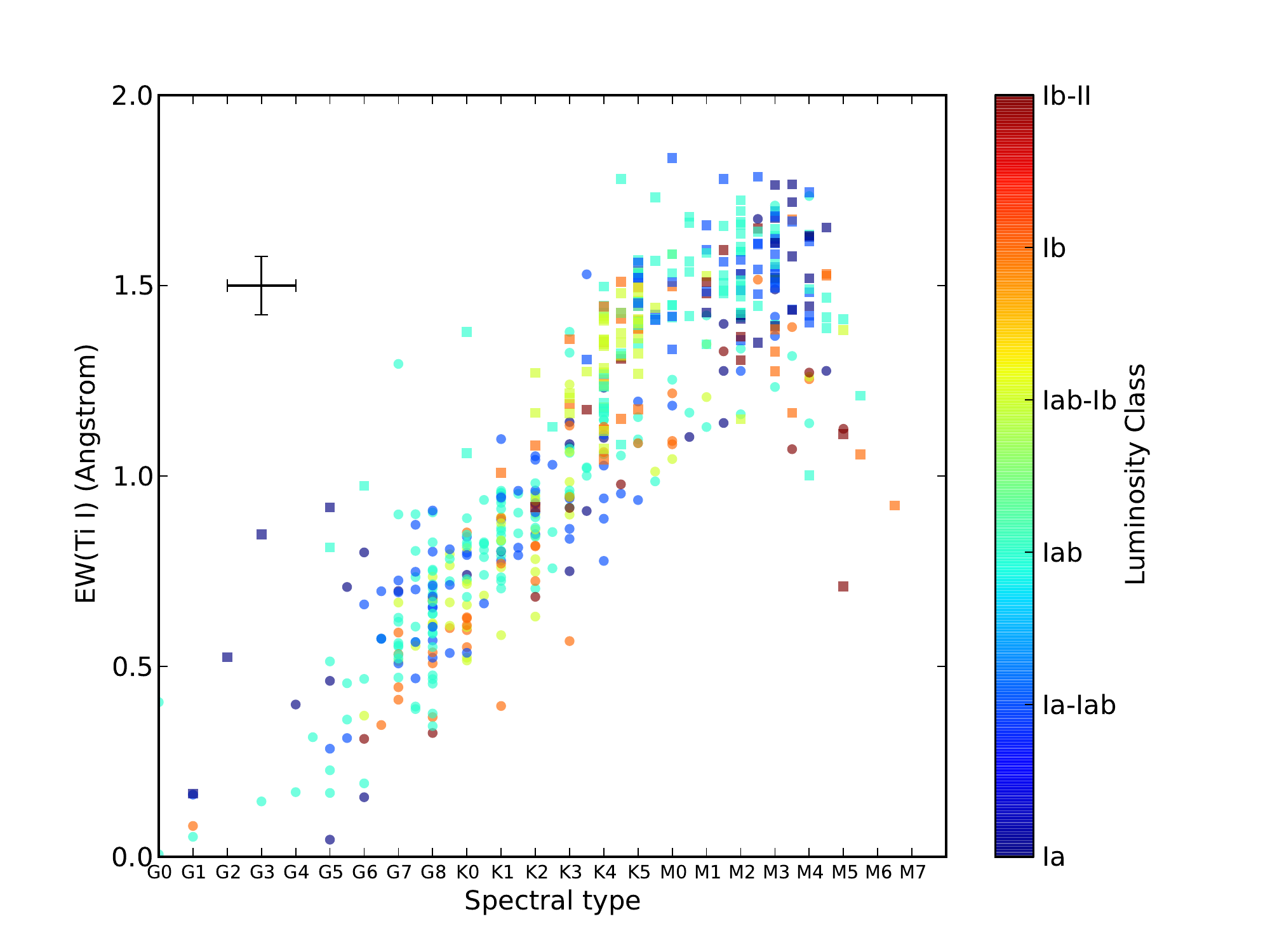}
   \caption{Spectral type as a function of the mean sum of the equivalent widths of Ti\,{\sc{i}}. The circles are CSGs from the SMC, the squares from the LMC. The single error bars represent the median uncertainties. The colour indicates the LC. This figure illustrates how, for stars later than M3, the TiO bands quickly affect their atomic lines.}
   \label{spt_example}
\end{figure}

\begin{figure}[ht!]
   \centering
   \includegraphics[trim=1cm 0.5cm 2cm 1.2cm,clip,width=9cm]{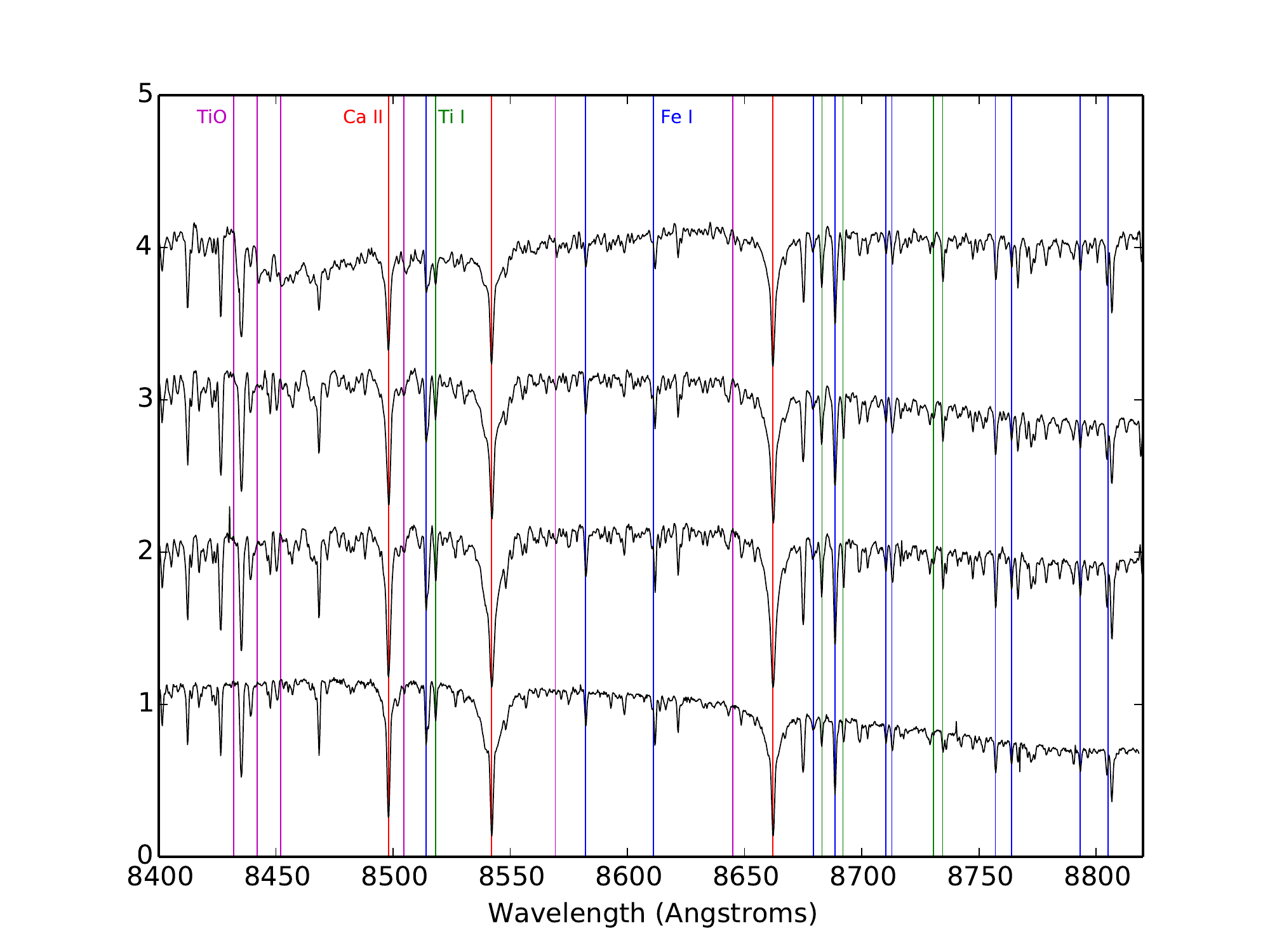}
   \caption{Example of the spectra used. This is a SpT sequence displaying atomic features measured, and also the position of the main TiO bands which may affect our measurements. The stars shown have all (except one) the same luminosity class and similar metallicity (they are all from the LMC). From bottom to top: [M2002]159974 (K2\:Iab), SP77~46-40 (K5\:Iab), [M2002]149767 (M3\:Iab), and [M2002]148035 (M5\:Ia). The dashed lines indicate the spectral features measured. Their colour represents the dominant chemical species in each feature: red for Ca\,{\sc{ii}}, blue for Fe\,{\sc{i}}, green for Ti\,{\sc{i}}, and magenta for the TiO bands. For more details, see the text.}
   \label{sec_spt}
\end{figure}

There are many interesting atomic and molecular features in the CaT spectral region that have often been used for spectral classification \citep[e.g.][]{kir1991,car1997}. In spite of this, we decided to use the optical range to perform the classification according to classical criteria, while we performed systematic measurements of the main atomic features in the CaT spectral region, which was observed at higher resolution. With this approach we made sure that none of the features measured is directly related to the SpT and LC assigned, because the classification was done in a completely different wavelength range. There is a strong reason to obtain quantitative measurements of atomic features in the CaT region rather than in the optical range: TiO bands do not appear in the infrared region until the M1 subtype, while in the optical range they start to grow from K0 (becoming dominant around M0), eroding both continuum and atomic lines. In consequence, our atomic line measurements are not affected by molecular bands for subtypes M3 or earlier (which is most of our sample). Beyond this subtype, the effect of the bands on both continuum and features is unavoidable over the whole range, causing a quick decrease in the EWs of all atomic lines as the SpT increases (see Figs.~\ref{spt_example} and~\ref{sec_spt}). In view of this, in this work we have not used any values measured on stars later than M3. Since the vast majority ($\sim92\%$) of stars in our sample are M3 or earlier, we still have a statistically significant sample to describe the CSG behaviour.

For this work we have selected a number of lines, most of them often used for spectral classification, such as the CaT itself (luminosity marker) and many lines of Ti\,{\sc{i}} and Fe\,{\sc{i}}, whose ratios are classical criteria for SpT and/or LC. We measured the equivalent widths of all these lines in all our stars through an automated and uniform method. However, because the central wavelength in the 2010 observations was slightly different, there is a small number of lines that lie outside the spectral range observed in that run. Table~\ref{lines} contains the list of all the lines measured, together with relevant information.

To measure the EWs, we defined for each atomic line a wavelength range (at rest frame) that covers the line itself, and two continuum regions, one on the red side and the other on the blue side of this range. We brought all the spectra to the rest wavelength by correcting for their observed RVs, and we used the continuum points at both sides of the atomic line to calculate through a linear regression the theoretical continuum. We then used this continuum to calculate the EW. We have not used a global continuum for our whole spectral range because our spectra are not calibrated in flux, and thus we do not know the real shape of the continuum. Our local continua provide readily comparable measurements for all stars. For late-M types, the EWs are a measurement of what remains of the line when the TiO bands have eroded away the continuum. For earlier subtypes, it is a true measurement of the EW of the line. The uncertainties in the EWs were calculated through the method proposed by \cite{vol2006}. For those stars observed more than once on the same epoch, we have obtained the EWs for each spectrum, calculating then the mean EW for the star on that epoch weighted by the signal-to-noise of each spectrum. Because of the expected spectral variability, the EWs from spectra taken on different epochs have not been combined.
	
From these measurements, we have defined three indices. EW(CaT) is the sum of the EWs for the three lines in the triplet. EW(Ti\,{\sc{i}}) is the sum of EWs for the five Ti\,{\sc{i}} lines measured.  EW(Fe\,{\sc{i}}) is the sum of the EWs of the twelve Fe\,{\sc{i}} lines present in the spectra of all epochs except those from 2010 (for this epoch, four Fe\,{\sc{i}} lie outside the spectral range observed).

\subsection{Bolometric magnitudes}
\label{mbols}

To calculate the bolometric magnitudes ($m_{\textrm{bol}}$) of our stars, we have chosen the bolometric correction (BC) proposed by \cite{bes1984}, because it is given as a function of $(J-K)$, and our data show a clear trend between SpT and this colour (see Fig.~10 in Paper~I).

The reddening to the clouds is relatively small, with typical values around $E(B-V)\sim0.1$ \citep{sos2002,kel2006}. Some CSGs exhibit heavy mass-loss, and so we might expect some amount of circumstellar extinction in these objects. However, the CSGs with larger mass-losses are those with later SpTs, as shown in Paper~I. Since most of our sample is M3 or earlier, we should expect few stars to present significant self-absortion. In addition, the reddening for the $J$ and $K$ bands is much lower than for optical bands ($A_{K}\sim0.1A_{V}$; \citealt{car1989}). Therefore, we do not expect the $m_{\textrm{bol}}$ calculated to be significantly affected by extinction for any of our stars. In any case the effect of the reddening would be lesser than the effect of the position of our stars inside the clouds \citep[e.g. the SMC has a depth of 0.15 mag;][]{sub2012}. Thus, we have not corrected our magnitudes for this effect.

Many of our CSGs are variable stars, but the typical photometric amplitudes decrease with wavelength \citep{rob2008}, and we may expect very small variations for 2MASS bands, e.g.\ \cite{woo1983} found that RSGs do not have amplitudes larger than 0.25 magnitudes in the $K$ band. In consequence, photometric variability should not affect in a significant way the value of $m_{\textrm{bol}}$ and so we may use these bolometric magnitudes in combination with information derived from spectra or other IR photometric bands even if they were taken at different epochs.

The photometric data used for this analysis are the $J$ and $K_{\textrm{S}}$ magnitudes from 2MASS \citep{skr2006}. We transformed the 2MASS magnitudes to the AAO system used by \cite{bes1984}, and then calculated the corresponding $m_{\textrm{bol}}$. Finally, the distance moduli to both MCs are well known, and so we have calculated the absolute bolometric magnitudes ($M_{\textrm{bol}}$), using $\mu=18.48\pm0.05$~mag for the LMC \citep{wal2012} and $\mu=18.99\pm0.07$~mag for the SMC \citep{gra2014}.

\begin{figure*}[th!]
   \centering
   \includegraphics[trim=1cm 0.5cm 2.3cm 1.2cm,clip,width=9cm]{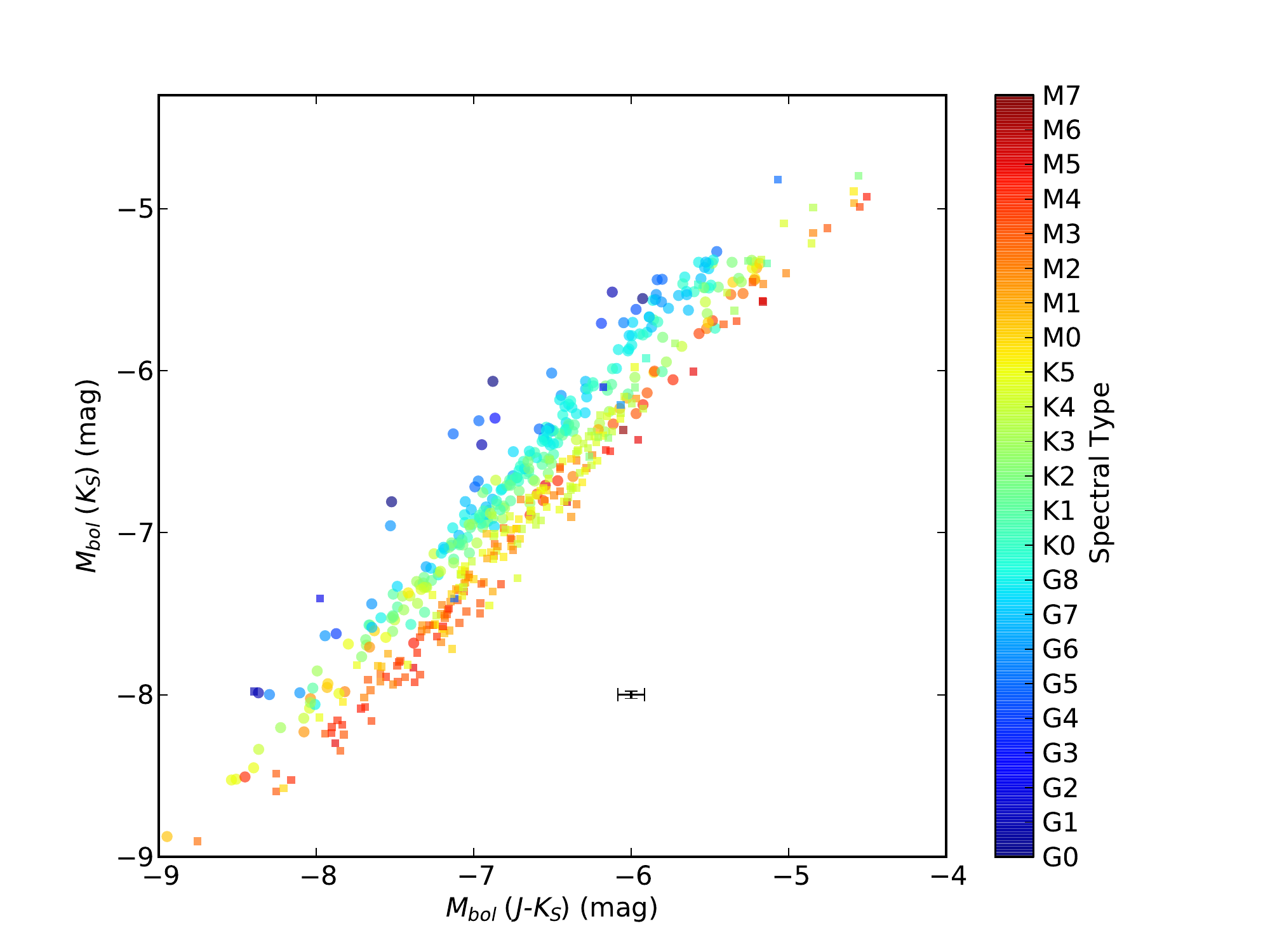}
   \includegraphics[trim=0.8cm 0.5cm 2.5cm 1.2cm,clip,width=9cm]{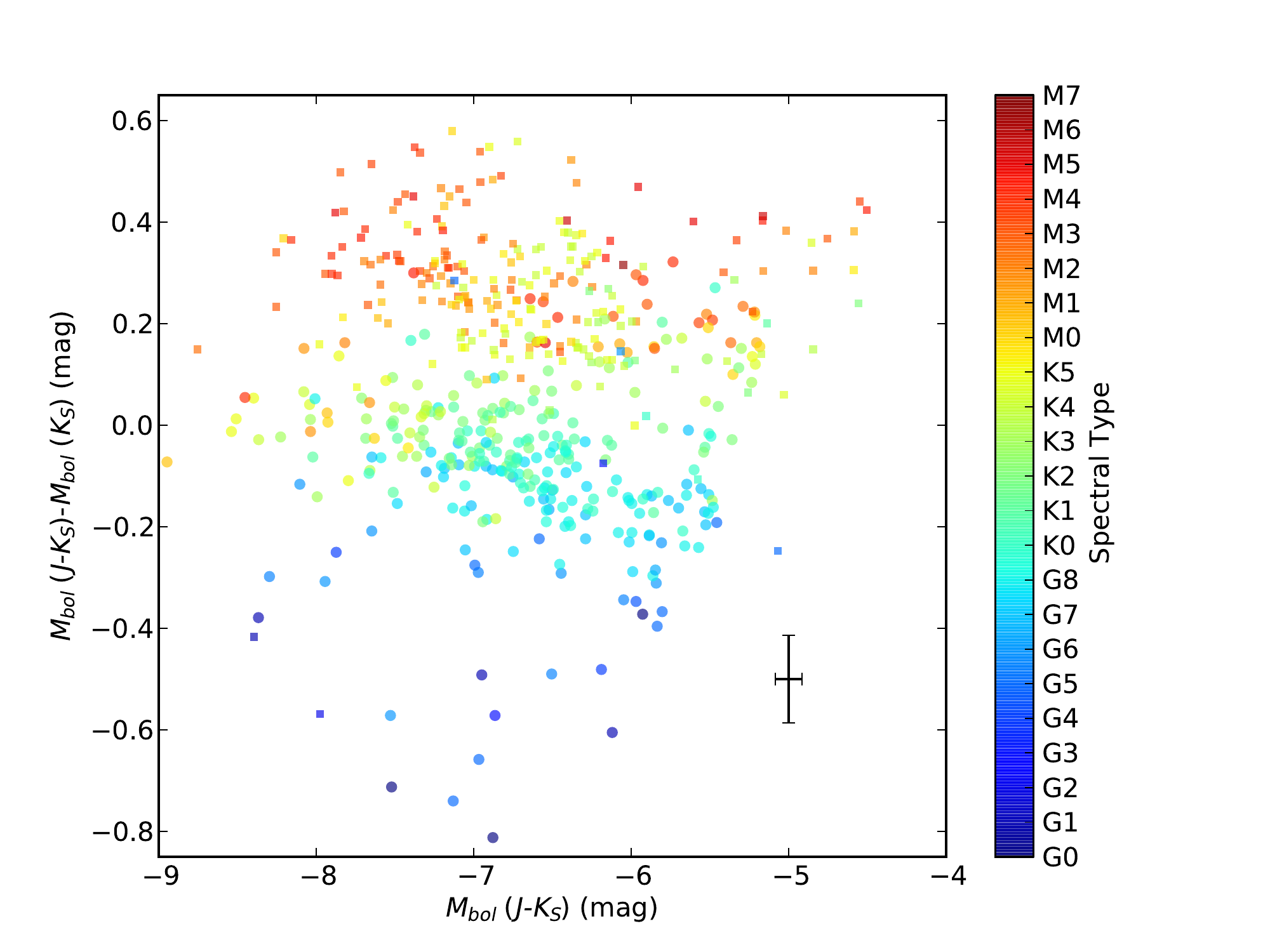} 
   \caption{Comparison of different bolometric magnitudes. The adopted magnitudes were calculated through the $J-K$ colour, following \cite{bes1984}, and are put in abscissa in both figures. Colour indicates the SpT. Shape identifies the galaxy: circles are from the SMC, squares from the LMC. The black cross represents the median uncertainties. {\bf Left (\ref{Mbol}a):} The ordinate show the bolometric magnitudes calculated from the $K_{\textrm{S}}$ band through   through the constant BC$=2.69$\:mag, as was proposed by \cite{dav2013}. {\bf Right (\ref{Mbol}b):} The ordinate show the difference between the bolometric magnitudes calculated through the two analysed ways.}
   \label{Mbol}
\end{figure*}

We have studied the relation between these BCs, calculated after \citet{bes1984}, and the constant BC$=2.69$\:mag for the $K_{\textrm{S}}$ band proposed by \cite{dav2013}, independent of the observed colours.  We have to note that there is a systematic difference between the SMC and the LMC in both cases, because \cite{bes1984} used two separate formulas to calculate the BCs, one for CSGs from the MW and the LMC, and the second for those from the SMC. The only difference between both formulas is a constant of $0.12$~mag. We find a very good agreement ($r^{2}=0.998$; see Fig.~\ref{Mbol}) between our values and those obtained through the BC of \cite{dav2013}. This is because most of our stars have a $(J-K_{\textrm{S}})\sim1$\:mag, and then the BC from \citet{bes1984} is about $2.7$\:mag ($2.7$\:mag for the LMC and $2.6$\:mag for the SMC, applied to $K_{\textrm{S}}$). However, the $(J-K_{\textrm{S}})$ colour for stars in our sample ranges from $0.6$\:mag (early~G) to $1.5$\:mag (mid~M). For the extreme values of this range, the BC correction of \citet{bes1984} differs significantly from a constant value of $2.69$\:mag -- at $(J-K_{\textrm{S}})=0.60$\:mag, BC is $1.9$\:mag for the SMC, while at $(J-K_{\textrm{S}})=1.5$\:mag BC is $3.2$\:mag in the LMC.  As we show in Fig.~\ref{Mbol}b, the differences are specially large for those CSGs with SpTs specially early or late (reaching up to $0.8$~mag of difference). Thus, we have chosen to use throughout this work the expression from \citet{bes1984} to calculate $M_{\textrm{bol}}$, because it takes into account the changes in the SED due to temperature, which is reflected in the $(J-K_{\textrm{S}})$ colour.

\subsection{Synthetic spectra}
\label{syntetic}
Synthetic spectra were generated using two sets of one-dimensional LTE atmospheric models, namely: ATLAS-APOGEE (KURUCZ) plane-parallel  models \citep{mes12}, and MARCS spherical models with 15~M$_{\odot}$ \citep{gus08}. The radiative transfer code employed was \emph{spectrum} \citep{gra1994}. Although MARCS atmospheric models are spherical, \emph{spectrum} treats them as if they were plane-parallel. Therefore, the plane parallel transfer treatment might produce a small inconsistency in the calculations of synthetic spectra based on MARCS atmospheric models. However, the study of \citet{hei2006} concluded that any difference introduced by the spherical models in a plane-parallel transport scheme is small.

As line-list, we employed a selection of atomic lines from the VALD database \citep{pis1995,kup2000}, taking into account all the relevant atomic and molecular features that can appear in RSGs. In addition, as Van der Waals damping prescription we employed the Anstee, Barklem, and O'Mara theory, when available in VALD \citep[see ][]{bar2000}. The grid of synthetic spectra was generated for two different surface gravities (i.e.\ $\log{g}$~$=0$ and $1$\:dex). Effective temperature ($T_{\textrm{eff}}$) ranges from $3\,500$\:K to $4|,500$\:K with a step of 250\:K for the spectra generated using KURUCZ atmospheric models, whereas for the MARCS-based synthetic models, the $T_{\textrm{eff}}$ varies between $3|,300$\:K and~$4\,500$\:K. In this second case, the step is 250\:K above $4\,000$\:K, and 100\:K otherwise. The microturbulence ($\xi$) was fixed to $3\:$km~s$^{-1}$. Finally, the metallicity ranges from [M/H]~$=$~$-1$\:dex to [M/H]~$=0$\:dex in 0.25~dex steps.

\subsection{Correlations}
\label{corr}

A central point of the discussion in this work is the existence of certain correlations between different variables. To test each of them we used the procedure described in this section.

When SpT is used as one of the variables correlated, we assigned numerical values to the spectral subtypes: G0 is 0, and then one by one until G8, which is 8, then K0 is 9, K5 is 14, M0 is 15 and so on until M7 which is 22.

We used two different correlation coefficients, Pearson ($r$) and Spearman ($r_{\textrm{S}}$). $r$ is a sensitive coefficient that responds well to linear correlations, but is not very robust. On the other hand, $r_{\textrm{S}}$ is not so sensitive, but it is a very robust method that also can deal with non-linear correlations. To reduce the effect of outliers and also to ensure that the results obtained would not be driven by the stochasticity of our particular sample, we used a Montecarlo process to evaluate realistic uncertainties. Thus, for each correlation that was tested, we randomly generated $10\,000$ subsamples from the original sample, calculating both $r$ and $r_{\textrm{S}}$ for each subsample. To summarise all this information, we present the mean and sigma values obtained from the $10\,000$ subsamples for both correlation coefficients. This method is very robust and also gives a good measurement of the uncertainty associated to our correlation coefficients. We also provide the correlation coefficients for the whole original sample (i.e. the results of a standard statistical test), which are systematically higher than those calculated through the Montecarlo process, because this process is evaluating all the uncertainties hidden in our data. These results for the whole samples can be considered as upper boundaries. 

\section{Results}

\subsection{Spectral type and atomic features}
\label{SpT_atomic}

The three main physical magnitudes that determine the presence and intensity of atomic lines in any of our spectra are temperature, luminosity and metallicity. Traditionally, for most stars, SpT has been used as a direct proxy for temperature, with later type stars being cooler than those with earlier types. However, under the light of the works discussed above, this idea has to be revisited for CSGs. \cite{dav2013} proposed that the SpT is mainly determined by luminosity. Since the depth of TiO bands has a dependence on luminosity, and SpT is assigned on the basis of the strength of these bands, there must certainly be a dependence between SpT and luminosity. Moreover, TiO bands form in the upper layers of the atmosphere, where extension, molecular opacities and other effects that are poorly understood result in complex radial temperature profiles \citep[and references therein]{dav2013}. For this reason, in this work we concentrate on atomic features, which form in deeper layers, where 3D models suggest that the temperature structure is close to that assumed in the simpler 1D models that are generally used to simulate the atmospheres of RSGs. Naturally, if SpT and luminosity are related, we cannot count on finding any spectral feature that will react only to effective temperature. However, not all lines depend on luminosity, temperature and metallicity in the same way. This implies that the global behaviour of a given set of lines will not be the same if a combination of luminosity and metallicity is the main effect behind atomic line behaviour along the SpT sequence (thus determining SpT, as \citealt{dav2013} suggest) or if temperature is the main contributor to SpT. Moreover, \citet{gaz2014} have shown that a few diagnostic atomic lines (of Si\,{\sc{i}}, Fe\,{\sc{i}}, and Ti\,{\sc{i}}) are enough to derive accurate parameters for RSGs.

\begin{figure*}[th!]
   \centering
   \includegraphics[trim=1cm 0.3cm 2cm 1.2cm,clip,width=9cm]{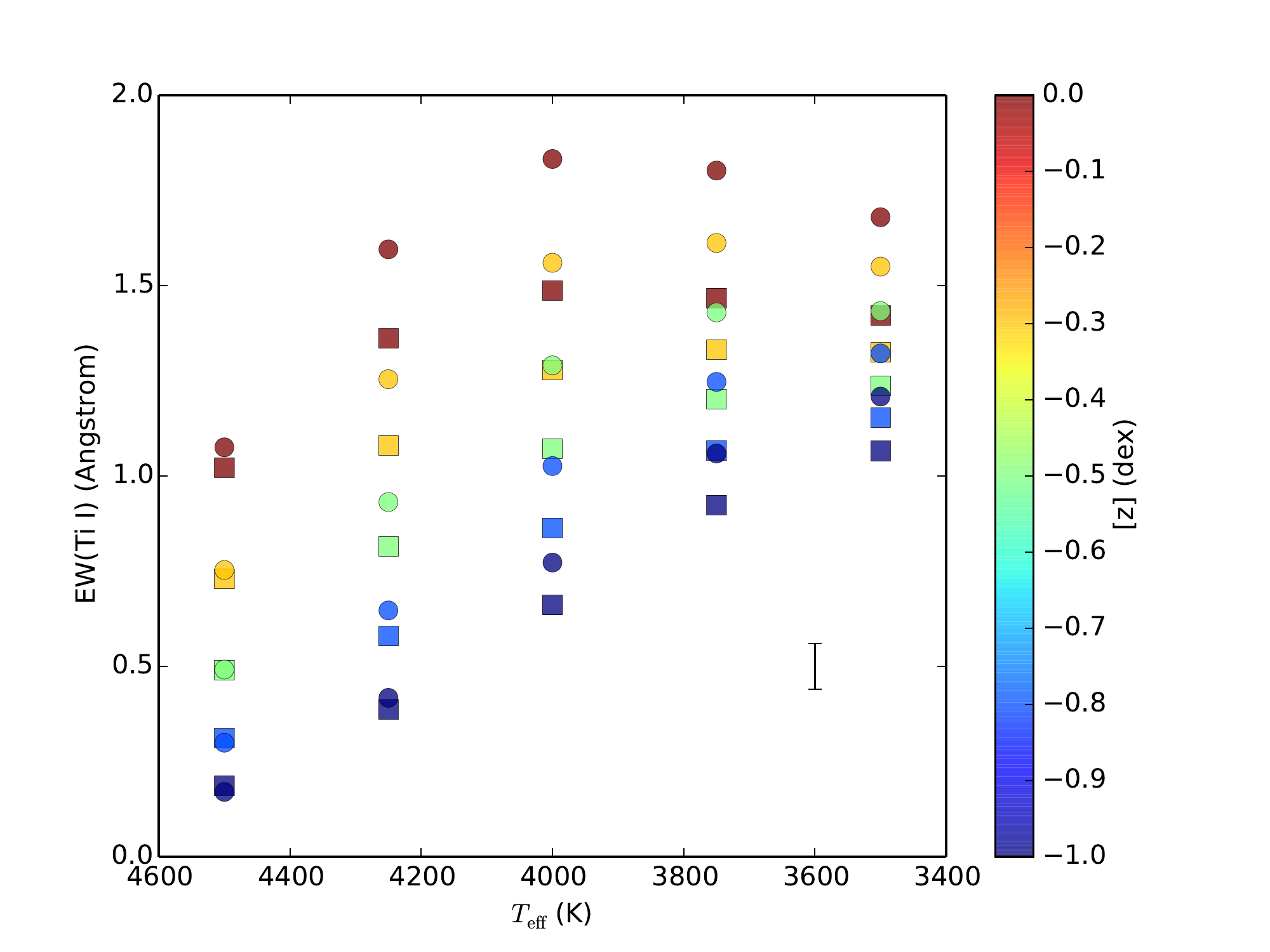}
   \includegraphics[trim=1cm 0.3cm 2cm 1.2cm,clip,width=9cm]{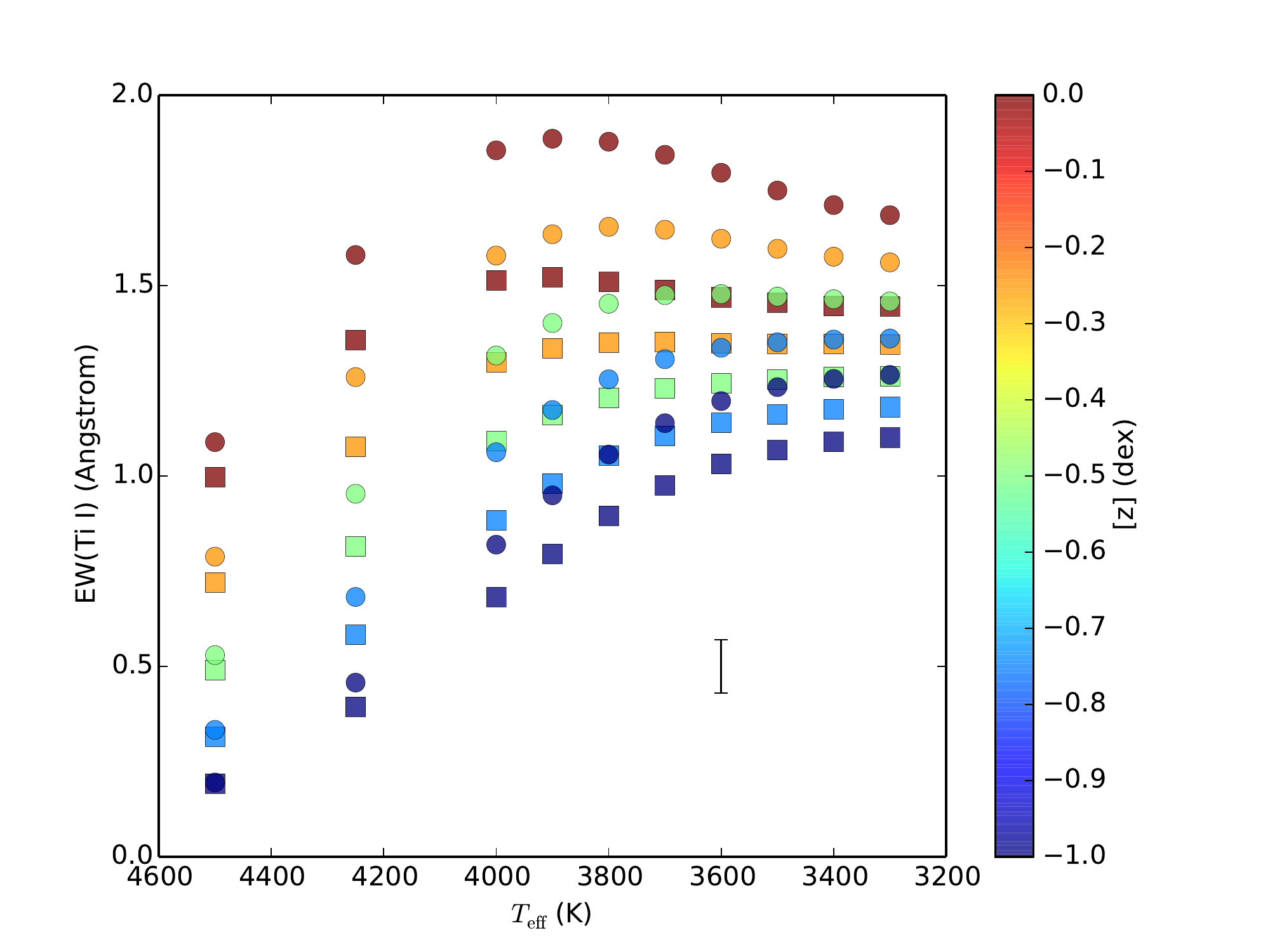}
   \caption{EW(Ti\,{\sc{i}}) index measured in two grids of synthetic spectra based on KURUCZ (left) and MARCS (right) atmospheric models. The $x$-axis shows effective temperature (with inverted scale to ease the comparison with the SpT sequence), the colours indicate metallicity [$Z$] and the shapes indicate surface gravity (circles are $\log\,g=0.0$ and squares $\log\,g=1.0$~dex) of each synthetic spectrum from the grid. The vertical bar represents the median error in EW(Ti\,{\sc{i}}) measurements.} 
   	\label{syn_Ti}
\end{figure*}

\begin{figure*}[th!]
   \centering
   \includegraphics[trim=1cm 0.3cm 2cm 1.2cm,clip,width=9cm]{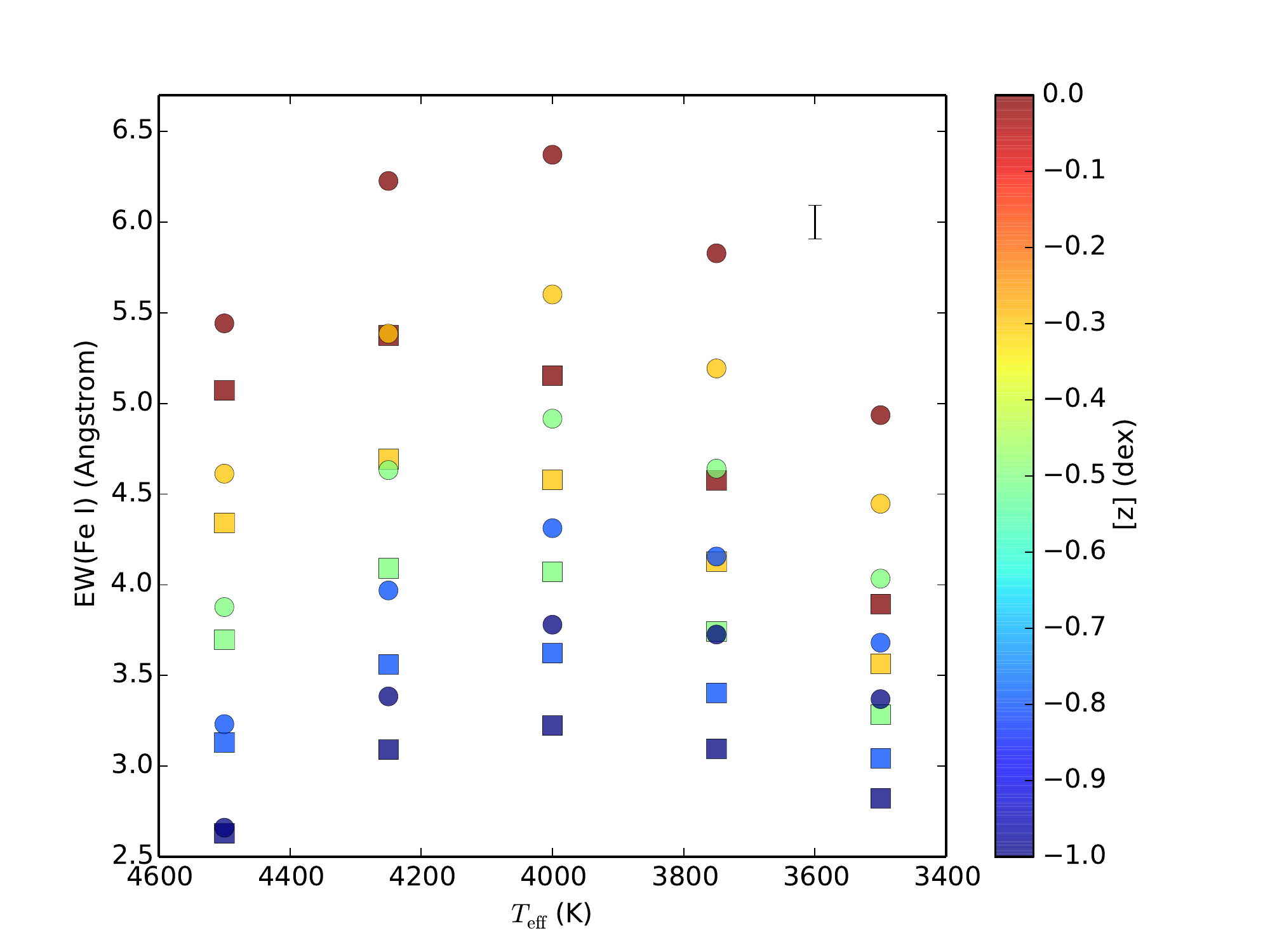}
   \includegraphics[trim=1cm 0.3cm 2cm 1.2cm,clip,width=9cm]{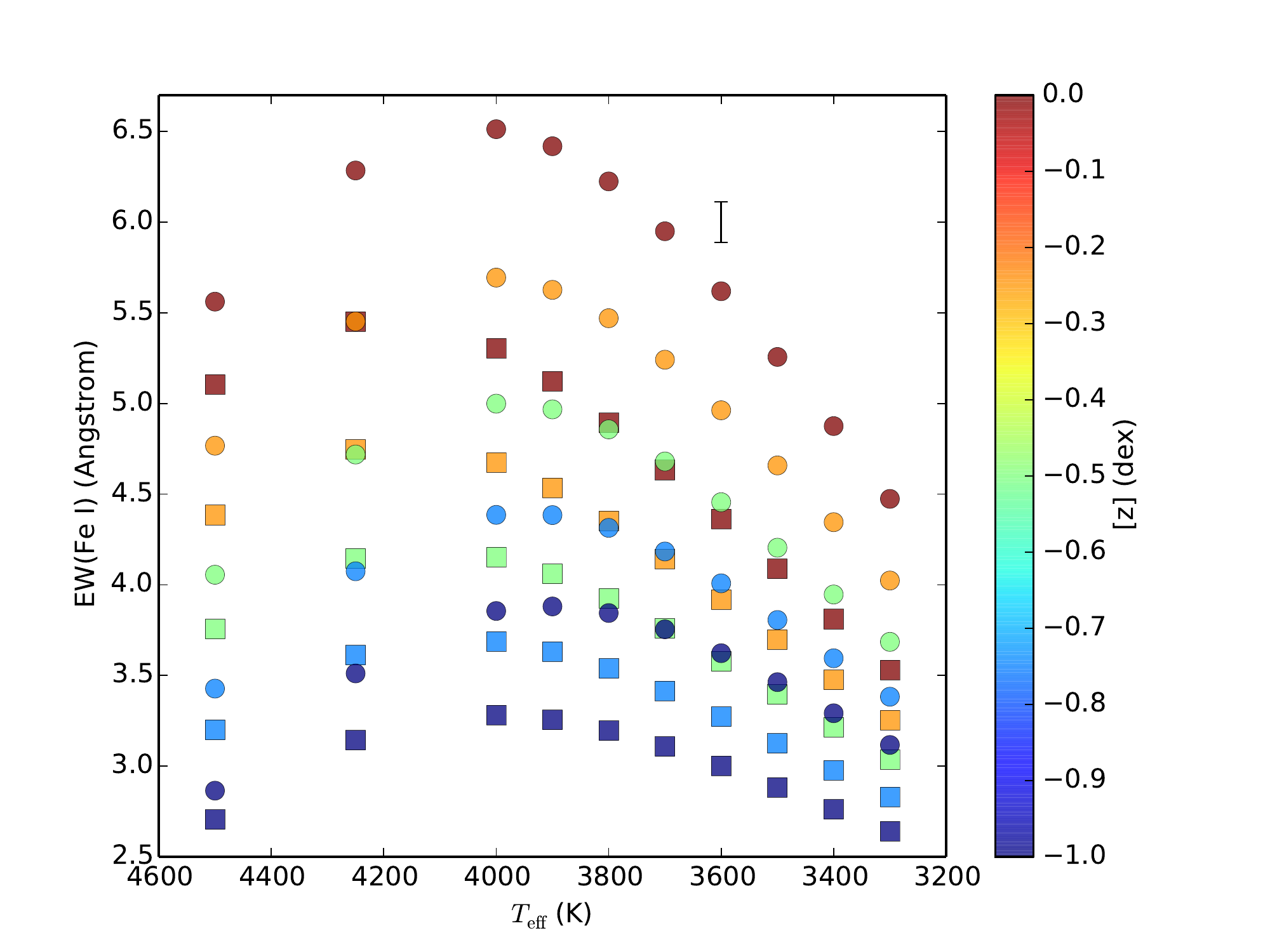}
   \caption{EW(Fe\,{\sc{i}}) index measured in two grids of synthetic spectra based on KURUCZ (left) and MARCS (right) atmospheric models. The display is the same as in Fig.~\ref{syn_Ti}.}
   \label{syn_Fe}
\end{figure*}

In the past, Ti\,{\sc{i}} lines in the CaT spectral region have been used as SpT indicators \citep[e.g.][]{gin1994} because they are very sensitive to temperature, since Ti is a light element and they have low excitation potentials ($\chi_{e}\leqslant1$\:eV). Fe\,{\sc{i}} lines have also been used as SpT indicators, even if they are less sensitive to temperature than Ti\,{\sc{i}} (as Fe is heavier than Ti and its lower excitation potential is closer to the fundamental level), because there are many more intense Fe\,{\sc{i}} lines than \ion{Ti}{i} lines in the CaT spectral region. These lines, on the other hand, are not very sensitive to luminosity (i.e.\ surface gravity), although Fe\,{\sc{i}} lines are more sensitive than Ti\,{\sc{i}} ones. Because of this, some ratios of nearby Fe\,{\sc{i}} and Ti\,{\sc{i}} lines have been used as luminosity criteria \citep[e.g.\ Fe\,{\sc{i}}~$8514\:$\AA{} to Ti\,{\sc{i}}~$8518\:$\AA{} in][]{kee1945}. \textcolor{blue}{To help us understand} how the Ti\,{\sc{i}} and Fe\,{\sc{i}} lines that we have used for our indices depend on temperature, surface gravity (i.e.\ luminosity) and metallicity, we have measured these lines in a grid of synthetic spectra generated using KURUCZ and MARCS stellar atmospheric models (see Section~\ref{syntetic}), following the same procedure used for the observed spectra. The results are shown in Figs.~\ref{syn_Ti} and \ref{syn_Fe}.

\begin{figure*}[th!]
   \centering
   \includegraphics[trim=1cm 0.3cm 2cm 1.2cm,clip,width=9cm]{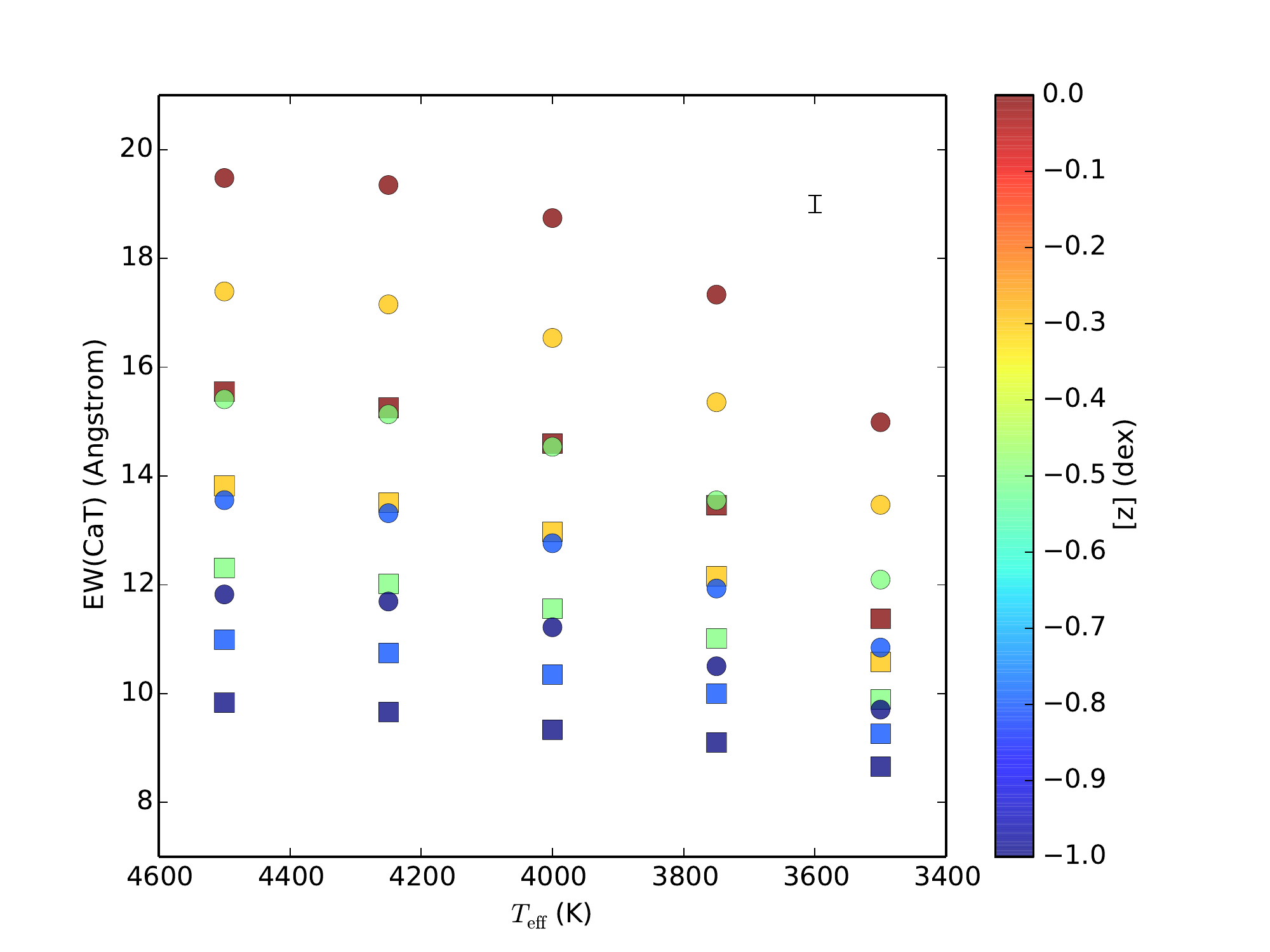}   \includegraphics[trim=1cm 0.3cm 2cm 1.2cm,clip,width=9cm]{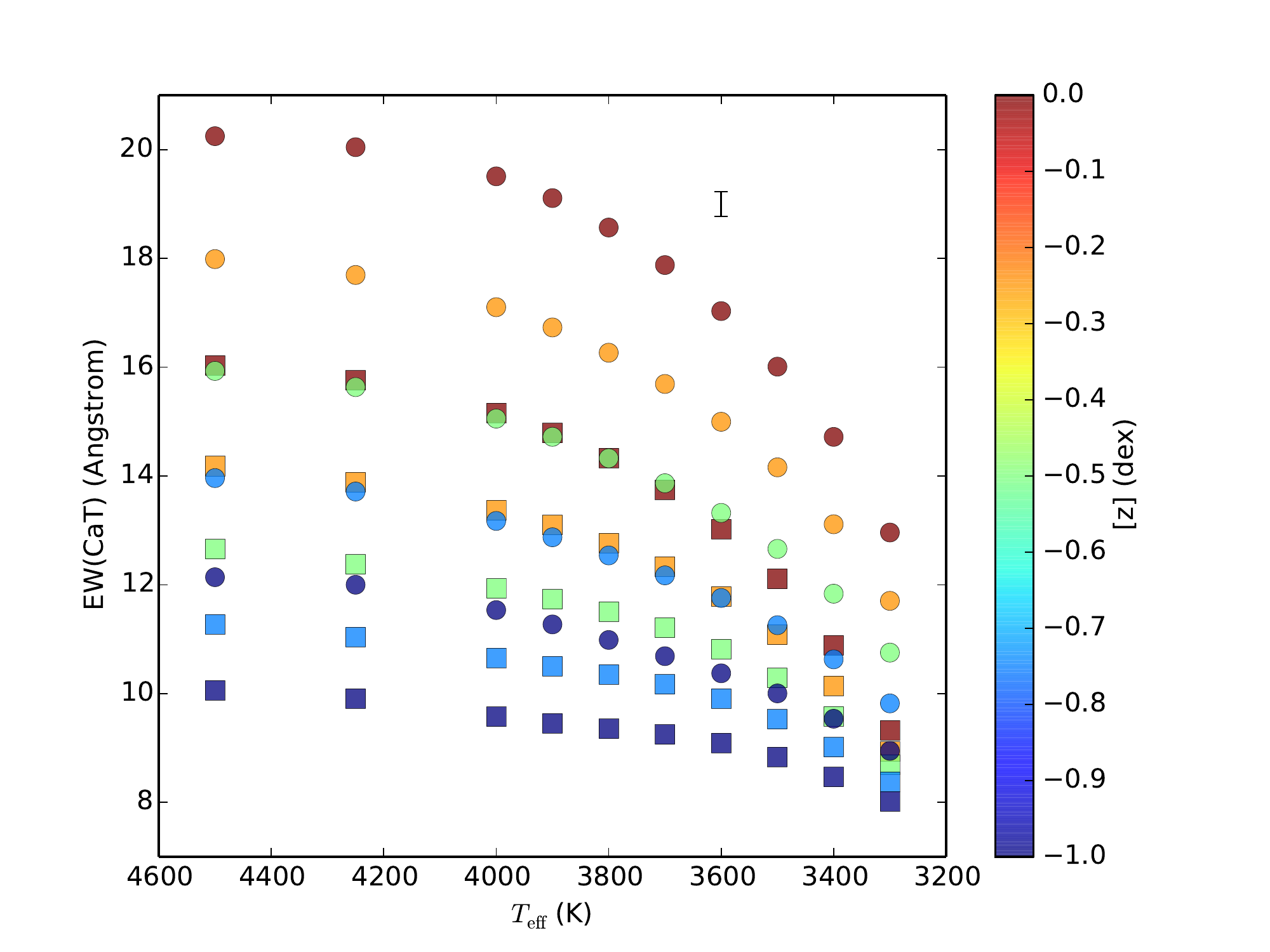}
   \caption{EW(CaT) index measured in two grids of synthetic spectra based on KURUCZ (left) and MARCS (right) atmospheric models. The display is the same as in Fig.~\ref{syn_Ti}.}
   \label{syn_CaT}
\end{figure*}

For the grid of synthetic models, we have chosen a temperature range based on the typical temperatures derived for SGs in previous works (see Section~\ref{intro}). Nevertheless, we have not reached temperatures below $3300$\:K in MARCS and $3500$\:K in KURUCZ because of model limitations (see Section~\ref{syntetic}). Moreover, temperatures lower than these would correspond (according to the published effective temperature scales, see Section~\ref{intro}) to mid- to late-M SpTs. We have not measured the intensity of lines in stars later than M3 in the observed spectra, because the continuum becomes heavily affected by TiO bands and atomic lines do not display their true behaviour, but only the effect of the molecular bands over them.

Finally, we remark that we have not related the model temperatures with the SpTs predicted by any of the effective temperature scales discussed before, as we only want to explore the behaviour of the lines.

Our synthetic spectra include molecular features, which have a small effect over our lines at temperatures higher than $4\,000$\:K, but introduce more important differences with respect to models without molecular features at lower temperatures. To understand these effects, we also evaluated synthetic spectra generated without molecular features, finding that the behaviour of the lines measured does not change qualitatively. Their addition affects the EWs measured by changing slightly their values, specially decreasing them at temperatures lower than $4000$\:K, and increasing sightly their sensitivity to luminosity. As can be seen in Fig.~\ref{syn_Ti}, EW(Ti\,{\sc{i}}) in synthetic spectra has a clear linear dependence with temperature down to $\sim4\,000$\:K at Solar metallicity, and down to lower temperatures at lower metallicities. From there the slope starts to decrease as the temperature drops further, because the lines used are coming close to the saturation part of the curve of growth, and also because the effect of the molecular bands. The EW(Fe\,{\sc{i}}) index also has a linear trend with temperatures down to $\sim4\,000$\:K, but with a slope lower than EW(Ti\,{\sc{i}}). Moreover, EW(Fe\,{\sc{i}}) starts to decrease for temperatures lower than $\sim4\,000$\:K. EW(Ti\,{\sc{i}}) shows little dependence on surface gravity and it is clear that the Fe\,{\sc{i}} lines are more sensitive than the Ti\,{\sc{i}} ones, justifying the use of \ion{Fe}{i}/\ion{Ti}{i} ratios as luminosity indicators.

The CaT is very sensitive to luminosity \citep[e.g.][]{dia1989}, and it has been widely used to separate SGs from other less luminous stars \citep[e.g.][]{gin1994}. In synthetic spectra, the EW(CaT) index shows a strong dependence on surface gravity (see Fig.~\ref{syn_CaT}). Its dependence on effective temperature is much weaker, and can be described as a slow decrease of EW(CaT) as temperature drops.

\begin{figure*}[th!]
   \centering
   \includegraphics[trim=1cm 0.5cm 2cm 1.2cm,clip,width=9cm]{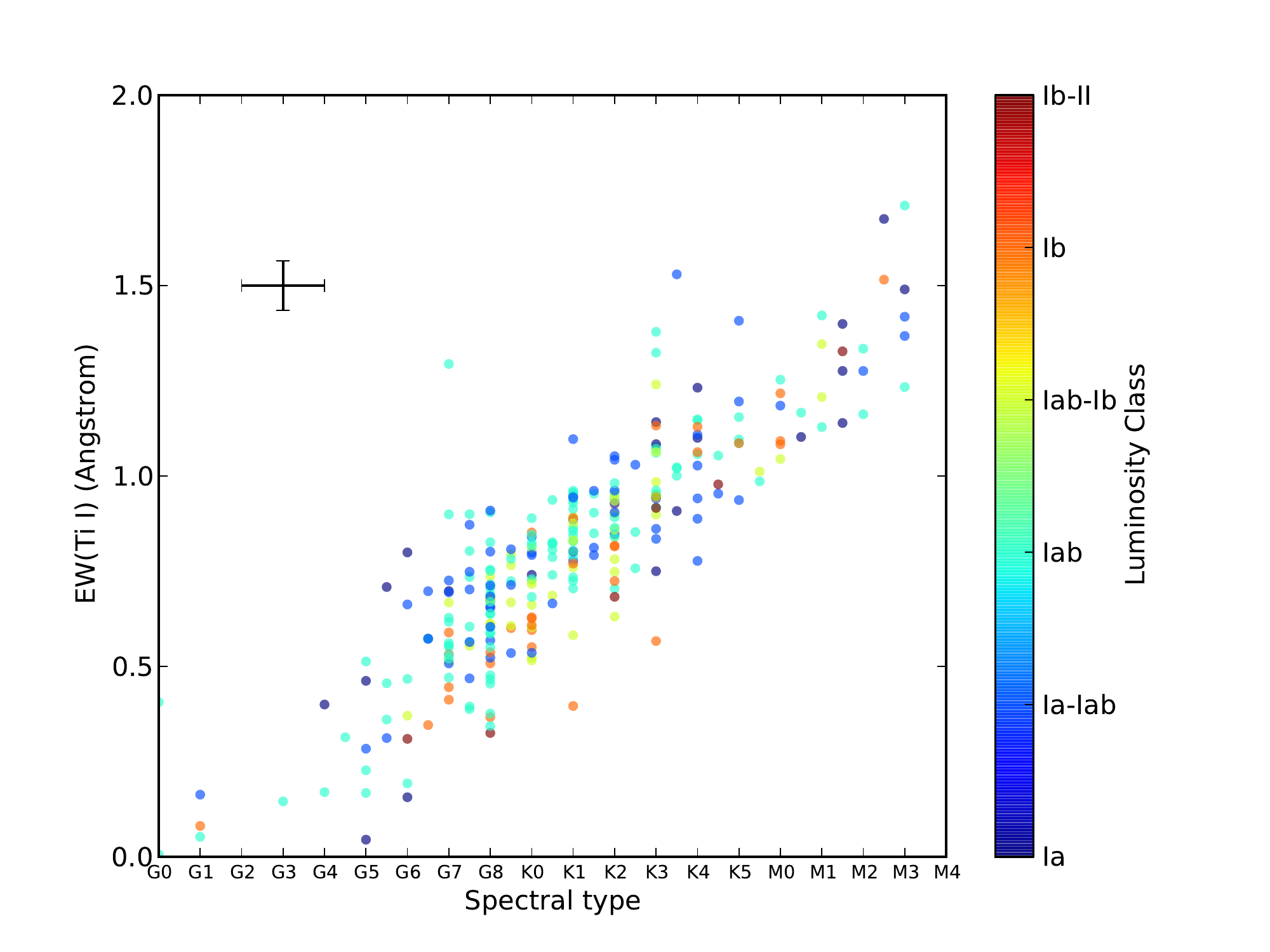}
   \includegraphics[trim=1cm 0.5cm 2cm 1.2cm,clip,width=9cm]{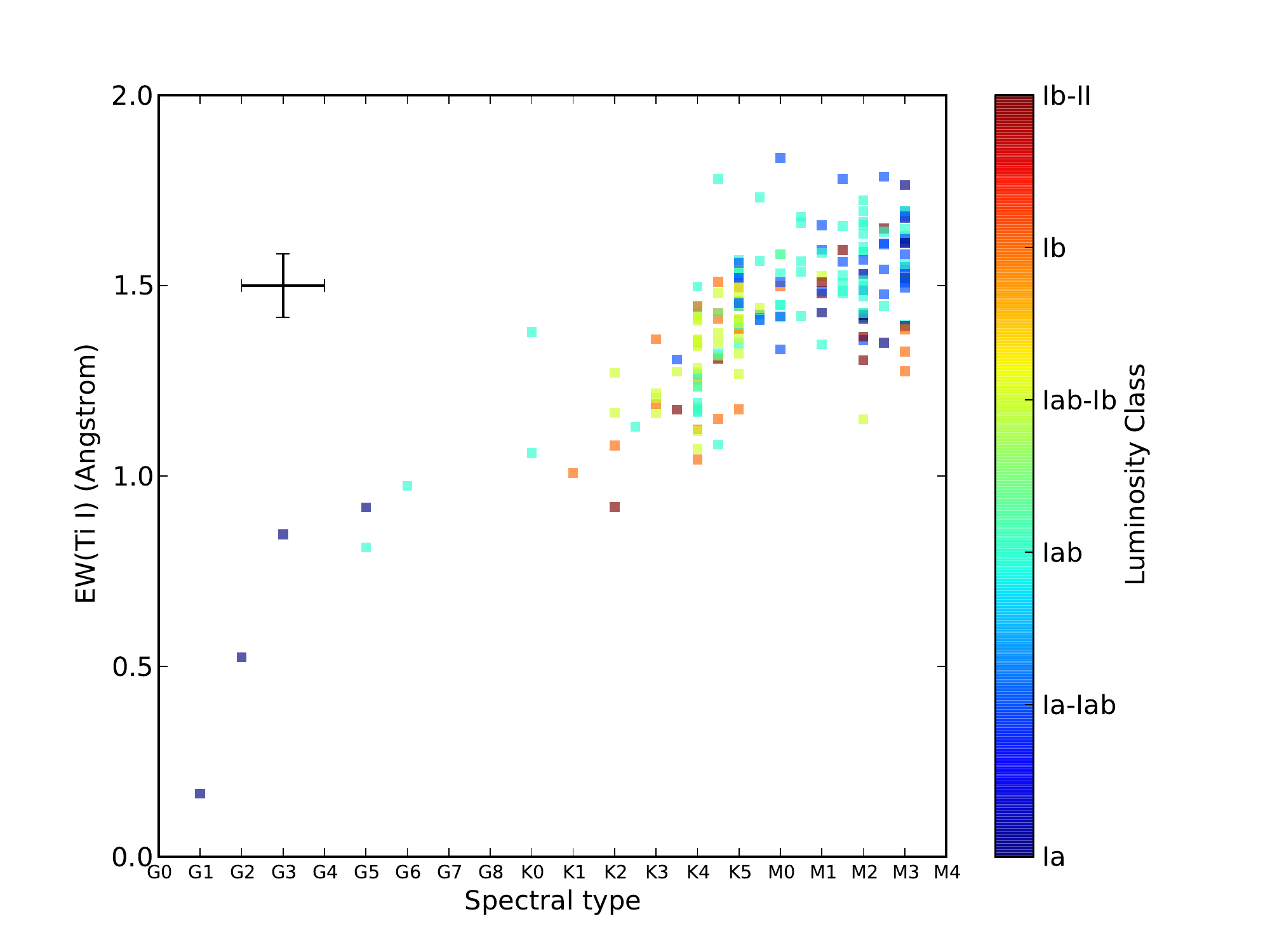}
   \caption{Sum of Ti\,{\sc{i}} equivalent widths against spectral type. The colour indicates luminosity class. The black cross represents the median uncertainties. The LMC data correspond to 2013 and the SMC data correspond to 2012, because all the stars observed in 2010 and 2011 were also observed in 2012 and 2013, and so each star is represented only once. Note that these figures present the same variables than in Fig.~\ref{spt_example}, but here we have split the data from each galaxy for clarity, easing the comparison with Figs.~\ref{SpT_Fe} and Figs.~\ref{SpT_CaT}, and they do not include those SGs later than M3, as their measurements are compromised by the TiO bands. Note also that both figures are on the same scale to make comparison easier. {\bf Left (\ref{SpT_Ti}a):} CSGs from the SMC.  {\bf Right (\ref{SpT_Ti}b):} CSGs from the LMC.}
   \label{SpT_Ti}
\end{figure*}

\begin{figure*}[th!]
   \centering
   \includegraphics[trim=1cm 0.5cm 2cm 1.2cm,clip,width=9cm]{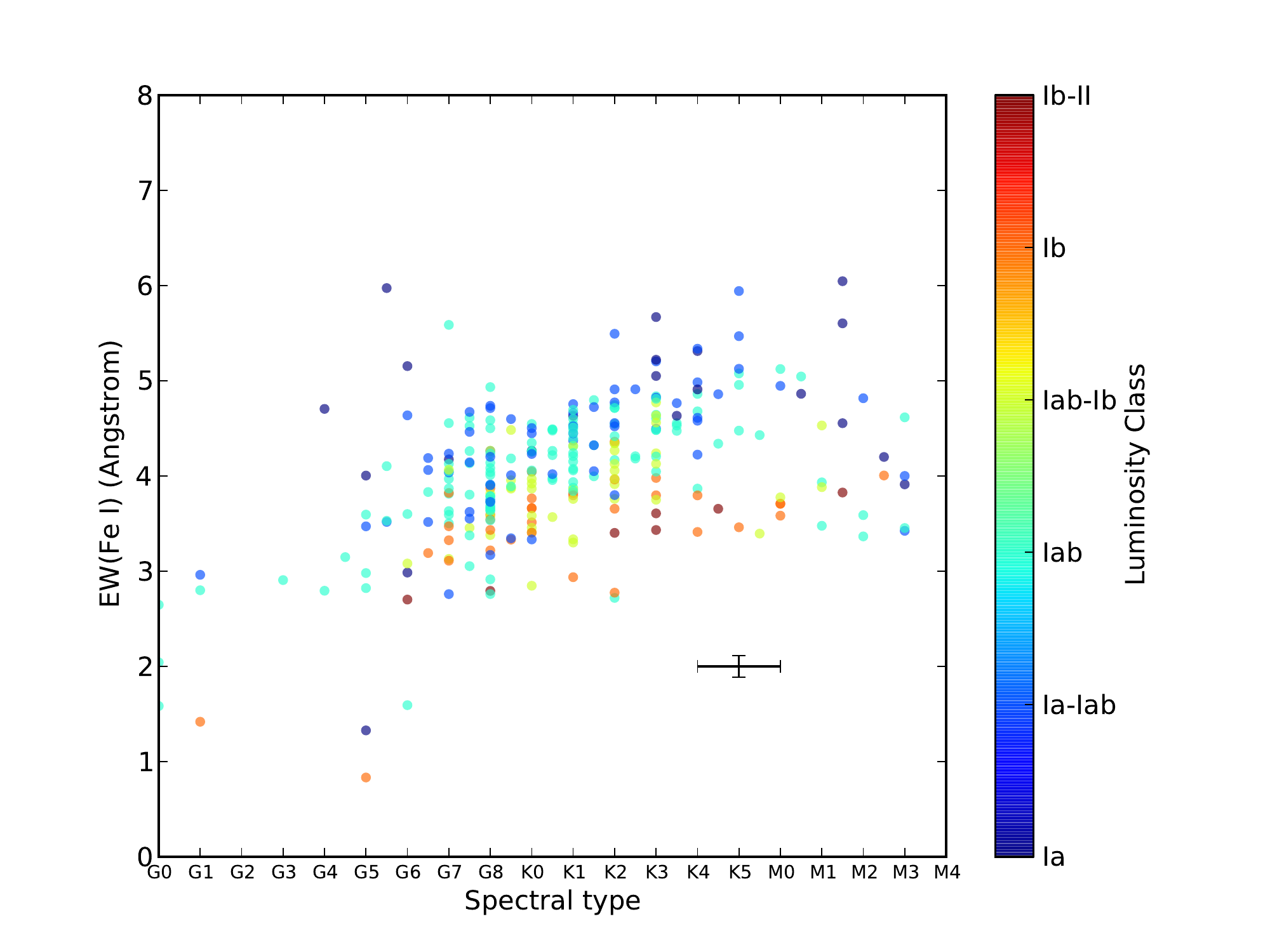}
   \includegraphics[trim=1cm 0.5cm 2cm 1.2cm,clip,width=9cm]{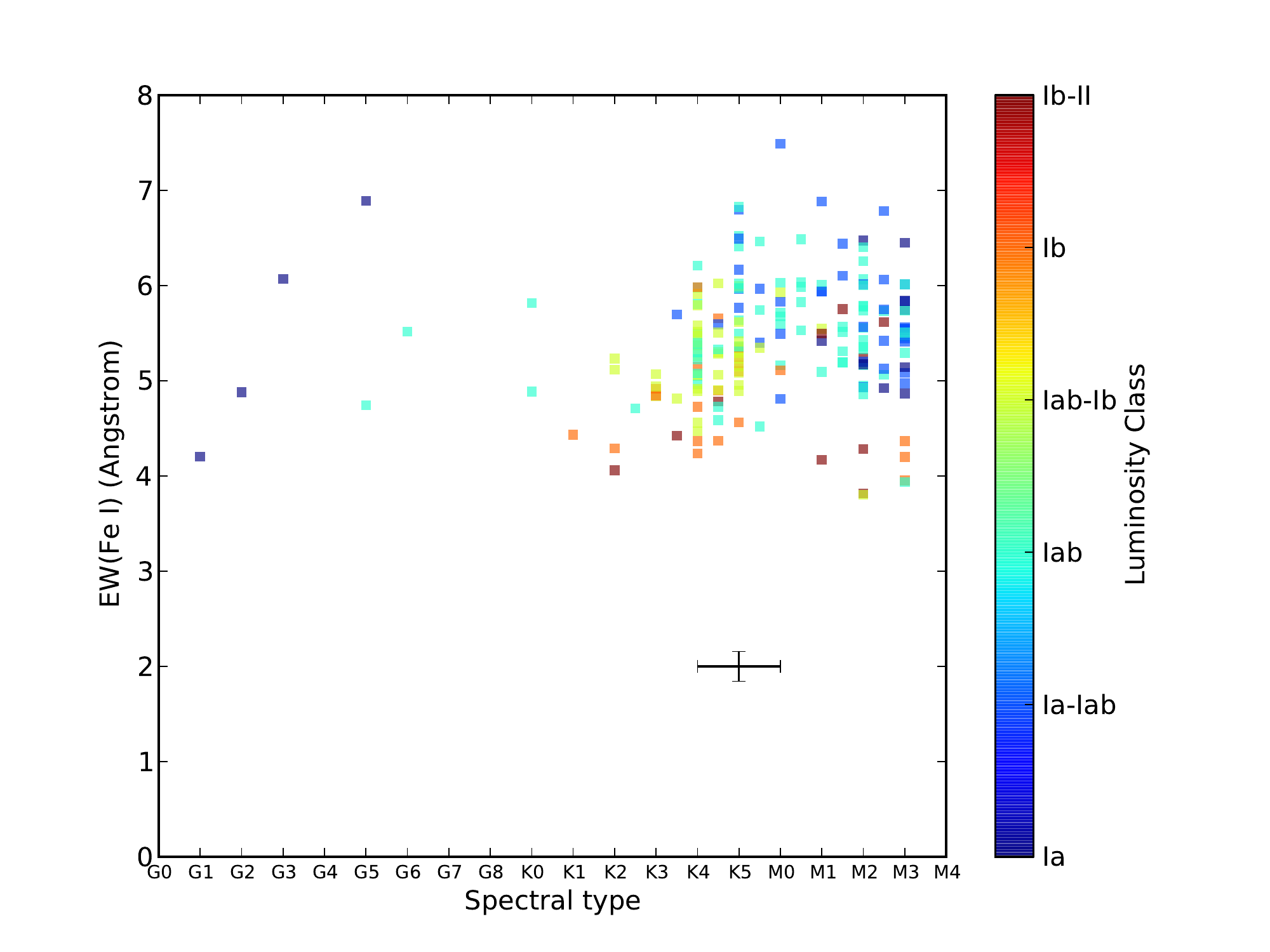}
   \caption{Sum of Fe\,{\sc{i}} equivalent widths against spectral type. The display is the same as in Fig.~\ref{SpT_Ti}. {\bf Left (\ref{SpT_Fe}a):} CSGs from the SMC.  {\bf Right (\ref{SpT_Fe}b):} CSGs from the LMC.}
   \label{SpT_Fe}
\end{figure*}

\begin{figure*}[th!]
   \centering
   \includegraphics[trim=1cm 0.5cm 2cm 1.2cm,clip,width=9cm]{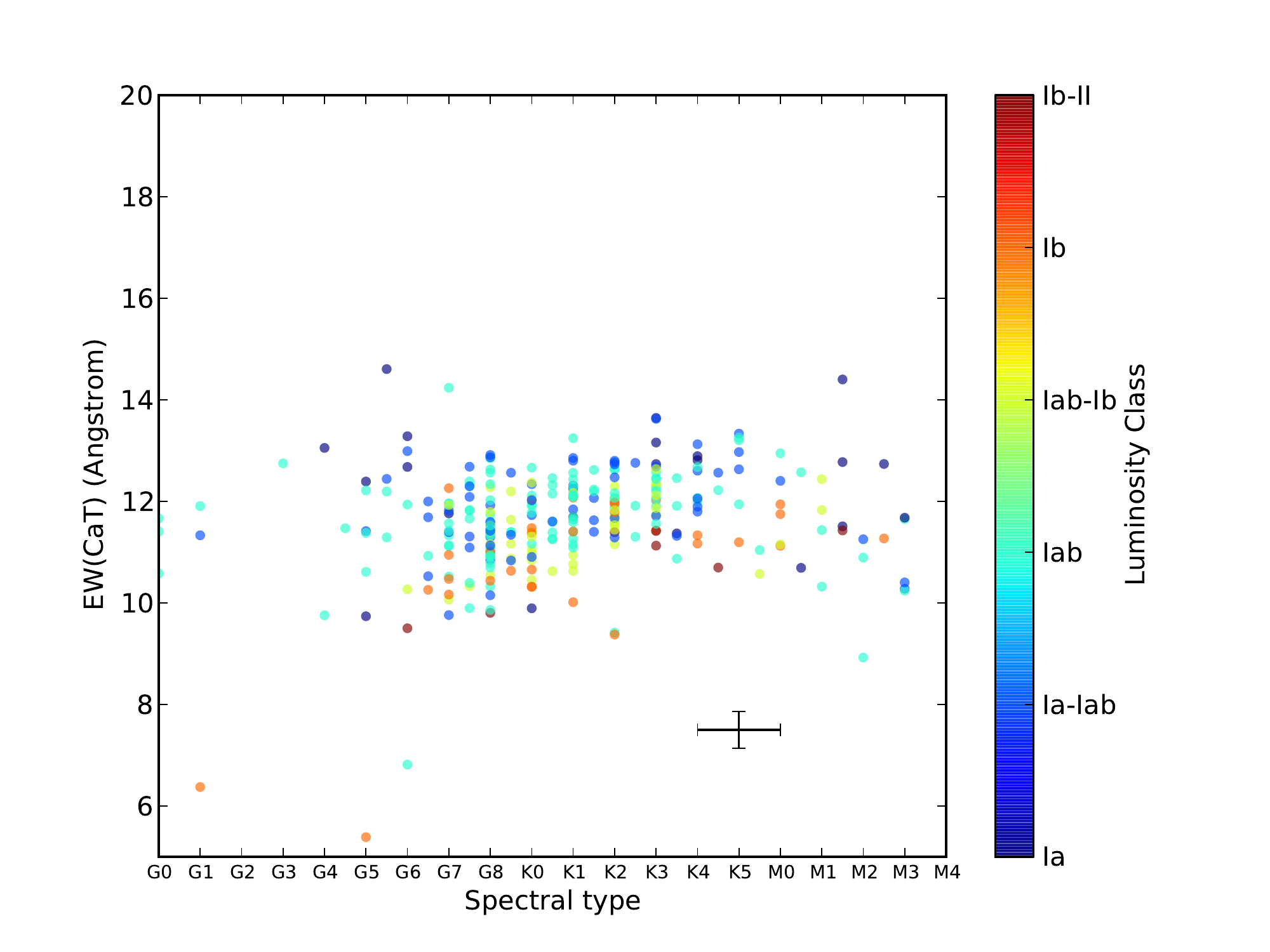}
   \includegraphics[trim=1cm 0.5cm 2cm 1.2cm,clip,width=9cm]{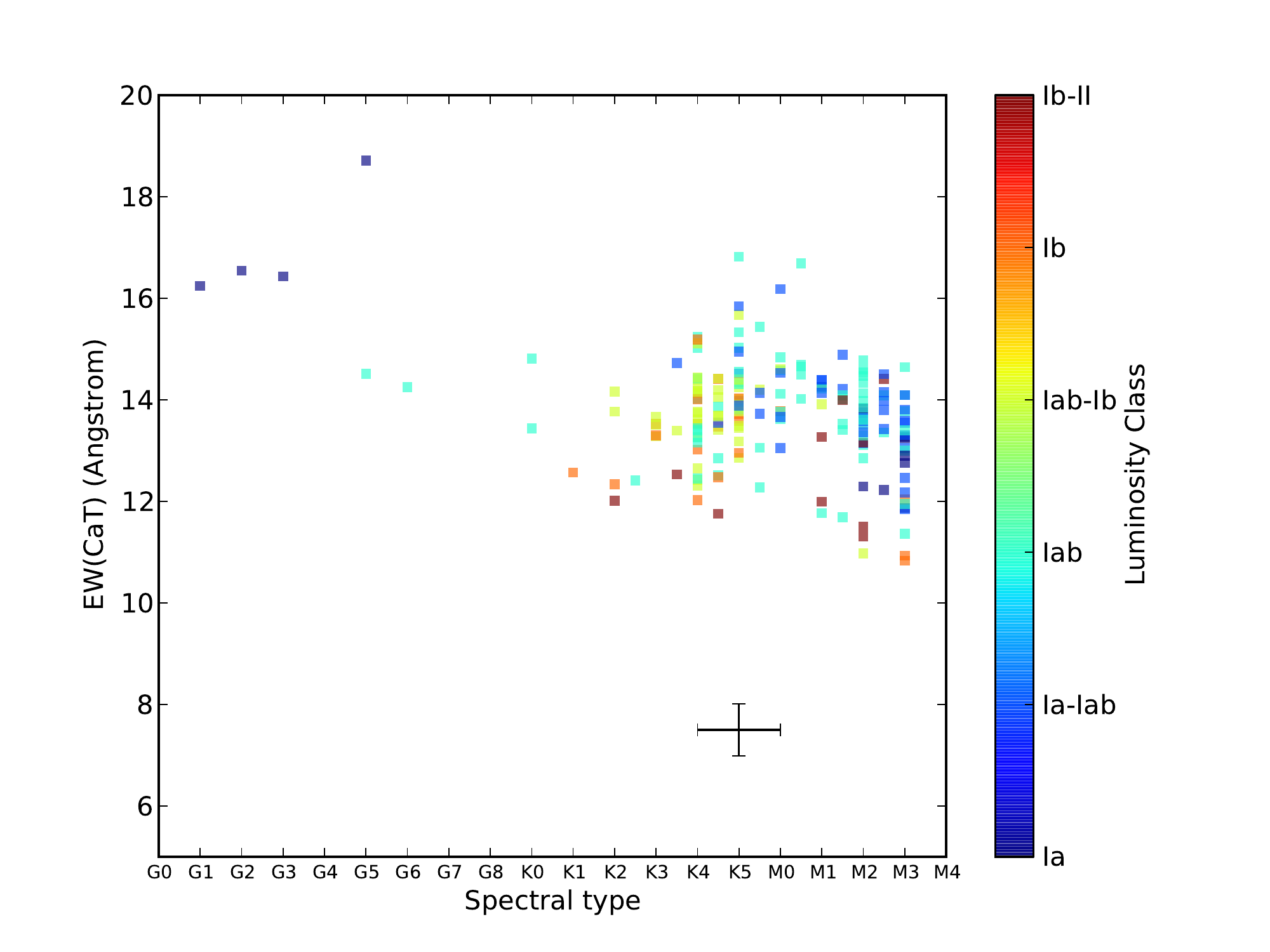}
   \caption{Sum of CaT equivalent widths against spectral type. The display is the same as in Fig.~\ref{SpT_Ti}. {\bf Left (\ref{SpT_CaT}a):} CSGs from the SMC. {\bf Right (\ref{SpT_CaT}b):} CSGs from the LMC.}
   \label{SpT_CaT}
\end{figure*}

The measurements of EW(Ti\,{\sc{i}}) and EW(Fe\,{\sc{i}}) indices derived from our observed spectra are shown in Figs.~\ref{SpT_Ti} and \ref{SpT_Fe}. We calculated the correlation coefficients between the SpT and these indices for each galaxy through the method explained in Section~\ref{corr}. The results are shown in Table~\ref{corr_spt_atom}. The EW(Ti\,{\sc{i}}) index presents a very clear linear positive trend with SpT from G0 down to the point where the lines become too affected by TiO bands at $\sim$M3 to be correctly measured. EW(Fe\,{\sc{i}}) presents a not very strong, but still significant, linear positive trend with SpT in the SMC, while in the LMC the trend is almost flat. Such trends would be in good accord with the behaviour observed in the synthetic spectra, if the SpT sequence depends mainly on temperature. In fact, if the SpT sequence should depend mostly on luminosity, we would find no clear correlation between EW(Ti\,{\sc{i}}) and SpT, as Ti\,{\sc{i}} lines are quite insensitive to surface gravity. Moreover, EW(Fe\,{\sc{i}}) should have a stronger correlation with SpT than EW(Ti\,{\sc{i}}), as it is more sensitive to luminosity. Finally, the behaviour of EW(CaT) with SpT is almost flat (Fig.~\ref{SpT_CaT}), -- its correlation coefficients are low and positive for the SMC, but low and negative for the LMC -- implying that it does not depend strongly on SpT, as it should do if the SpT sequence would be determined by luminosity.

\begin{table*}[th!]
\caption{Pearson ($r$) and Spearman ($r_{\textrm{S}}$) coefficients obtained for the correlations between different pairs of variables from the data of each galaxy. The values given by Montecarlo are the mean ones and their corresponding standard deviations. The details of the Montecarlo process we used are explained in Sect.~\ref{corr}. We also provide the correlation coefficients obtained for the original samples (without Montecarlo).}
\label{corr_spt_atom}
\centering
\begin{tabular}{ c c | c | c c | c c }
\hline\hline
\noalign{\smallskip}
\multicolumn{2}{c|}{Variables correlated}&Galaxy&\multicolumn{2}{c|}{Mean coefficients from Montecarlo}&\multicolumn{2}{c}{From the original sample}\\
X&Y&&$r\pm\sigma_{\textrm{P}}$&$r_{\textrm{S}}\pm\sigma_{\textrm{S}}$&$r$&$r_{\textrm{S}}$\\
\noalign{\smallskip}
\hline
\noalign{\smallskip}
Spectral type&EW(Ti\,{\sc{i}})&SMC&$0.815\pm0.012$&$0.793\pm0.015$&$0.876$&$0.863$\\
Spectral type&EW(Ti\,{\sc{i}})&LMC&$0.69\pm0.02$&$0.57\pm0.04$&$0.79$&$0.70$\\
Spectral type&EW(Fe\,{\sc{i}})&SMC&$0.462\pm0.019$&$0.42\pm0.02$&$0.490$&$0.44$\\
Spectral type&EW(Fe\,{\sc{i}})&LMC&$0.20\pm0.03$&$0.18\pm0.03$&$0.21$&$0.22$\\
Spectral type&EW(CaT)&SMC&$0.18\pm0.03$&$0.20\pm0.03$&$0.21$&$0.23$\\
Spectral type&EW(CaT)&LMC&$-0.23\pm0.04$&$-0.15\pm0.04$&$-0.26$&$-0.17$\\
\noalign{\smallskip}
\hline
\noalign{\smallskip}
EW(Ti\,{\sc{i}})&$M_{\textrm{bol}}$&SMC&$-0.02\pm0.04$&$-0.223\pm0.015$&$-0.129$&$-0.234$\\
EW(Ti\,{\sc{i}})&$M_{\textrm{bol}} (<-6$~mag$)$&SMC&$-0.214\pm0.019$&$-0.304\pm0.019$&$-0.222$&$-0.323$\\
EW(Ti\,{\sc{i}})&$M_{\textrm{bol}}$&LMC&$-0.30\pm0.03$&$-0.46\pm0.03$&$-0.33$&$-0.53$\\
EW(Fe\,{\sc{i}})&$M_{\textrm{bol}}$&SMC&$-0.09\pm0.11$&$-0.674\pm0.012$&$-0.527$&$-0.704$\\
EW(Fe\,{\sc{i}})&$M_{\textrm{bol}} (<-6$~mag$)$&SMC&$-0.408\pm0.14$&$-0.619\pm0.014$&$-0.417$&$-0.645$\\
EW(Fe\,{\sc{i}})&$M_{\textrm{bol}}$&LMC&$-0.556\pm0.019$&$-0.529\pm0.019$&$-0.580$&$-0.556$\\
EW(CaT)&$M_{\textrm{bol}}$&SMC&$-0.063\pm0.08$&$-0.47\pm0.03$&$-0.392$&$-0.524$\\
EW(CaT)&$M_{\textrm{bol}} (<-6$~mag$)$&SMC&$-0.29\pm0.02$&$-0.43\pm0.03$&$-0.32$&$-0.48$\\
EW(CaT)&$M_{\textrm{bol}}$&LMC&$-0.36\pm0.03$&$-0.29\pm0.04$&$-0.40$&$-0.32$\\
\noalign{\smallskip}
\hline
\noalign{\smallskip}
$\Delta($EW(Ti\,{\sc{i}})$)$&$\Delta($Spectral type$)$&Both&$0.48\pm0.03$&$0.42\pm0.03$&$0.62$&$0.57$\\
$\Delta($EW(Fe\,{\sc{i}})$)$&$\Delta($Spectral type$)$&Both&$0.04\pm0.04$&$0.04\pm0.04$&$0.05$&$0.03$\\
$\Delta($EW(CaT)$)$&$\Delta($Spectral type$)$&Both&$-0.01\pm0.04$&$-0.07\pm0.04$&$-0.01$&$-0.09$\\
\noalign{\smallskip}
\hline
\noalign{\smallskip}
Spectral type&$M_{\textrm{bol}} (<-6$~mag$)$&SMC&$-0.17\pm0.02$&$-0.20\pm0.02$&$-0.18$&$-0.22$\\
Spectral type&$M_{\textrm{bol}} (<-6.7$~mag$)$&SMC&$-0.28\pm0.03$&$-0.31\pm0.03$&$-0.30$&$-0.34$\\
Spectral type\tablefootmark{a}&$M_{\textrm{bol}} (<-6$~mag$)$&LMC&$-0.37\pm0.03$&$-0.47\pm0.04$&$-0.40$&$-0.53$\\
\noalign{\smallskip}
\hline
\noalign{\smallskip}
\end{tabular}
\tablefoot{
\tablefoottext{a}{For this calculation, the few CSGs from the LMC earlier than G7 were treated as outliers and removed (see Fig.~\ref{SpT_Mbol}b).}
}
\end{table*}

\begin{figure*}[th!]
   \centering
   \includegraphics[trim=1cm 0.5cm 2cm 1.2cm,clip,width=9cm]{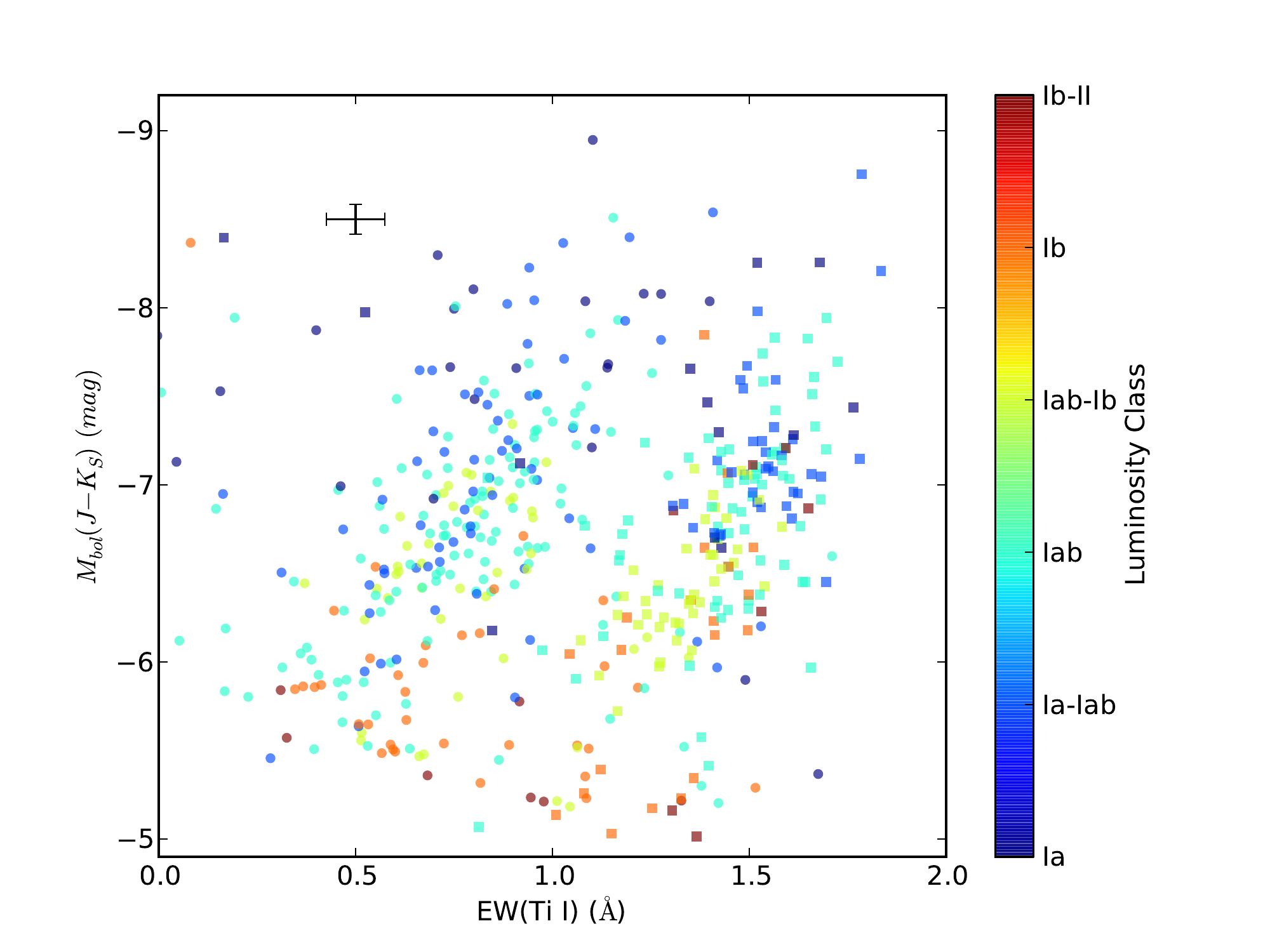}
   \includegraphics[trim=1cm 0.5cm 2cm 1.2cm,clip,width=9cm]{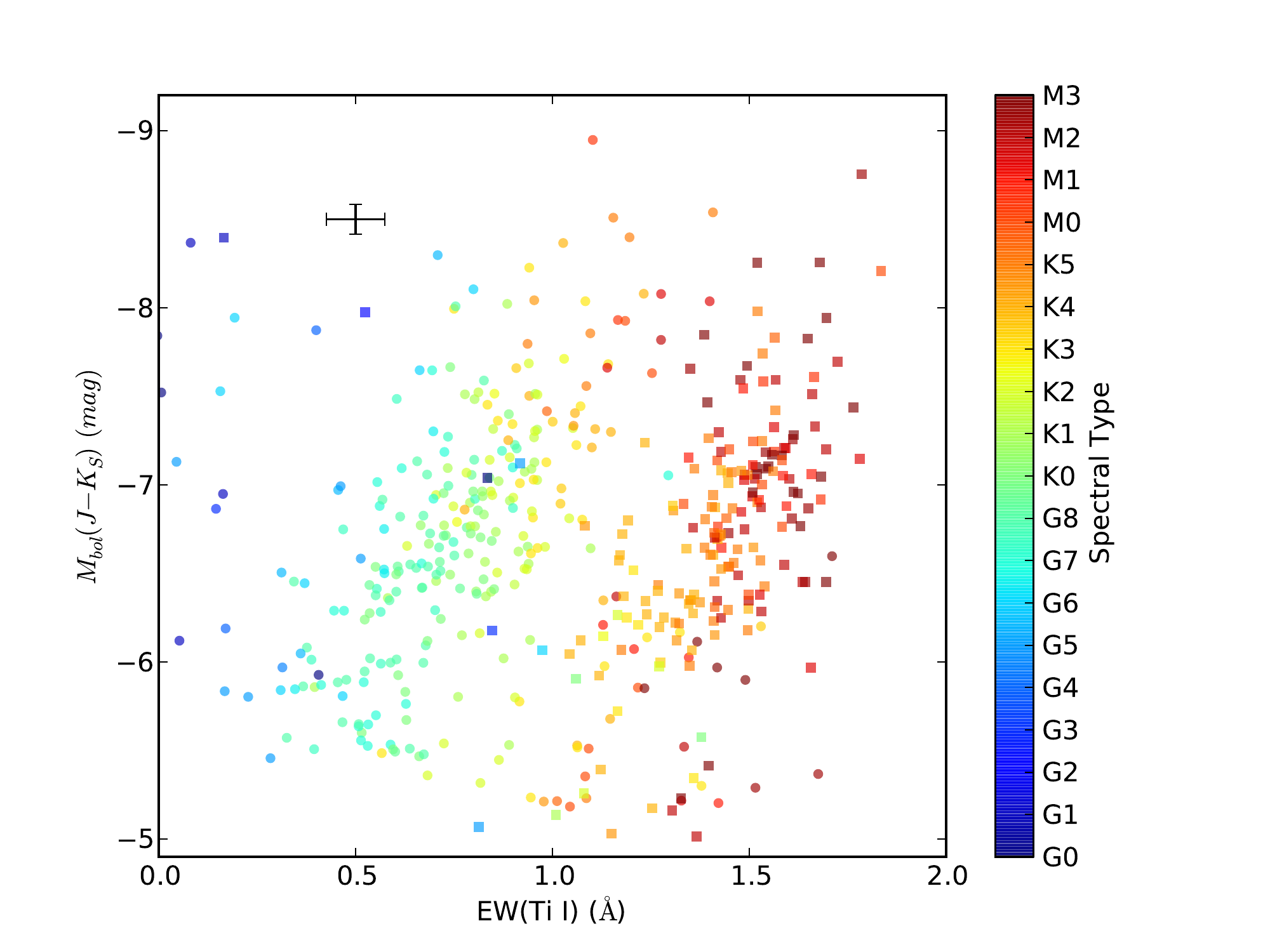}
   \caption{Sum of Ti\,{\sc{i}} equivalent widths against bolometric magnitude. The shapes indicate the host galaxy: LMC stars are squares, and SMC stars are circles. The LMC data correspond to 2013 and the SMC data correspond to 2012, as in Fig.~\ref{SpT_Ti}. The black cross represents the median uncertainties. {\bf Left (\ref{Ti_Mbol}a):} The colour indicates the LC.  {\bf Right (\ref{Ti_Mbol}b):} The colour indicates the SpT.}
   \label{Ti_Mbol}
\end{figure*}

\begin{figure*}[th!]
   \centering
   \includegraphics[trim=1cm 0.5cm 2cm 1.2cm,clip,width=9cm]{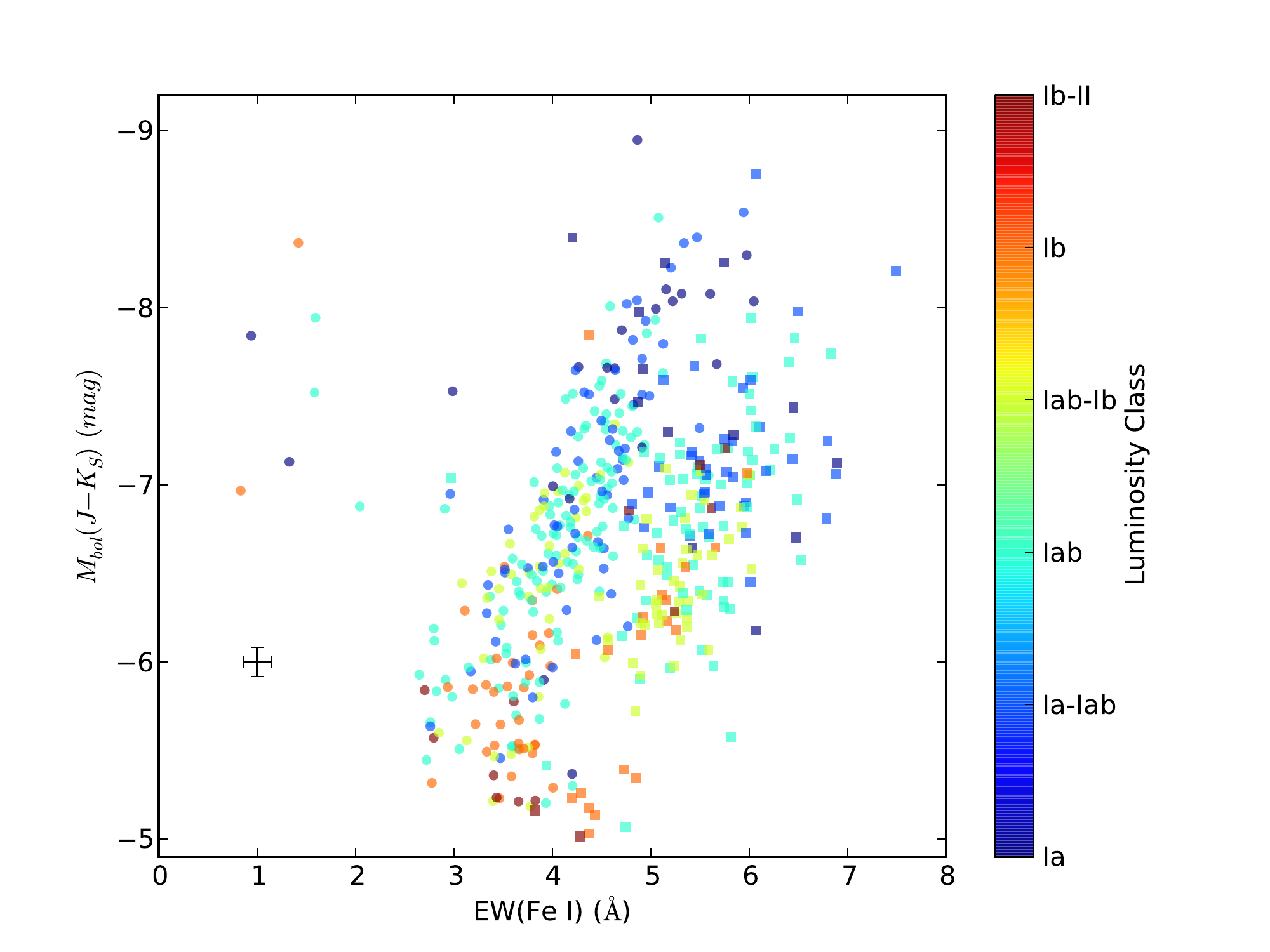}
   \includegraphics[trim=1cm 0.5cm 2cm 1.2cm,clip,width=9cm]{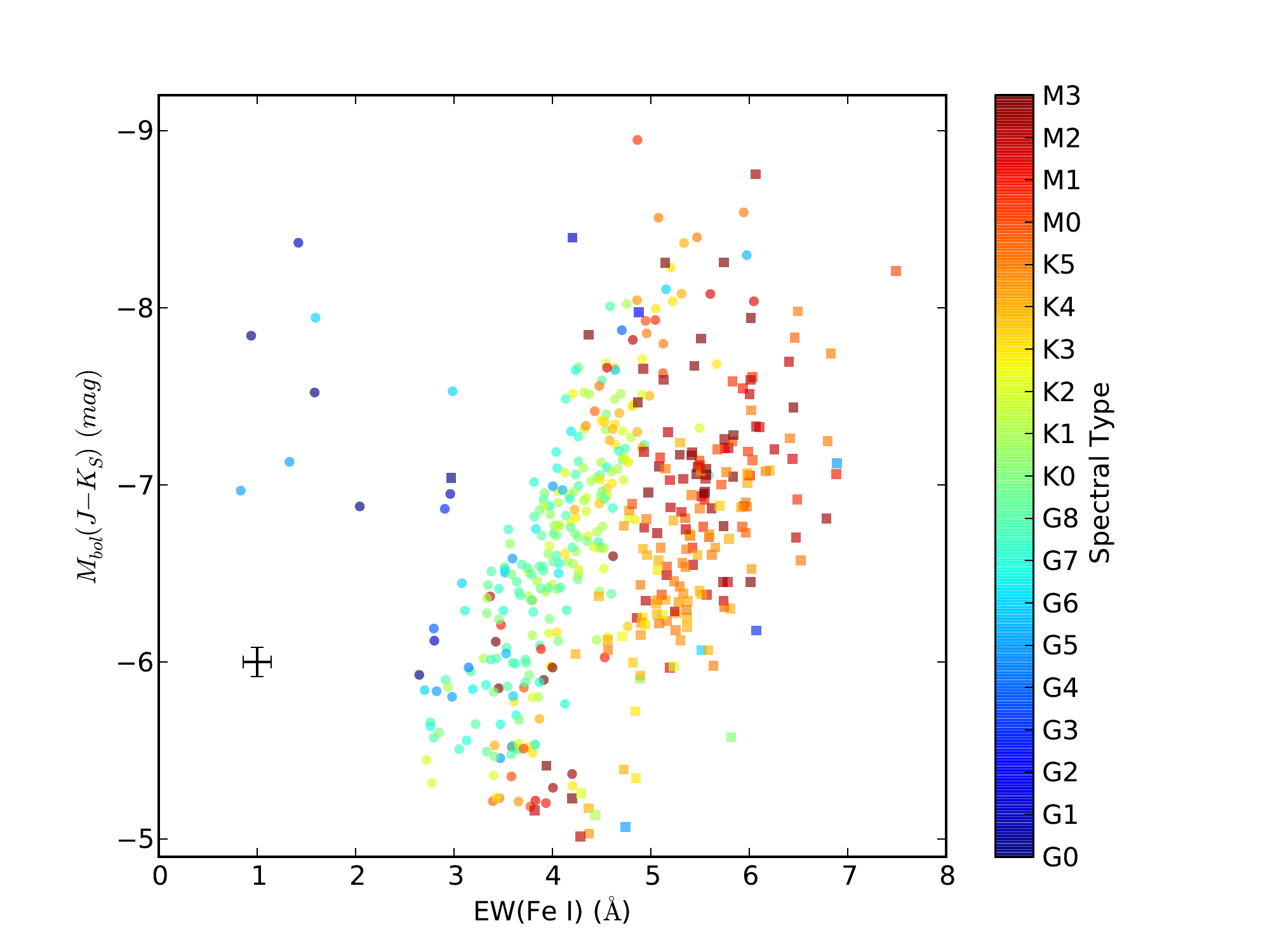}
   \caption{Sum of Fe\,{\sc{i}} equivalent widths against bolometric magnitude. The display is the same as in Fig.~\ref{Ti_Mbol}. {\bf Left (\ref{Fe_Mbol}a):} The colour indicates the LC. {\bf Right (\ref{Fe_Mbol}b):} The colour indicates the SpT.}
   \label{Fe_Mbol}
\end{figure*}

\begin{figure*}[th!]
   \centering
   \includegraphics[trim=1cm 0.5cm 2cm 1.2cm,clip,width=9cm]{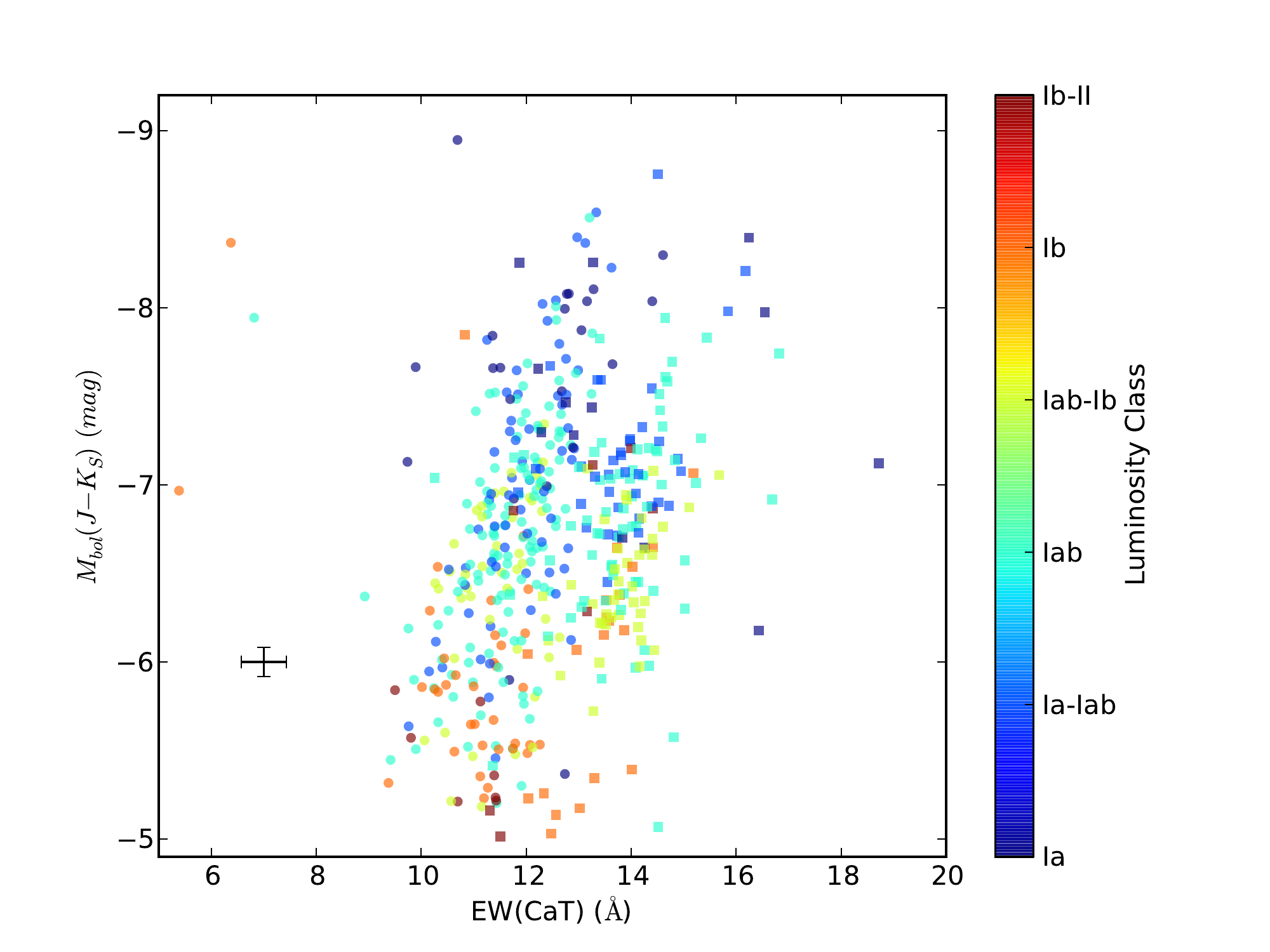}
   \includegraphics[trim=1cm 0.5cm 2cm 1.2cm,clip,width=9cm]{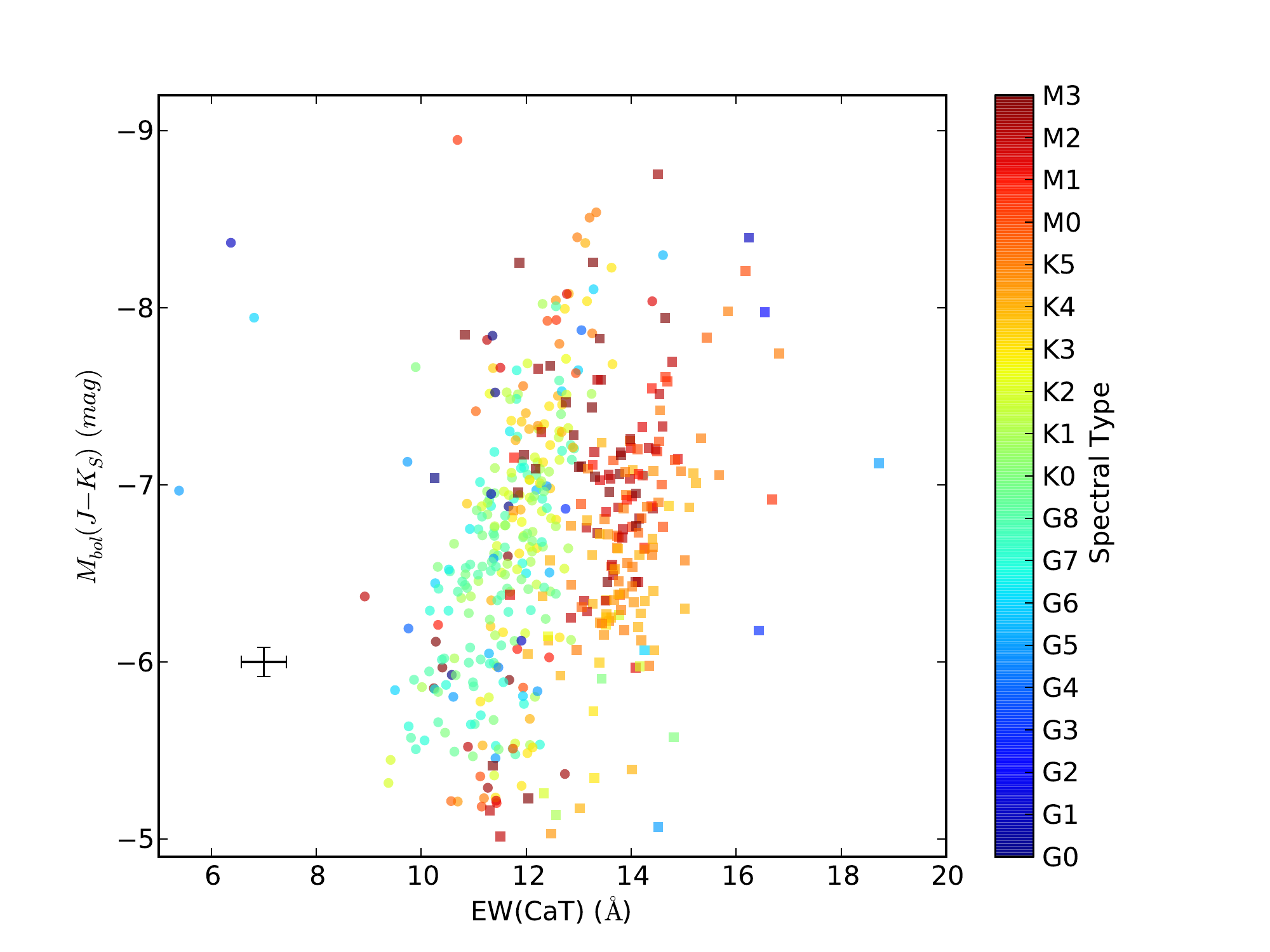}
   \caption{The equivalent width of the CaT against bolometric magnitude. The display is the same as in Fig.~\ref{Ti_Mbol}.{\bf Left (\ref{CaT_Mbol}a):} The colour indicates the LC. {\bf Right (\ref{CaT_Mbol}b):} The colour indicates the SpT.}
    \label{CaT_Mbol}
\end{figure*}

In Figs.~\ref{Ti_Mbol}, \ref{Fe_Mbol} and \ref{CaT_Mbol}, we show the relation between the EW indices and $M_{\textrm{bol}}$ (i.e.\ luminosity) for the observed spectra. All indices show a linear positive relation with luminosity, and both MC populations display the same slope, although the EWs are shifted by a constant, which may be attributed to metallicity. We calculated the correlation coefficients for these trends (shown in Table~\ref{corr_spt_atom}), and found that for the SMC the values of $r$ were close to $0$, while values of $r_{\textrm{S}}$ are not close to zero, indicating the presence of a correlation under the noisy effect of many outliers. The underlying reason can be seen in the figures themselves: there is a large number of low luminosity supergiants (LCs Ib or Ib-II, mostly with $M_{\textrm{bol}}>-6$~mag) present in the sample which do not follow the main linear trend (see Sect.~\ref{SpT_lum}). These objects, at the boundary with bright red giants, are morphologically classified as RSGs, but some of them may well be red giants (see Paper~I for a discussion). We checked this hypothesis by excluding all the stars fainter than $M_{\textrm{bol}}=-6$~mag and repeating the fits. In Table~\ref{corr}, we also display the correlation coefficients for the sample containing only mid-and high luminosity CSGs. Among the trends of the three indices, the EW(Ti\,{\sc{i}}) ones present the coefficients closest to $0$. Their $r$ values indicate that this index hardly presents any linear correlation with  $M_{\textrm{bol}}$, but $r_{\textrm{S}}$ values are higher, suggesting that there is some non-linear correlation, though not very strong. The EW(CaT) index present slightly clearer correlations than EW(Ti\,{\sc{i}}), but the best correlations are found for EW(Fe\,{\sc{i}}). According to synthetic spectra, increasing surface gravity should have a weak effect on EW(Ti\,{\sc{i}}), a stronger effect on EW(Fe\,{\sc{i}}) and the clearest effect on EW(CaT). Contrarily, we found that EW(Fe\,{\sc{i}}) displays a stronger correlation than EW(CaT). If we assume the hypothesis that all RSGs have roughly the same temperature (i.e.\ the SpT does not depend mainly on temperature), we should see a stronger correlation for the EW(CaT) than for EW(Fe\,{\sc{i}}), which is not the case. On the other hand, if we assume that temperature decreases towards later subtypes, this situation may be explained because of the behaviour that EW(CaT) exhibits in the synthetic spectra, with lower values toward lower temperatures. As can be seen in Fig.~\ref{CaT_Mbol}b, the most luminous stars tend to be those with latest SpTs. Thus, the increase of EW(CaT) towards higher luminosities would be partially compensated by the effect of the decreasing temperatures. The correlations presented in the previous paragraphs are very difficult to reconcile with the hypothesis that all RSGs have the same temperature, with their SpTs being determined by luminosity. The atomic lines that display a stronger correlation with SpT (Ti\,{\sc{i}}) are those having the strongest dependence on temperature and the weakest dependence on luminosity, while the Fe\,{\sc{i}} lines, which are expected to be more sensitive to luminosity than to temperature, show a clearly stronger correlation with $M_{\textrm{bol}}$ than with SpT. In addition, CaT lines, which are expected to be the most sensitive to luminosity and the less sensitive to temperature, have a flat trend with SpT, but a weak though significant correlation with $M_{\textrm{bol}}$.

Another factor we have to take into account in these correlations is the role of the metallicity. As mentioned before (see Section~\ref{intro}), differences in metallicity cause a shift in the mean SpT of a population. Metallicity thus has a clear impact on SpTs, affecting them in two ways. On one hand, metallicity may constrain the evolution of CSGs, causing them to stop moving towards lower temperatures at different values of $T_{\textrm{eff}}$, as predicted by evolutionary models (for further discussion see Sect.~\ref{SpT_vs_lum}). But metallicity also affects directly the EW of lines and the strength of bands. Under the hypothesis that all RSGs have approximately the same temperature independently of metallicity, and given that we find no evidence for EW(Ti\,{\sc{i}}) being driven by luminosity, its behaviour with SpT could only be explained through the effect of metallicity. In this case, given that EW(Ti\,{\sc{i}}) presents a strong correlation with the SpT, the SpT sequence would become a metallicity sequence. As can be seen in Figs.~\ref{syn_Ti} and~\ref{syn_Fe}, the range of variation for metallic lines due to changes in metallicity is similar to the range due to changes in temperature.

\begin{figure}[h]
   \centering
   \includegraphics[trim=1cm 0.5cm 2cm 1.2cm,clip,width=9cm]{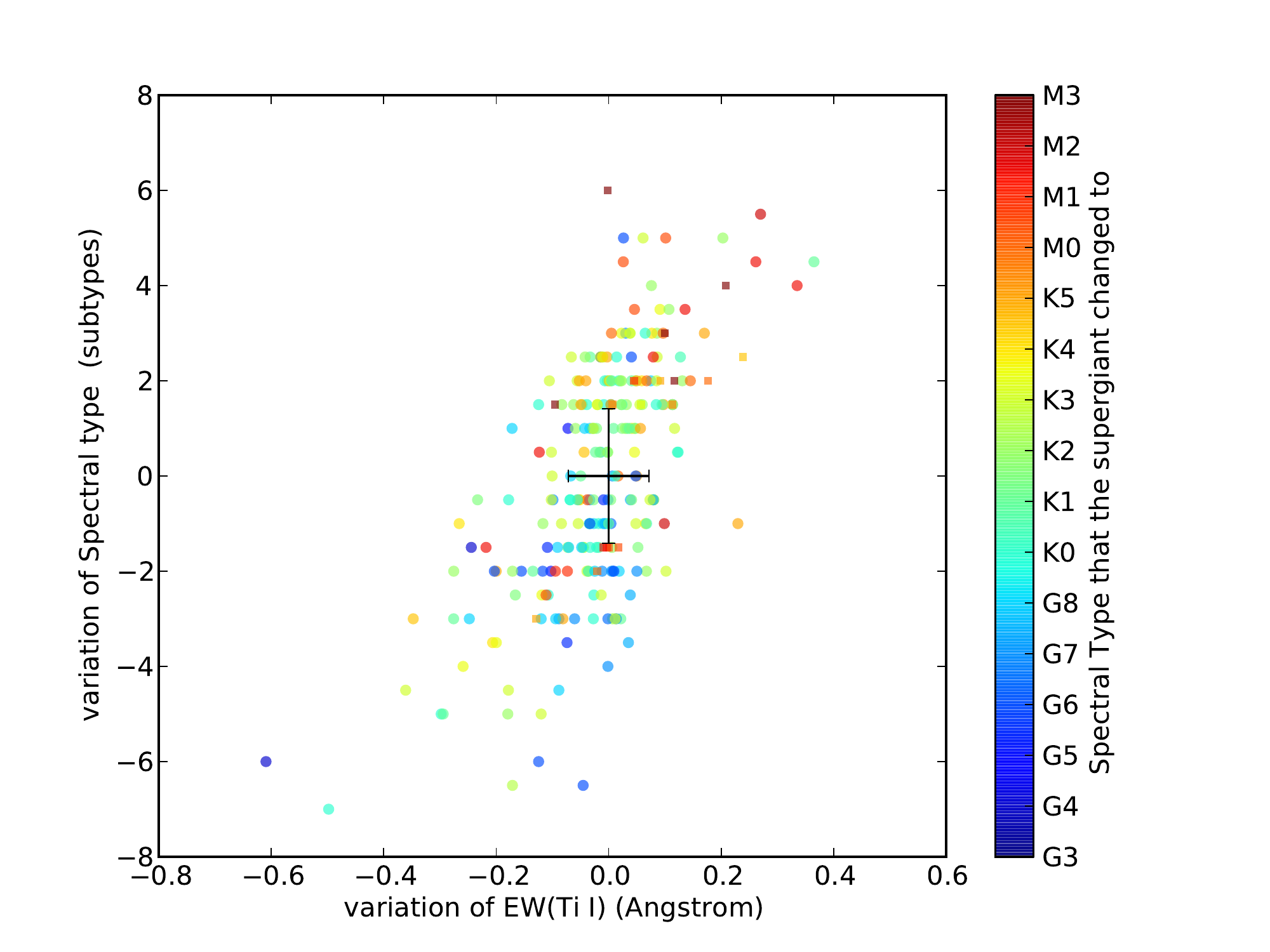}
   \caption{Variations in the EW(Ti\,{\sc{i}}) index against variations in SpT. Each point is the difference for a given star between two epochs. The colour indicates the SpT that the CSG changed to. Squares are LMC CSGs; circles are SMC CSGs. The black cross at (0,0) shows the median error. Epochs when a star moved to SpTs later than M3 are not used.}
   \label{var_spt_ti}
\end{figure}

\begin{figure}[h]
   \centering
   \includegraphics[trim=1cm 0.5cm 2cm 1.2cm,clip,width=9cm]{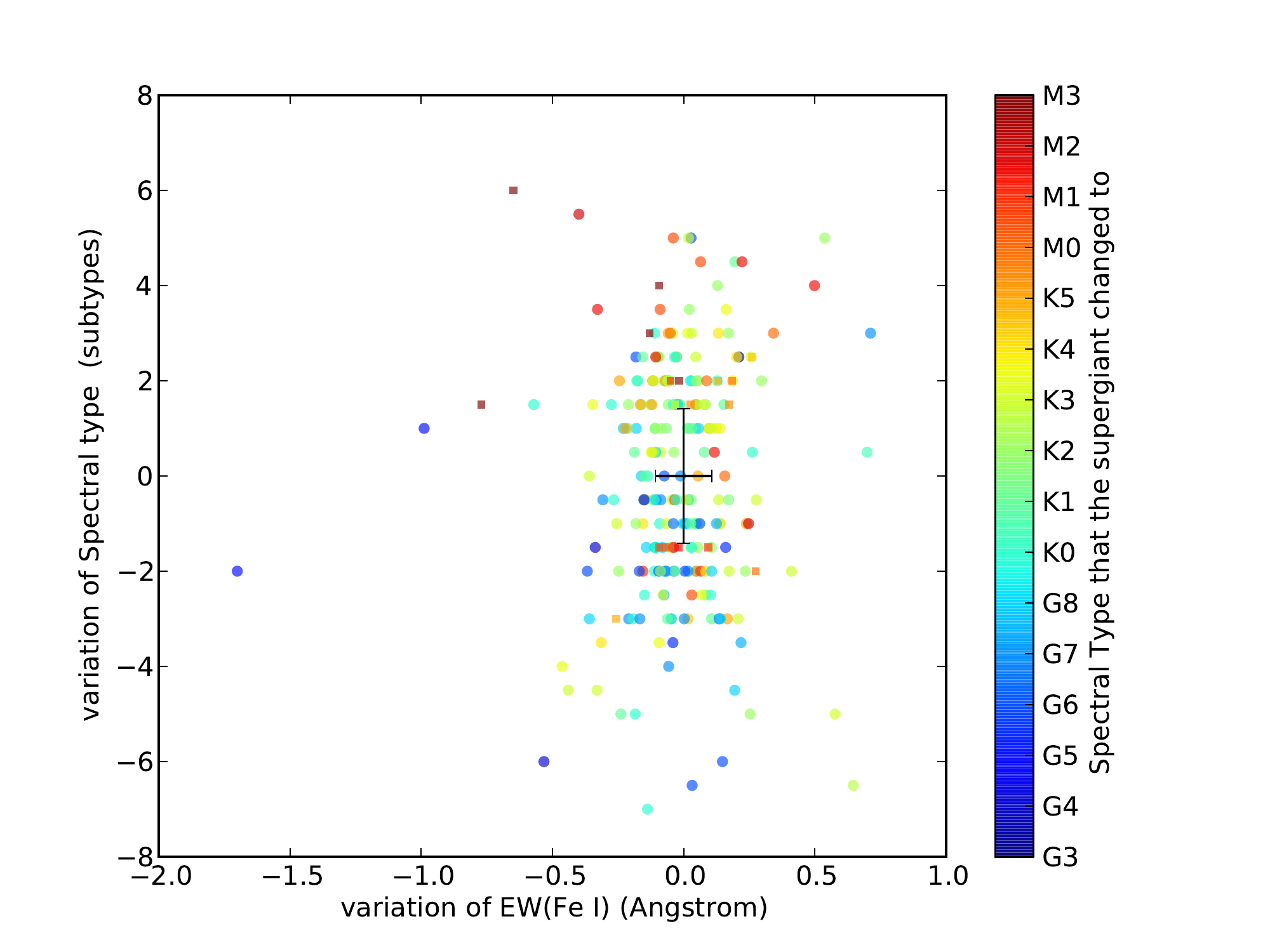}
   \caption{Variations in EW(Fe\,{\sc{i}}) against variations in SpT. The display is the same as in Fig.~\ref{var_spt_ti}.}
   \label{var_spt_fe}
\end{figure}

\begin{figure}[h]
   \centering
   \includegraphics[trim=1cm 0.5cm 2cm 1.2cm,clip,width=9cm]{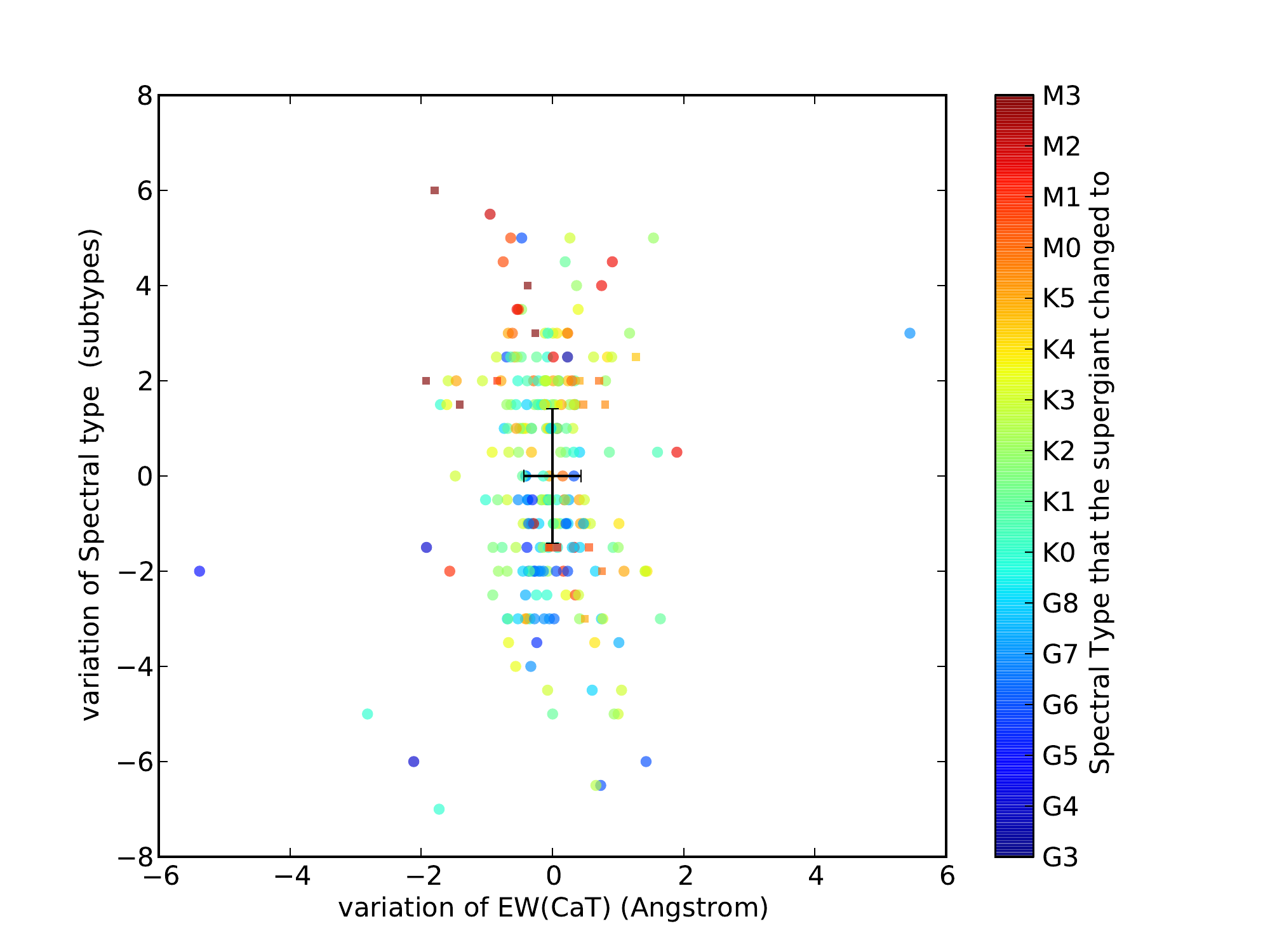}
   \caption{Variations in EW(CaT) against variations in SpT. The display is the same as in Fig.~\ref{var_spt_ti}.}
   \label{var_spt_cat}
\end{figure}

There are, however, two strong objections to this interpretation. The first one is the relatively wide range of spectral types for RSGs observed in a given population. For example, RSGs in the solar neighbourhood are all expected to have about the same metallicity, but display SpTs spanning the whole K and M ranges. Moreover, a significant fraction of the RSGs in the MCs are variable, in the sense that they present different SpTs at different epochs. In the next section, we give a detailed description of spectroscopic and photometric variability in our sample. For the stars that we have classified as spectroscopic variables, we have measured the EW indices at both extremes of the SpT variation seen. In Fig.~\ref{var_spt_ti}, we plot the variation of EW(Ti\,{\sc{i}}) against the difference in spectral subtypes. We also calculated the coefficients obtained for this pair of variables (shown in Table~\ref{corr_spt_atom}), and we found a not very strong but still significant correlation between them. This figure clearly demonstrates that EW(Ti\,{\sc{i}}) varies in a given star as its spectral type changes by an amount similar to the difference between two stars of different spectral types. The highest change in EW(Ti\,{\sc{i}}) is about $0.4$~\AA{}, which in synthetic spectra correspond to a change of $\sim0.3$\,dex. Since the metallicity of a star is not expected to change {\em at all} along its variability cycle, this rules out metallicity as the main driver of the SpT sequence. Moreover, in Figs.~\ref{var_spt_cat} and~\ref{var_spt_fe}, we see that EW(CaT) or even EW(Fe\,{\sc{i}}), which are much less sensitive to temperature, do not change when the SpT of the star changes. In fact, their correlation coefficients (both $r$ ans $r_{\textrm{S}}$) are $\sim0$, indicating that these lines are clearly insensitive to SpT variations. If the changes in EW(Ti\,{\sc{i}}) would be caused by changes in metallicity, we should also expect some changes in these other lines. 

To summarize, the presence of SpT variations in a given star is one further argument against all RSGs having the same temperature or the SpT sequence being determined by luminosity. Only EW(Ti\,{\sc{i}}), the index with a strongest dependence on temperature, shows coherent variations with significant correlation coefficients when a star changes its SpT by several subtypes. Contrarily, EW(CaT), the index most sensitive to luminosity, does not change at all along SpT variations (its correlation coefficients are $\sim0$), against what we should expect if SpT is determined by luminosity changes. Thus, the evidence again points to temperature as the explanation for the SpT sequence. However if SpT variability is caused by temperature changes, why are EW(Fe\,{\sc{i}}) changes not correlated at all with SpT variations? There is an explanation in the fact that EW(Fe\,{\sc{i}}) does not have a monotonic behaviour with temperature, but shows a maximum around $4\,000$\:K (see Fig.~\ref{syn_Fe}). If we accept that this temperature roughly corresponds to mid-K subtypes in the MCs, as suggested by the different $T_{\textrm{eff}}$ scales, our spectrum variables are moving across this maximum in both directions, and so temperature changes, either to higher or lower temperatures, can increase or decrease the EW(Fe\,{\sc{i}}) in a manner that will look random when only two or three epochs are available.

If so, all the observed changes in line strengths could be explained by temperature changes of only a few hundred Kelvin, which are entirely compatible with the variability in SpT, but cannot be explained within the current framework by any other of the major physical magnitudes.

The correlations we have found are hardly compatible with luminosity and metallicity as the main effects to explain the behaviour of atomic lines (represented by the EW indices) along the SpT sequence of CSGs. This result, however, does not imply that SpT is unrelated to luminosity and metallicity. Several authors have suggested a correlation between luminosity and SpT, and this will be discussed at length in Sect.~\ref{SpT_lum}. Likewise, metallicity determines the mean SpT of a population and the strength of metallic lines at a given SpT. Therefore, the correlations found seem to indicate that both luminosity and metallicity have an indirect effect on spectral type, but at a given metallicity, the SpT sequence seems to be a temperature sequence, modulated to some degree by luminosity, as happens across the whole MK system.

Our analysis has been confined to stars with SpTs earlier than M4, and it may be argued that for the second half of the M sequence, which is determined exclusively by TiO bands, this temperature dependence may become negligible, and the SpT of late-M stars may be determined mainly by luminosity. This is not impossible, as there are very few such late-M supergiants (most of them in the MW), and they are all characterised by very heavy mass loss \citep[e.g.][]{hum1974b}. There are, however, no compelling reasons to take this view either. In the optical range, the TiO bands arise at early-K subtypes, and our classification was done attending to the growth of these bands. Despite this, our atomic lines in the CaT range, which are not affected at all by TiO bands down to M2, show a behaviour along the SpT sequence dominated by temperature. The increasing strength of the TiO bands along the SpT sequence, which is determined by decreasing temperature more than by any other physical parameter, can only be explained if the intensity of the TiO bands has a non-negligible dependence on temperature, at least down to SpT M3 (the range that we have probed). There does not seem to be a strong reason to believe that this dependence stops for types later than M3 simply because \ion{Ti}{i} loses its sensitiveness to temperature. In any case, as RSGs with SpT later than M3 are very rare, even in the Milky Way \citep{eli1985,lev2012}, and almost absent in galaxies with an average metallicity similar to that of the SMC, our results describe the generality of CSGs in the MCs (and, by extension, presumably, in most galaxies), not only a peculiar minority.

\subsection{Spectral variability}
\label{variab}

We observed a group of luminous CSGs from each galaxy on more than one epoch, which allows the study of their spectral variability. These targets, about a hundred per cloud, are known RSGs from the lists of \citet{eli1985} or \citet{mas2003}. Most of the targets from the SMC were observed on three epochs (2010, 2011 and 2012), but a small number of them were observed only in two epochs, with a time interval between them of about a year. The LMC group was observed only on two epochs (2010 and 2013). We tagged as variable any CSG whose SpT has changed significantly between epochs, i.e.\ more than our mean error (one subtype). The details about these CSGs and their observation are given in Paper~I, but we provide a brief summary in Table~\ref{var_table}.

\begin{table*}[th!]
\caption{Spectral type variability among CSGs observed on multiple epochs. The number of CSGs tagged as variable, the fraction with respect to the total number $n$, and the 2-$\sigma$ confidence intervals for the fractions ($\Delta f$), which are equal to $1/\sqrt[]{n}$. CSGs are tagged as variable if their SpT changed between any two of the epochs indicated by more than 1 subtype. The SMC sample is showed twice, one using all three available epochs to check its variability, and the other using only the 2010 and 2012 epochs. We also show the mean SpT change among the CSGs tagged as variables (in the case of three epochs, we only show the largest change measured for each star.)}
\label{var_table}
\centering
\begin{tabular}{c | c | c c c | c }
\hline\hline
\noalign{\smallskip}
Galaxy&CSGs observed&\multicolumn{3}{c|}{Variable CSGs found}&$\overline{\Delta {\rm SpT}}$\\
(and epochs)&in all these epochs& Number& Fraction & $\pm \Delta f$&(subtypes)\\
\noalign{\smallskip}
\hline
\noalign{\smallskip}
SMC (all three epochs)&108&88&0.84&0.10&3.0\\
SMC (2010-2012)&102&48&0.47&0.10&2.9\\
LMC (2010-2013)&79&26&0.33&0.11&2.3\\
\noalign{\smallskip}
\hline
\end{tabular}
\end{table*}

At the sight of Table~\ref{var_table}, it is striking that the fractions of variable CSGs found in each galaxy are very different (33\% for the LMC and 84\% for the SMC). Initially, we attributed this difference to the fact that we observed the SMC on three epochs, while the LMC was observed only twice. To test if the difference could be caused by the different number of observation epochs, we checked the variability for the SMC CSGs using only two epochs (2010 and 2012, as this is the pair of epochs with most CSGs in common). The resulting fraction of variable CSGs in the SMC in this case is 47\%, lower than when we use three epochs, but still significantly larger than for the LMC. Moreover, the maximum SpT changes detected for SMC CSGs are larger than those for LMC CSGs.

\cite{eli1985} studied the photometric variability of the RSGs in both MCs. They find that RSGs from the LMC show larger variations than those from the SMC. However, at a given temperature, RSGs from both galaxies show similar variations. They also found a correlation between most of the photometric colours that they studied and the brightness changes. They estimated, through their calculated colour-SpT relations, that if the observed colour changes are matched by SpT variations, then RSGs from the SMC would have a typical variation of about 1.5 subtypes, while those from the LMC would vary by about 1 subtype. This is because colour differences between typical subtypes of LMC RSGs are larger than those between typical subtypes of SMC RSGs.

To explore the relation between photometric and spectral variations in both MCs, we have used the works of \cite{yan2011,yan2012} (Y\&J onward). They studied the photometric variability of a large number of RSGs taken mainly from the same source as our sample \citep{mas2003}. Thus, our multi-epoch sample has a high overlap with them. Unfortunately, Y\&J could not analyse their whole initial sample. We have used for the comparison only those RSGs tagged as "long secondary period" (LSP) or "semi-regular" (SR) in their papers, i.e.\ those from Tables~2 and~4 from Y\&J2011 and Tables~3 and~6 from Y\&J2012. In Table~\ref{var_yan} we show the result of this cross-match. Even though we have a similar number of CSGs in common with them (about 40) in each MC, the number of objects that we tag as spectral variables is significantly different between the LMC and the SMC samples, with the SMC RSGs again demonstrating a higher degree of variability. This difference confirms the findings of \cite{eli1985} that a similar photometric change implies a different spectral variation, depending on the mean SpT of the stars, and therefore, in statistical terms on the metallicity of the host galaxy.

\begin{table*}[th!]
\caption{Result of the cross-match between CSGs observed on multiple epochs, and those RSGs tagged as LSP or SR in the works of Y\&J. We indicate how many stars are in common ($n$), and how many of them were identified as spectral variables, and the corresponding fractions. The 2-$\sigma$ confidence intervals for the fractions ($\Delta f$), which are equal to $1/\sqrt[]{n}$, are also shown. The SMC sample is showed twice, one using all three available epochs to check its variability, and the other using only the 2010 and 2012 epochs.}
\label{var_yan}
\centering
\begin{tabular}{c | c | c c c}
\hline\hline
\noalign{\smallskip}
Galaxy&Number of CSGS tagged&\multicolumn{3}{c}{LSP and SR with spectral variation detected}\\
(and epochs)& as LSP or SR in common&Number&Fraction& $\pm \Delta f$\\
\noalign{\smallskip}
\hline
\noalign{\smallskip}
SMC (all three epochs)&41&34&0.83&0.16\\
SMC (2010-2012)&40&23&0.58&0.16\\
LMC (2010-2013)&42&11&0.26&0.15\\
\noalign{\smallskip}
\hline
\end{tabular}
\end{table*}

Indeed, this trend seems to extend to the MW as well. \cite{whi1978} studied the spectral variations of a large sample (128) of RSGs in the Galaxy. They tagged as spectral variables those RSGs with changes larger than half a subtype, while we are considering as being variable those with changes larger than 1 subtype. Almost all the RSGs in their sample were observed on more than two epochs (on average, each one of their stars was observed on 3.8 epochs). Despite this, they tagged only 28 of their RSGs as spectral variables. From these, only 9 have changes larger than 1 subtype, and thus would be considered as spectral variables in the present work. Therefore the fraction of spectral variables among the galactic RSGs ($0.07\pm0.09$) is significantly smaller than for the LMC RSGs ($0.33\pm0.11$).

From our data, we conclude that spectral variability among CSGs seems to be more frequent and implies larger spectral changes at lower metallicities. From the bibliography commented above, it seems that photometric variations are common among RSGs, even if their spectral variations are not noticeable. Following \cite{eli1985}, we may consider a simple explanation for the relations between spectral variability and metallicity: the spectra have a weaker response to colour changes at higher metallicities (because of their later average SpTs). However, a trend between the pulsation mode and the metallicity of the host galaxy was found by Y\&J2012. Thus, there may be a relation between spectral variation and pulsation mode, but this possibility cannot be asserted or rejected with our data.

\section{Discussion}
\subsection{Spectral type, luminosity and mass loss}
\label{SpT_lum}

All our results, and those in the literature, show that the relation between spectral type, luminosity and mass loss is complex, and these variables cannot be treated individually. In this section, we analyse how they are intertwined.

\subsubsection{Spectral type and luminosity}
\label{SpT_vs_lum}

As mentioned above, in the literature there are many hints of a relation between SpT and luminosity, with later stars being typically more luminous \citep[e.g.][and references therein]{dav2013}. Such a relation can also be seen in Fig.~1 from \cite{lev2006}, if indirectly (this figure plots temperature and not SpT, but temperature can be considered equivalent to SpT in this work, because their effective temperature scale was calculated from the fit to TiO bands). Nevertheless, the relation between SpT and luminosity has not been further investigated, neither its possible connection with evolutionary state (i.e.\ mass-loss).

\begin{figure*}[th!]
   \centering
   \includegraphics[trim=1cm 0.5cm 2cm 1.2cm,clip,width=9cm]{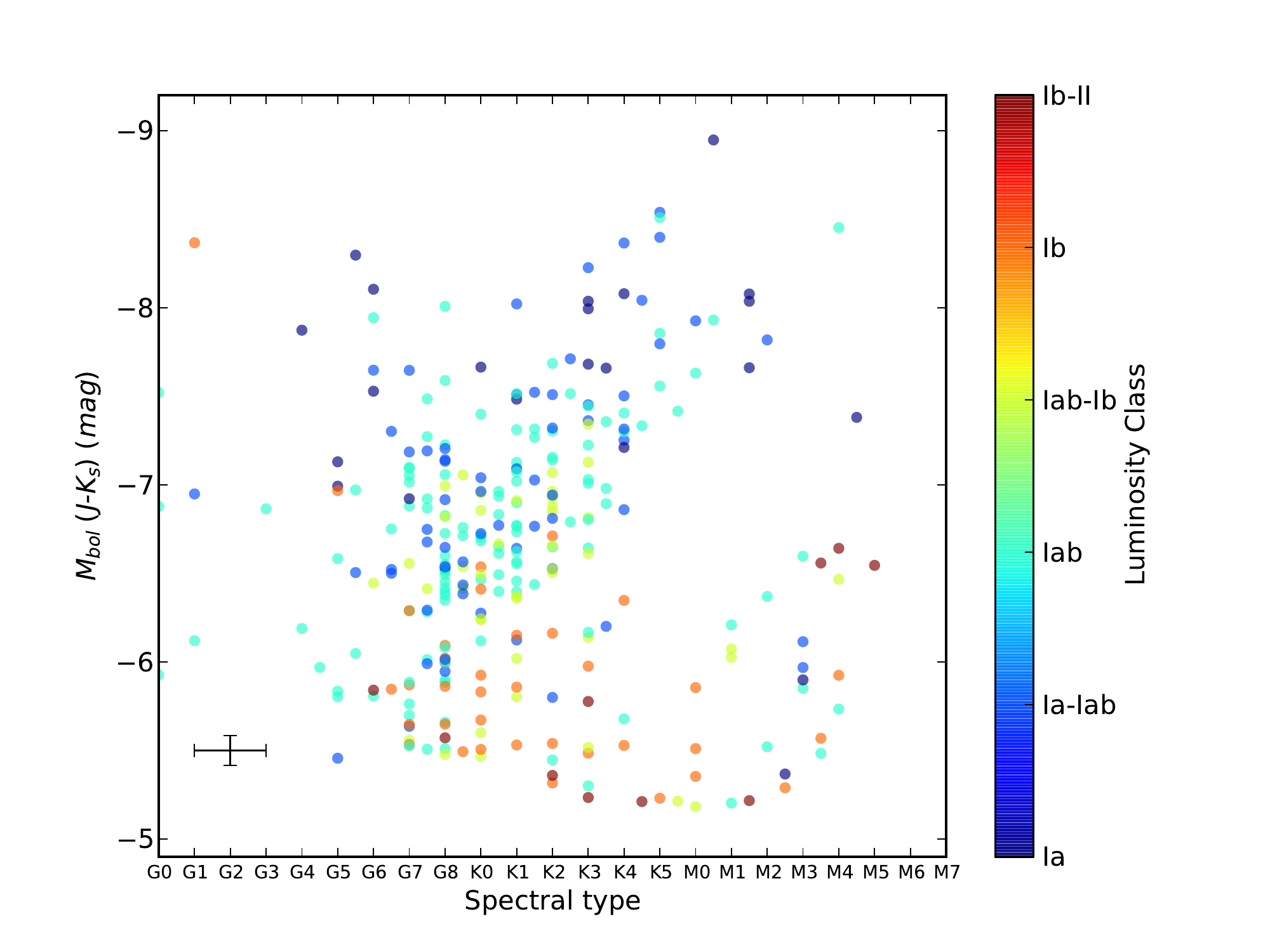}
   \includegraphics[trim=1cm 0.5cm 2cm 1.2cm,clip,width=9cm]{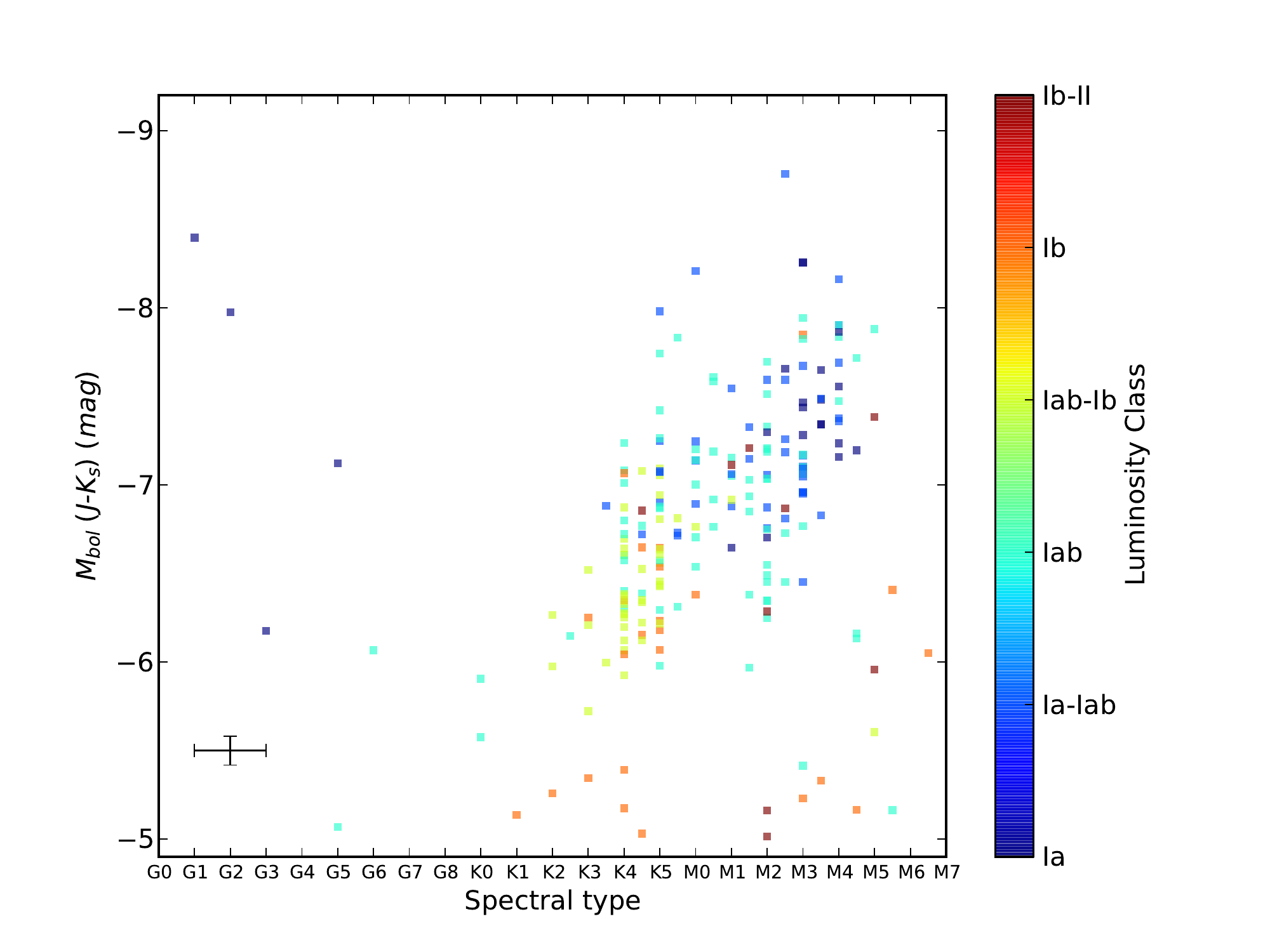}
   \caption{Spectral type against $M_{\textrm{bol}}$ (derived from ($J-K_{\textrm{S}}$)). Colour indicates the luminosity class. The LMC data correspond to 2013 and the SMC data correspond to 2012, because all the stars observed in 2010 and 2011 were also observed in 2012 and 2013 and we wanted to avoid to represent the same CSG more than once. The black cross represents the median uncertainties. Note that both figures have the same colour scale, to make clear the comparison. {\bf Left (\ref{SpT_Mbol}a):} CSGs from the SMC. {\bf Right (\ref{SpT_Mbol}b):} CSGs from the LMC.}
      \label{SpT_Mbol}
\end{figure*}

In Figs.~\ref{SpT_Mbol}a and~\ref{SpT_Mbol}b, we plot SpT and $M_{\textrm{bol}}$ for both MCs. The populations represented in these diagrams may be divided in two different groups. The first group is formed by most of the low-luminosity SGs (Ib or dimmer), which are spread all over our SpT range. From G down to early-M subtypes, most of them have $M_{\textrm{bol}}$ between $\sim-5$ and $\sim-6$ (up to slightly higher luminosity in the SMC sample). For mid and late-M subtypes, these stars reach $M_{\textrm{bol}}\sim-6.5$, but they are clearly separated from the higher-luminosity M~SGs (which do not reach late-M types). In both galaxies, these lower-luminosity groups cannot be considered large enough to draw statistically significant conclusions on their properties, because of their limited numbers. Our exposure times were optimized for the observation of the bright population, and so priority in fibre assignment was given to this population. In addition, many of the fainter targets included did not result in usable spectra. Thus in this work we will limit our analysis to the second group (i.e.\ stars more luminous than $-6$~mag).

This second group is formed by most of the high and mid-luminosity CSGs (Iab-Ib or brighter). These stars are spread along a strip starting at early SpTs and low luminosities (slightly more luminous than $M_{\textrm{bol}}=-6$~mag) that extends toward later SpTs and up to the highest luminosities present in our samples. The range of SpTs covered differs between galaxies, because of the metallicity effect discussed previously. For both galaxies there is a correlation between SpT and $M_{\textrm{bol}}$ (their coefficients are shown in Table~\ref{corr_spt_atom}). It is much clearer for the LMC in part because of the smaller number of low luminosity SGs (specially later than M0) with $M_{\textrm{bol}}<-6$~mag. To check if faint outliers are causing the lower correlation coefficients for the SMC, we have also calculated the coefficients for the data from the SMC using a more restrictive luminosity boundary ($M_{\textrm{bol}}<-6.7$~mag). In this case, we obtain a clearer correlation, but still significantly weaker than for the whole bright group in the LMC.

\begin{figure*}[th!]
   \centering
   \includegraphics[trim=0.85cm 0.5cm 1.8cm 1.2cm,clip,width=9cm]{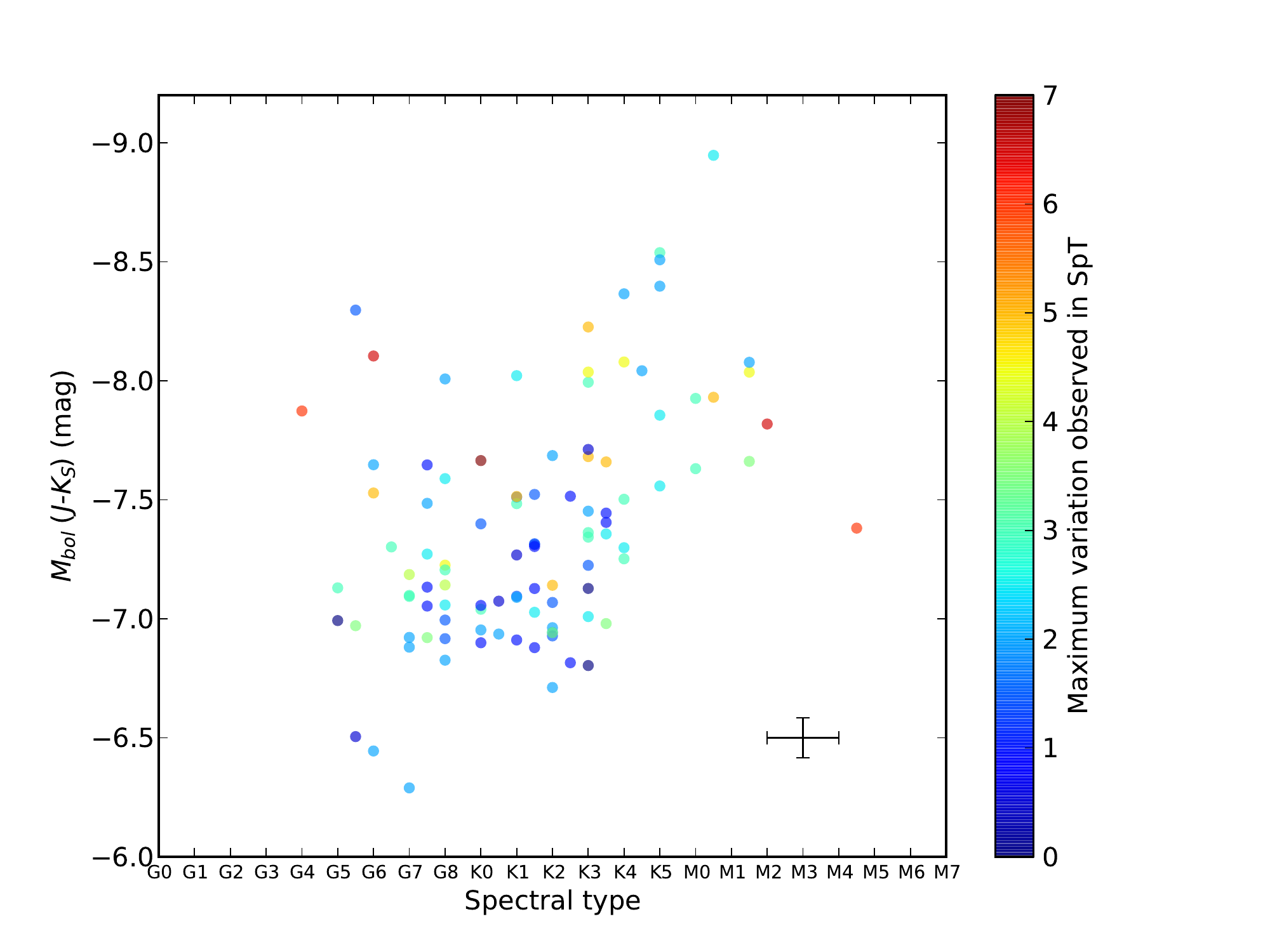}
   \includegraphics[trim=0.85cm 0.5cm 1.8cm 1.2cm,clip,width=9cm]{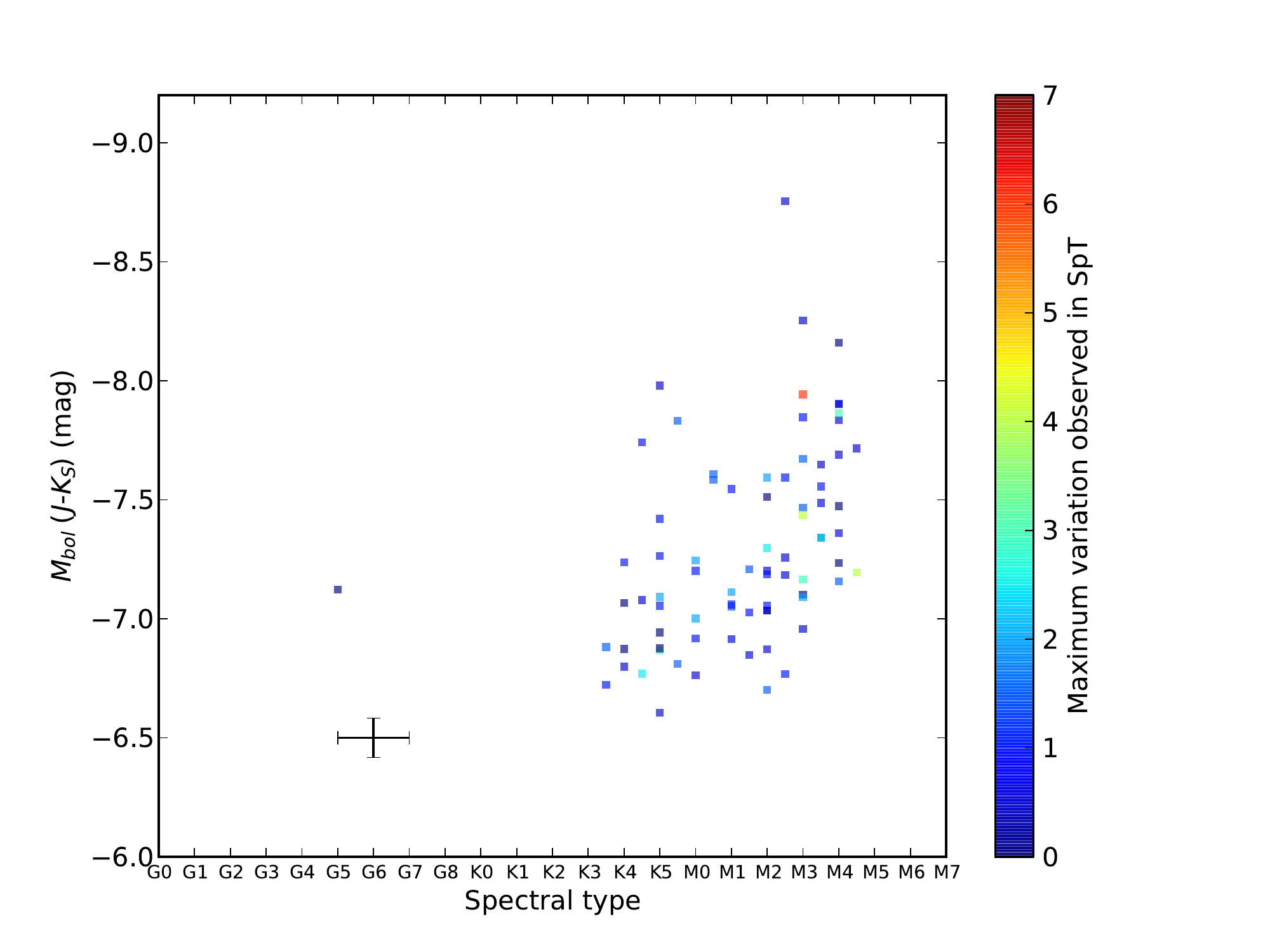}
   \caption{Spectral type against $M_{\textrm{bol}}$ (derived from ($J-K_{\textrm{S}}$)). Colour indicates the maximum SpT variation observed for each star among the epochs it was observed (see section~\ref{variab} for more details). The samples used here are the same as in Fig.~\ref{SpT_Mbol}, but we have represented only those stars observed on more than one epoch. The black cross represents the median uncertainties. The $x$-axis scale is the same as in Fig.~\ref{SpT_Mbol} to ease the comparison. {\bf Left (\ref{var_max}a):} CSGs from the SMC.  {\bf Right (\ref{var_max}b):} CSGs from the LMC.}
      \label{var_max}
\end{figure*}

The weaker correlation between SpT and $M_{\textrm{bol}}$ for the SMC, even when only RSGs more luminous than $M_{\textrm{bol}}<-6.7$~mag are included, is likely due to the presence of a moderate number of CSGs with $M_{\textrm{bol}}$ between $\sim-7$ and $-8$~mag spread along the whole SpT distribution. Many of them have SpTs earlier than expected from the trend. Such spread in SpT is not observed among LMC stars in the same magnitude range. A possible explanation for this difference lies in the SpT variability of CSGs, because most of the SMC CSGs observed on more than one epoch present significant SpT variability (up to 7 subtypes), while most of the LMC CSGs do not (see Fig.~\ref{var_max} and Section~\ref{variab}).

These correlations between SpT and $M_{\textrm{bol}}$ may seem contradictory with the behaviours of the indices discussed in Sect.~\ref{SpT_atomic}. From the correlations studied there, we concluded that SpT seems much more likely to depend mainly on $T_{\textrm{eff}}$ than on luminosity. However, the results found here indicate a correlation between SpT and $M_{\textrm{bol}}$. This correlation can already be anticipated in Figs.~\ref{Ti_Mbol}b, \ref{Fe_Mbol}b and \ref{CaT_Mbol}b, where the brightest stars concentrate strongly towards the latest SpTs. This is likely the reason why, in addition to its strong correlation with SpT, EW(Ti\,{\sc{i}}) also has a weak correlation with $M_{\textrm{bol}}$. In any case, the correlations between SpT and $M_{\textrm{bol}}$ are not very strong. They are significantly weaker than those between SpT and EW(Ti\,{\sc{i}}) or $M_{\textrm{bol}}$ and EW(Fe\,{\sc{i}}). When considered together, all these results suggest that the relation between SpT and luminosity is indirect, i.e.\ SpT is not directly determined by luminosity, but both are connected by the underlying physics.

\begin{figure*}[th!]
   \centering
   \includegraphics[trim=1cm 0.3cm 2cm 1.2cm,clip,width=9cm]{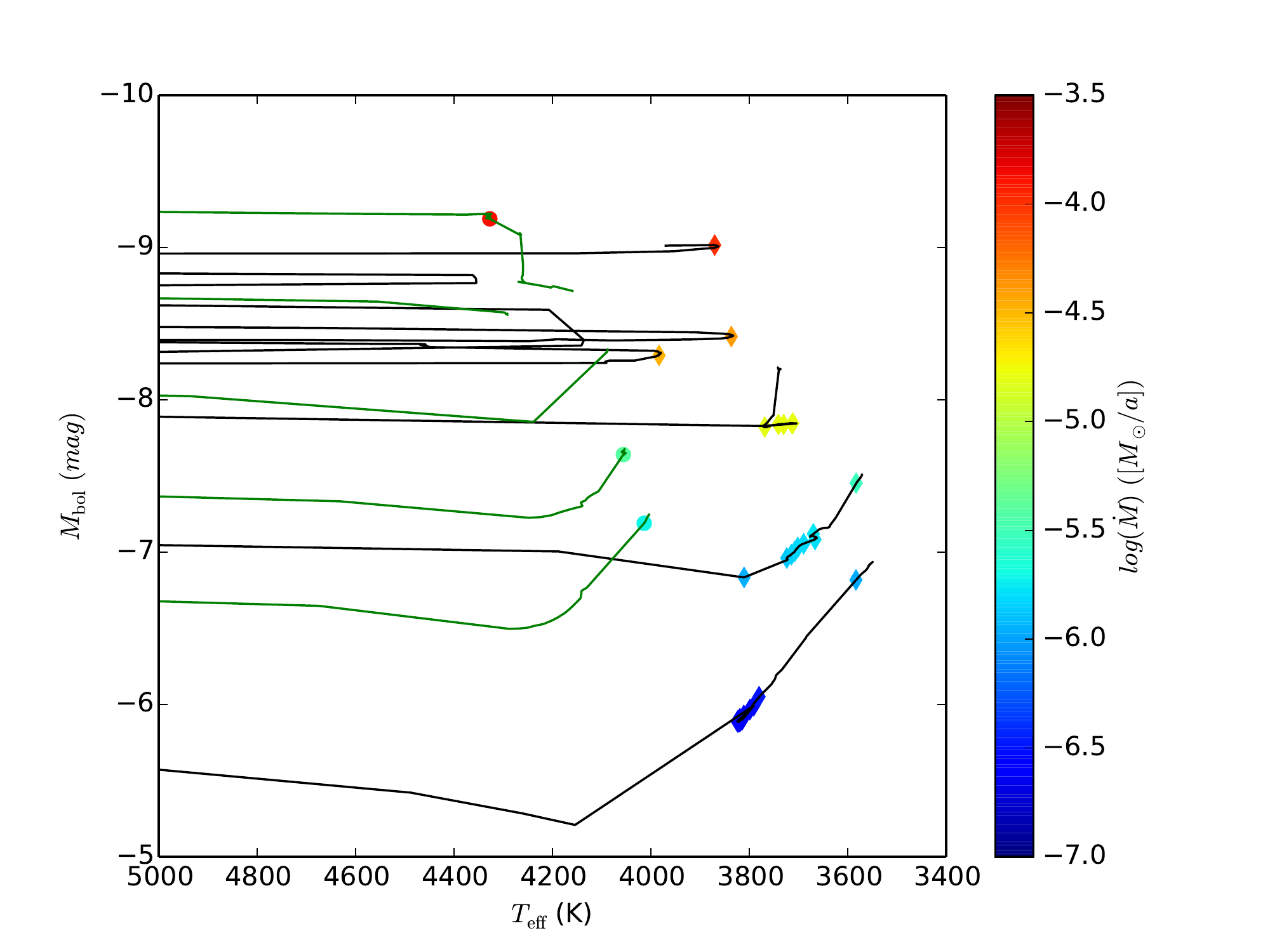}
   \includegraphics[trim=1cm 0.3cm 2cm 1.2cm,clip,width=9cm]{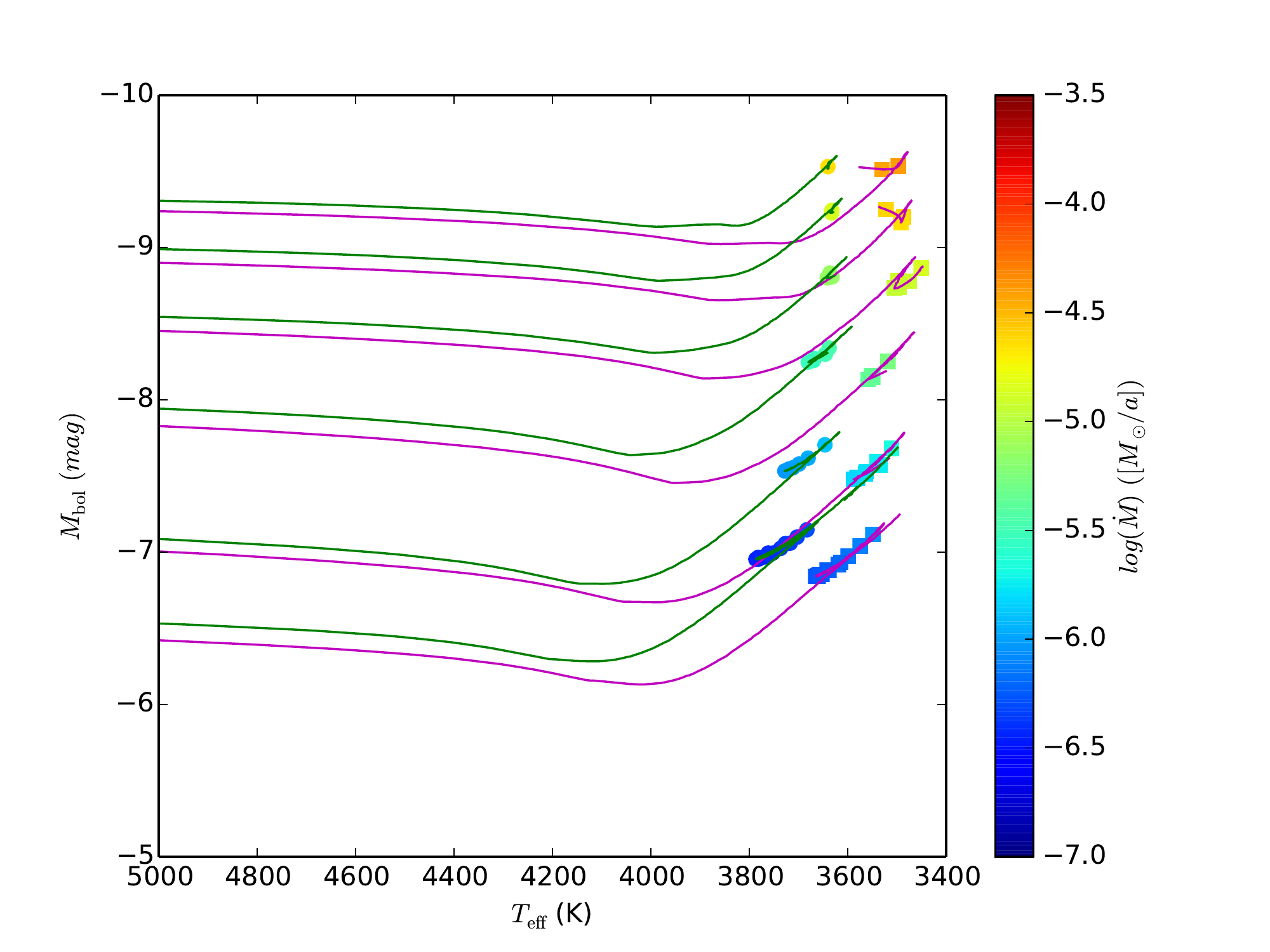}
   \caption{Theoretical evolutionary tracks, represented in the $T_{\textrm{eff}}$ vs $M_{\textrm{bol}}$ plane. The colour of the tracks indicates their metallicity: black for Solar metallicity, magenta for LMC typical metallicity, green for SMC typical metallicity. The coloured points along the tracks are separated by $0.1$~Ma, and their colours indicate mass-loss.
   {\bf Left (\ref{mod}a):} Geneva models, from \cite{eks2012,geo2013}. No tracks for LMC metallicty are available. Solar metallicity is $Z=0.014$. The tracks shown here correspond, from bottom to top, to stars of 12, 15, 20, 25 and $32\:M_{\sun}$. {\bf Right (\ref{mod}b):} Models from \cite{bro2011}. The evolutionary tracks shown here are, from bottom to top, those of stars with 12, 15, 20, 25, 30 and $35\:M_{\sun}$.}
   \label{mod}
\end{figure*}

In Fig.~\ref{mod} we show the evolutionary tracks (without rotation) generated using models from the Geneva group \citep{eks2012,geo2013}, and from \cite{bro2011}. As we do not know the exact temperature to SpT relation, we cannot directly overplot our data on these tracks. However, we have demonstrated above (Sect.~\ref{SpT_atomic}) that such a relation must exist, and this allows us to compare the behaviours seen for the observed samples and those expected from the evolutionary tracks.

In the Geneva models, we only see a decrease in temperature as we move to higher luminosity when we compare the 12 and $15\:$M$_{\sun}$ evolutionary tracks (see Fig.~\ref{SpT_lum}). The long-term stable regions along these tracks (marked by groups of coloured points) are cooler and more luminous for the $15\:$M$_{\sun}$ track than for the $12\:$M$_{\sun}$ one, spanning from $M_{\textrm{bol}}\:\sim-6$ to $\sim-7.5$. Nevertheless, the tracks for $20\:$M$_{\sun}$ and higher masses (which correspond to $M_{\textrm{bol}}< -7.5$~mag), tend to reach temperatures warmer than less massive (and less luminous) tracks. In our data (see Fig.~\ref{SpT_Mbol}), the trend to later SpT with increasing luminosity, extends from $M_{\textrm{bol}}\:\sim-6.5$ to the highest values observed. This lack of agreement between observations and evolutionary tracks is not new, and has also been found in previous works \citep[e.g.][]{lev2006}.

In the evolutionary models from \cite{bro2011} the long-term stable regions along the tracks for 12, 15 and $20\:M_{\sun}$ span almost the same temperature range, and only the stable regions for the $25\:M_{\sun}$ track seems to be slightly cooler than for the less luminous stars. For tracks of higher masses, their stable regions tend again to be slightly warmer than for the $25\:M_{\sun}$ star. Following these predictions, we should have about the same SpTs for the luminosity range between $M_{\textrm{bol}}\:\sim-7$ and $\sim-8.5\:$mag, and only find slightly later types about $\sim-9$. This scenario does not match the observations either.

The origin of the observed SpT-luminosity trend for mid- and high-luminosity CSGs is uncertain. We can envisage two possibilities: either more massive stars spend their time as RSGs at later SpTs (lower temperatures) than less massive objects or a given star evolves to higher luminosity and lower temperatures during its lifetime as an RSG (as suggested by \citealt{dav2013})\footnote{Under the hypothesis that all RSGs have the same $T_{\textrm{eff}}$, the second possibility is not fully incompatible with the evolutionary tracks. Contrarily if, as our data seem to support, SpT is correlated to $T_{\textrm{eff}}$, none of these scenarios is supported by current evolutionary tracks}. In the first case, the observed trend would be a direct consequence of the evolution of CSGs with different masses (and thus, different ages). Since CSGs are supposed to have evolved at roughly constant luminosity from the blue side of the Hertzsprung-Russel Diagram, more massive CSGs should have reached higher luminosities. This scenario seems a logical explanation for our observed distributions. Our samples should have a mixture of CSGs with different ages and masses because they were selected from all over both MCs. In addition, there is clear evidence among well-studied MW RSGs showing that late-M stars are intrinsically brighter and, in many cases, are quite massive \citep[e.g.][]{wit2012,arr2013}. However, there are also some observations that seem to support the second scenario. In MW open clusters containing RSGs, their SpT frequency distribution tends to peak around M1\,--\,M2, but there are also a number of RSGs with later SpTs that tend to be more luminous \citep{neg2013}. Since RSGs in each cluster should have roughly the same age and mass, the presence of much brighter RSGs of later SpT is difficult to explain. For example, in the cluster Stephenson~2 the latest RSG is about M7, with $M_{\textrm{bol}}=-8.3$~mag. It is about 2~mag brighter than the less luminous RSGs in the cluster \citep{dav2007}. According to the Geneva evolutionary models, this star should have about $\sim25\:M_{\sun}$, while most RSGs from the cluster would have between 12 and~$15\:M_{\sun}$.  In addition, the most luminous CSGs in clusters are also those with the highest mass-loss rates and most noticeable circumstellar envelopes. In consequence it has been proposed \citep{dav2013,neg2013} that, as a given RSG evolves, it goes from early-M subtypes down to later ones while its luminosity and mass-loss also grow. All the evolutionary tracks from \cite{bro2011} and those from Geneva for $12$ and $15\:M_{\sun}$ predict an increase of luminosity along the evolution at the end of the tracks, but the increment is, in the best case, about one bolometric magnitude, far less in other tracks. So, to explain the observed trends, which span more than 2 bolometric magnitudes in our data, but even more in previous works \citep{lev2005}, luminosity changes along the evolution far larger than predicted by the evolutionary models are needed. Since the trends that we observe are inconsistent with the models, these objections must be treated with caution, and so  both hypotheses seem consistent with our data at this point. Detailed studies of individual open clusters in different galaxies are needed to decide between them.

\subsubsection{Mass-loss}
A positive correlation between mass-loss and luminosity has been widely demonstrated for RSGs \citep[e.g.][]{bon2010,mau2011}. A dependence with SpT has also been noted \citep{coh1973}, with later types demonstrating higher mass loss rates \citep{hum1972}. This dependence with SpT is not as mere indirect consequence of the SpT-luminosity relation, as it also happens at a given luminosity. \cite{loo2005} found an empirical relation between the mass-loss rate and both luminosity and temperature for AGBs and RSGs in the LMC. This relation establishes that the mass-loss rate grows for increasing luminosity and decreasing temperature (i.e.\ later SpTs).

\cite{bon2010} found that all the RSGs that they studied in the SMC presented a lower mass-loss rate than those in the LMC. Thus, temperature seems to have a major role in the mass-loss rates of CSGs. However, they noted that the colour that they used to estimate mass loss ($K_{\textrm{S}}-[24]$) measures mainly mass-loss related to dust, and there could be a fraction of escaped gas that is not detected by this colour.  On the other hand, \citet{mau2011} conclude that the total estimated mass-loss for RSGs scales with metallicity. This result suggests that the dusty mass-loss among SMC CSGs is lower than in those from the LMC not only because of the less favorable conditions for dust condensation (lower metallicities and higher temperatures than in the LMC), but also because the total mass-loss is indeed lower in SMC CSGs than in LMC ones. As our samples of CSGs for both galaxies are significantly larger than those in these previous works, we can perform a more extensive study of relationships between mass-loss, luminosity and SpT (i.e.\ temperature).

\begin{figure*}[th!]
   \centering
   \includegraphics[trim=1cm 0.5cm 1.8cm 1.2cm,clip,width=9cm]{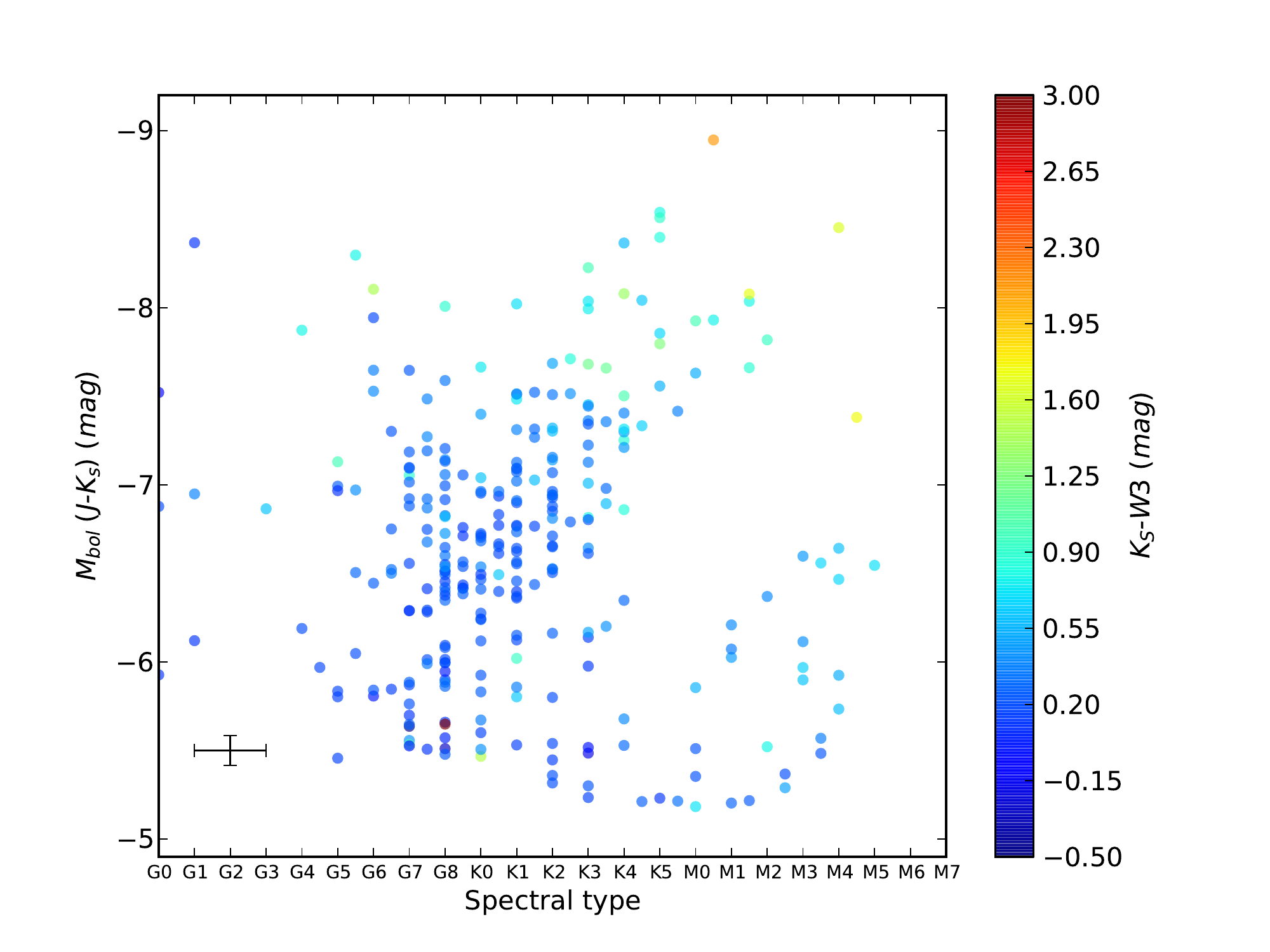}
   \includegraphics[trim=1cm 0.5cm 1.8cm 1.2cm,clip,width=9cm]{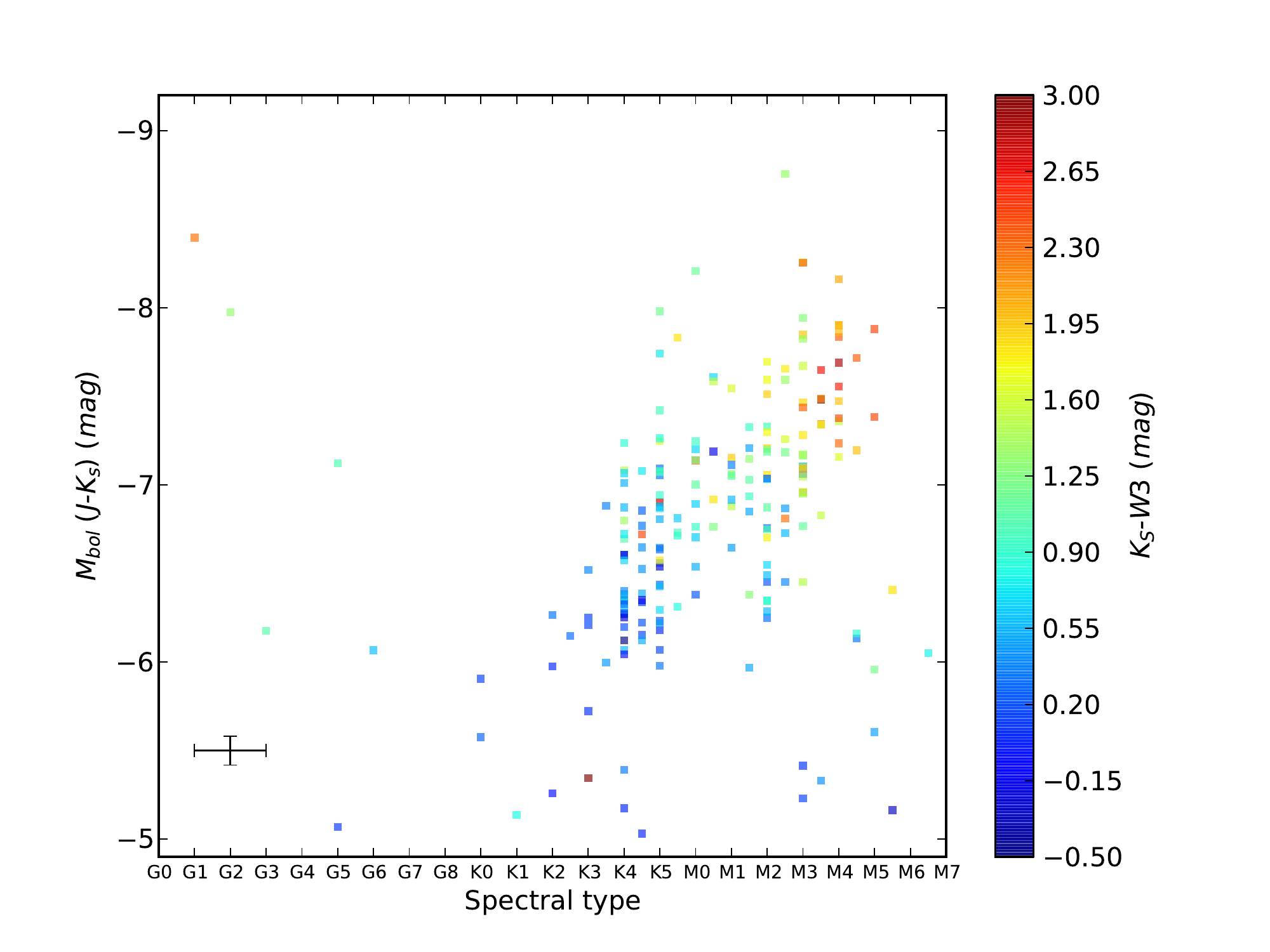}
   \caption{Spectral type to $M_{\textrm{bol}}$ (derived from ($J-K_{\textrm{S}}$). Colour indicates the value of $K_{\textrm{S}}-[W3]$, which is related to mass-loss (see text). The LMC data correspond to 2013 and the SMC data correspond to 2012, because all the stars observed in 2010 and 2011 were also observed in 2012 and 2013 and we wanted to avoid to alter the same CSG more than once. The black cross represents the median uncertainties. Note that both figures are in the same scale, to facilitate the comparison. {\bf Left (\ref{SpT_Mbol_ML}a):} CSGs from the SMC.  {\bf Right (\ref{SpT_Mbol_ML}b):} CSGs from the LMC.}
      \label{SpT_Mbol_ML}
\end{figure*}

The ($K_{\textrm{S}}-[12]$) colour was shown to be a good mass-loss indicator for RSGs by \cite{jos2000}. Unfortunately, IRAS photometry is not available for all our stars. In consequence, we have decided to use the WISE~$[W3]$ band \citep{wri2010}, as it is similar to IRAS-[12]. Fraser et al. (in prep) calculated that the difference between these bands is only an offset ($[12]=W3-0.435$). Thus, we have used the colour ($K_{\textrm{S}}-[W3]$), calculated through the 2MASS~$K_{\textrm{S}}$ and WISE~$[W3]$. The result is plotted in Fig.~\ref{SpT_Mbol_ML}. For the luminous group from the LMC, we can see that high values of ($K_{\textrm{S}}-[W3]$) clearly concentrate on the most luminous and latest CSGs. This behaviour matches with the prediction of the empirical formula proposed by \cite{loo2005}. The SMC CSGs exhibit the same behaviour, with the highest values of ($K_{\textrm{S}}-[W3]$) at highest luminosities and latest subtypes. Nevertheless, at a given luminosity the values of ($K_{\textrm{S}}-[W3]$) for the SMC sample are lower than those found for the LMC sample.

These results confirm the differences in dusty mass-loss rates (and thus in the dust production) between CSGs from the LMC and the SMC found by \cite{bon2010}, but as the colour ($K_{\textrm{S}}-[12]$) is also related to dusty mass-loss, we cannot give new hints about the total mass-loss rates. However, we have confirmed that at given luminosity CSGs from SMC are warmer than those from LMC. Thus the surface gravities of SMC CSGs have to be higher than those of LMC 
s, and this is likely to result in lower mass-loss rates.

\begin{figure*}[th!]
   \centering
   \includegraphics[trim=0.85cm 0.5cm 1.8cm 1.2cm,clip,width=9cm]{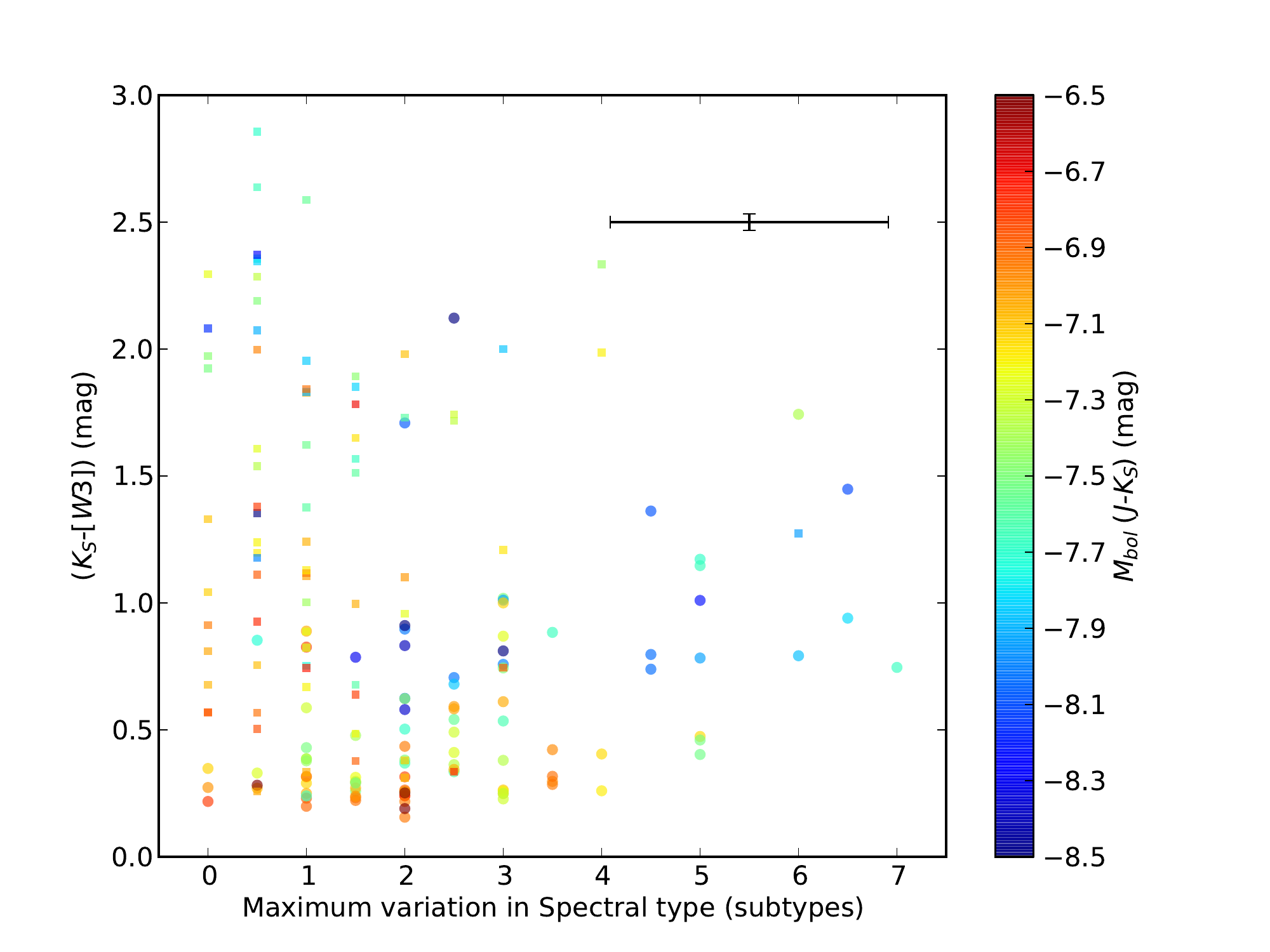}
   \includegraphics[trim=0.85cm 0.5cm 1.8cm 1.2cm,clip,width=9cm]{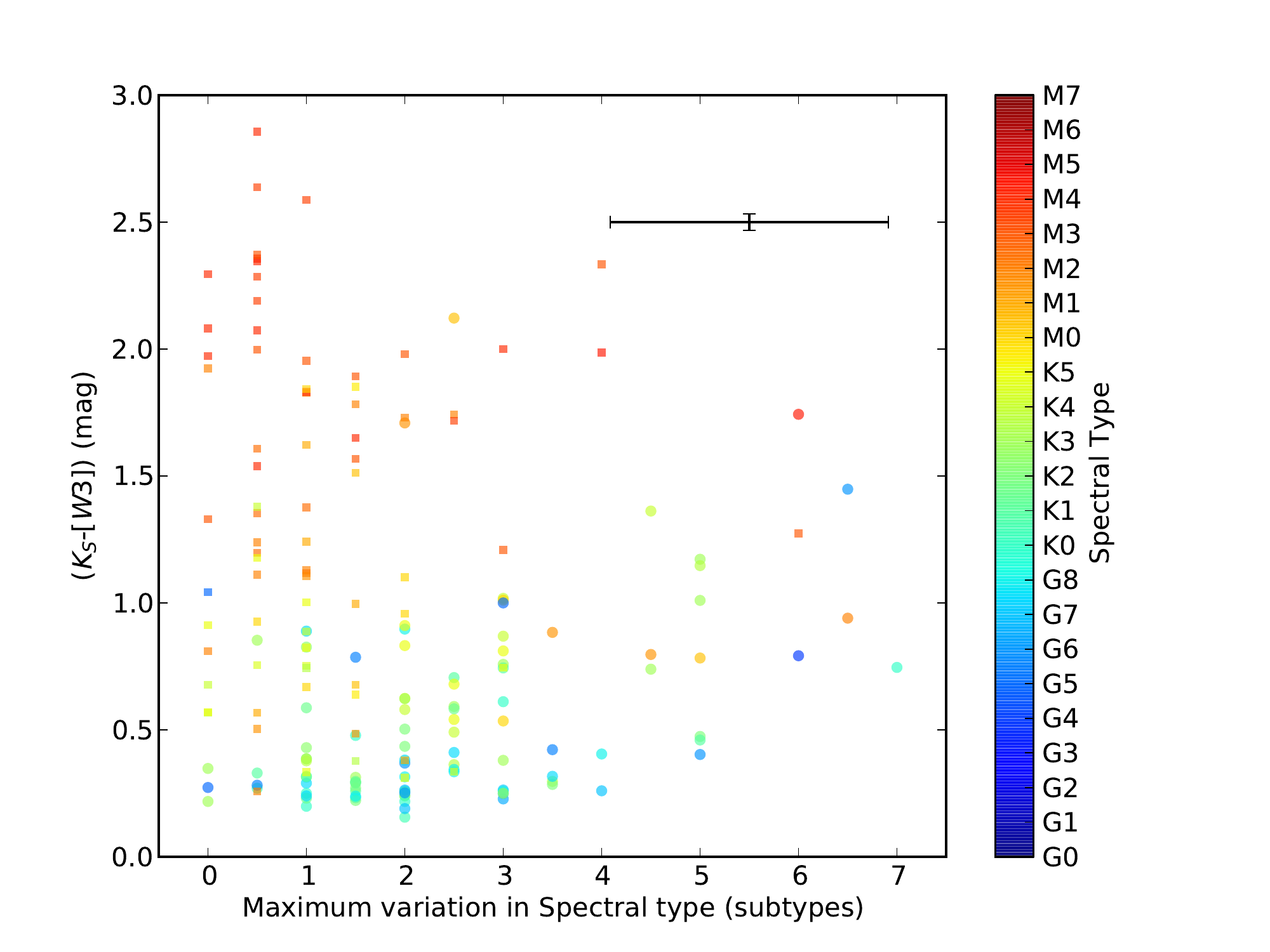}
   \caption{Maximum change observed in SpT to $K_{\textrm{S}}-[W3]$, which is related to mass-loss (see text). The samples used here are the same as in Fig.~\ref{SpT_Mbol}, but we have represented only those stars observed on more than one epoch. The black cross represents the median uncertainties. The circles are CSGs from the SMC, the squares from the LMC. {\bf Left (\ref{max_var_ml}a):} The colour indicates $M_{\textrm{bol}}$ (derived from ($J-K_{\textrm{S}}$))  {\bf Right (\ref{max_var_ml}b):} The colour indicates SpT.}
      \label{max_var_ml}
\end{figure*}

We have to note that while LMC CSGs present higher mass-loss than SMC ones, they have less spectral variability. To explore if there exists a relation between mass-loss and spectral variability, we have represented the maximum SpT change observed (Max($\Delta$SpT)) and the mass-loss indicator ($K_{\textrm{S}}-[W3]$) in Fig.~\ref{max_var_ml}. It includes only those stars from 2012 (SMC) and 2013 (LMC) that have been observed on more than one epoch. At first sight, there does not seem to be a clear relation. However, there could be a few different trends present in the plot. None of them is very clear, but we have to take into account that the measurement of the maximum change in SpT is the result of only two or three measurements at different random moments of their variation, i.e.\ even though Max($\Delta$SpT) is indicative of the maximum intrinsic spectral variation amplitude, it is not a definitive value. The Max($\Delta$SpT) values may change (to greater values of SpT variation) in future observations.

Most of the LMC CSGs identified as spectral variables (Max($\Delta$SpT)$>1$~subtype) seem to follow a trend (Trend~I). It begins around ($K_{\textrm{S}}-[W3]$)$\sim1.5$~mag, growing slowly up to ($K_{\textrm{S}}-[W3]$)$\sim2.5$~mag at Max($\Delta$SpT)$=4$~subtypes. There are also two CSGs from the SMC present in this trend. These two CSGs are more luminous than the LMC ones, but they share their late SpTs and they have two of the three highest ($K_{\textrm{S}}-[W3]$) values in our SMC sample. The behaviour of Trend~I (a positive correlation between mass-loss and the spectral variation amplitude) has not been reported before. However, there are a few RSGs known because of  higher mass-loss rates and larger spectral variability typical of RSGs. For examples, see \cite{lev2007} for the MCs and \cite{sch2006} for the Galaxy. The second trend (Trend~II) is dominated by high luminosity CSGs from the SMC with detected variability, plus a few objects from the LMC. It seems to begin around ($K_{\textrm{S}}-[W3]$)$\sim0.5$~mag, and to end about ($K_{\textrm{S}}-[W3]$)$\sim1.5$~mag, reaching the largest  Max($\Delta$SpT) values. This trend is broad and its lower ($K_{\textrm{S}}-[W3]$) values grow slightly ($\sim0.2$~mag) with Max($\Delta$SpT), while their upper ($K_{\textrm{S}}-[W3]$) values grow more clearly, reaching $\sim1.5$~mag. It is worth noting that most of the SMC CSGs in this trend have SpTs (from mid K to early-M) later than the average for their galactic population. They may have higher mass-loss rates than other SMC objects simply because lower temperatures favour mass-loss. There are a few early CSGs with large spectral changes, but they might have been caught at the early edge of their spectral variation range.

There are also two groups of CSGs which do not present a clear correlation between Max($\Delta$SpT) and ($K_{\textrm{S}}-[W3]$). One is formed by those CSGs without detected variability (Max($\Delta$SpT)$\leqslant1$), but with a minimum ($K_{\textrm{S}}-[W3]$) value of 0.5~mag. Most of them are from the LMC, and they are found with any ($K_{\textrm{S}}-[W3]$) value in our range (from 0.5 to almost 3 mag). There are also some CSGs from the SMC, but they are only found with low values of ($K_{\textrm{S}}-[W3]$) (less than 1~mag). The other group without correlation is formed by those CSGs with ($K_{\textrm{S}}-[W3]$)$<0.5$~mag. Most of these objects are mid-luminosity ($M_{\textrm{bol}}>-7.5$) CSGs from the SMC, with early SpTs (early K and G subtypes). Most CSGs in this group have detected spectral variation. However, they do not show any clear trend with mass-loss. It can be argued that this group is in fact part of  Trend~II. However we have decided to consider it a different group because it is comprised of CSGs with relatively early SpTs and mid luminosity, while Trend~II is composed by CSGs with relatively late SpTs (in terms of the SMC) and high luminosity.

\begin{table*}[th!]
\caption{Number of CSGs in each group described for Fig.~\ref{max_var_ml} (see text), and the number of them we have in common with Y\&J. As they grouped their RSGs in two groups, the LSP and LSP+SR (see text), we indicate how many of our stars belong to each of these groups. The 2-$\sigma$ uncertain intervals for the fractions ($\Delta f$) are shown, which are equal to $1/\sqrt[]{n}$ ($n$ is the total number of CSGs in common with Y\&J for the correspondent group).}
\label{trend_yan}
\centering
\begin{tabular}{c | c | c | c c c | c c c }
\hline\hline
\noalign{\smallskip}
&&Number of CSGs&\multicolumn{6}{c}{Groups from Yang~\& Jiang}\\
&Total number&in common with&\multicolumn{3}{c |}{LSP group}&\multicolumn{3}{c}{LSP+SR group}\\
Groups from Fig.~\ref{max_var_ml}&of CSGs&Yang~\& Jiang&Number&Fraction&$\pm \Delta f$&Number&Fraction&$\pm \Delta f$\\
\noalign{\smallskip}
\hline
\noalign{\smallskip}
Trend~I&16&6&5&0.83&0.41&1&0.17&0.41\\
Trend~II&42&22&9&0.41&0.21&13&0.59&0.21\\
Max($\Delta$SpT)$\leqslant1$&77&32&15&0.47&0.18&17&0.53&0.18\\
($K_{\textrm{S}}-[W3]$)$<0.5$&73&21&19&0.90&0.22&2&0.10&0.22\\
\noalign{\smallskip}
\hline
\end{tabular}
\end{table*}

Although Trends~I and~II seem similar, when we analyse the nature of the variation in the CSGs in each trend, we find significant differences. We have used the cross-match done in Sect.~\ref{variab} between our samples of CSGs observed on multiple epochs and the variable RSGs studied by Y\&J (see Table~\ref{trend_yan}). They classified their RSGs according to their dominant type of photometric variability in two groups. One is composed by those with short LSPs ($P<1000\:{\rm d}$) and SRs, which have in common periods of a few or several hundred days (LSP+SR group onward). The other group is composed by LSPs with periods of a few thousand days (LSP group onward). It seems accepted that the LSP+SR variability is caused by radial pulsations, while the origin of the LSP ones is not clear even though a number of different mechanisms have been proposed \citep[binarity, pulsation, convection cell, and surface hot spot; see][and references therein]{yan2012}. In any case, it seems clear that the underlying mechanism is different from that causing the radial pulsation of the LSP+SR group.

As we show in Table~\ref{trend_yan}, Trends~I and~II present significantly different fractions of CSGs classified as LSP or LSP+SR by Y\&J. While Trend~I is dominated by those with periods of thousands of days (LSP group), Trend~II presents a similar number of stars from both groups, LSP and LSP+SR (those with hundreds of days). In consequence, it seems unlikely that both trends may be manifestations of the same effect at different metallicities. On the other hand, the low  mass-loss group (($K_{\textrm{S}}-[W3]$)$<0.5$) is dominated by LSP stars from the SMC. In this, it might be related to Trend~I , which is mostly formed by LSP stars from the LMC. However, the two groups do not seem to share any other physical properties when compared to the rest of the population in their respective galaxies. While Trend~I stars present SpTs and mass-losses similar to the other LMC stars (the group without detected variability, i.e.\ Max($\Delta$SpT)$\leqslant1$), the low mass-loss components are earlier and with lower mass-losses than the CSGs from Trend~II.

In summary, even though we find some hints of correlations between mass-loss, spectral variability and the variability mechanisms, their interplay seems to be very complex. Further investigation is required to understand these connections.

\subsection{Spectral type distribution}
\label{SpT_distrib}

The SpT distribution of RSG populations has been studied in the past \citep{hum1979,eli1985,mas2003b}, for a few galaxies \citep[][and references therein]{lev2013}, even though only the MW and the MCs have been characterized with large observed populations. From all these works, two facts emerge: the SpT distribution has a bell-shape, and the centre and range of this distribution depends on the metallicity of the host galaxy.

\begin{figure*}[th!]
   \centering
   \includegraphics[trim=1cm 0.5cm 1.5cm 1.2cm,clip,width=9cm]{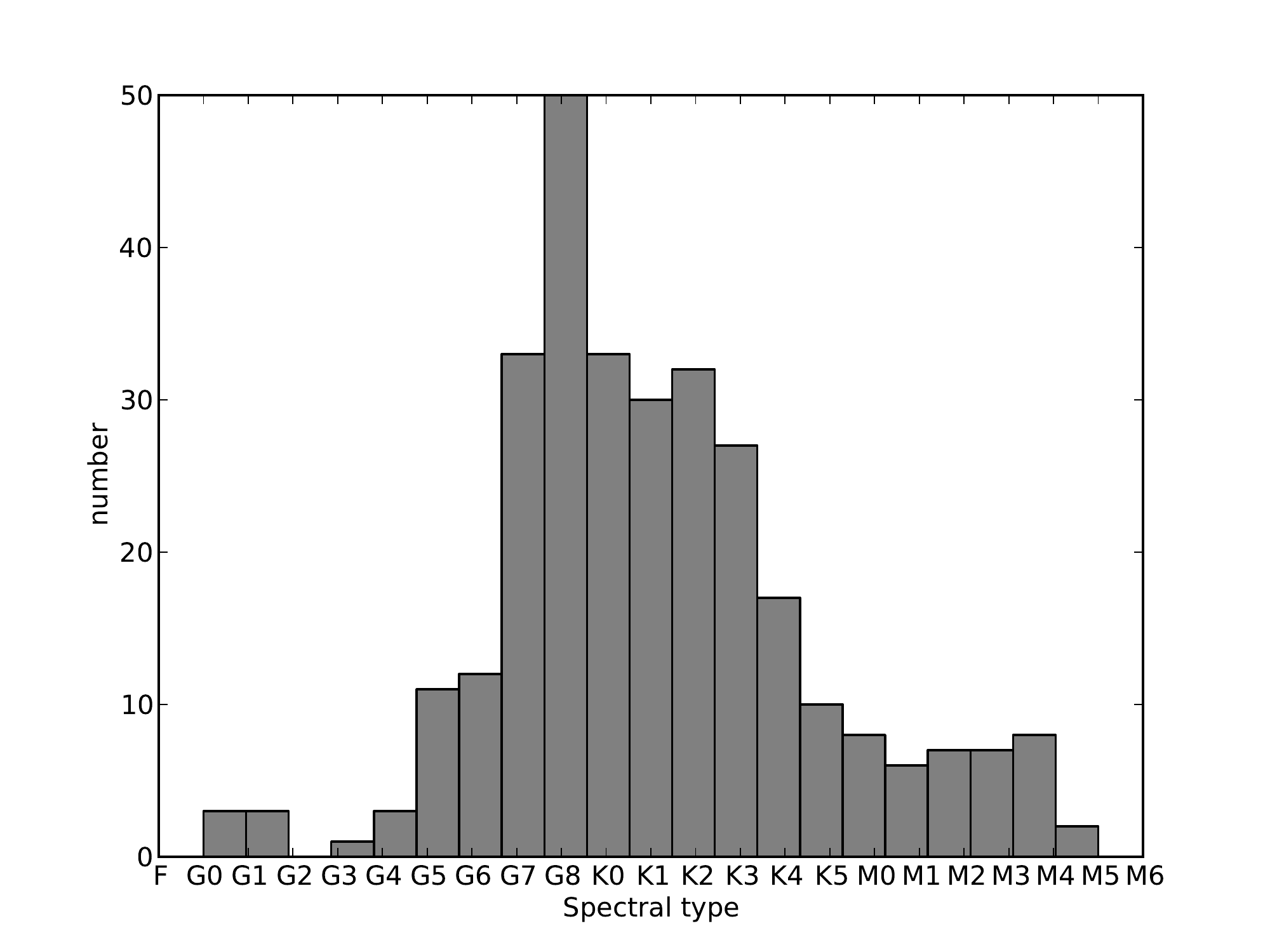}
   \includegraphics[trim=1cm 0.5cm 1.5cm 1.2cm,clip,width=9cm]{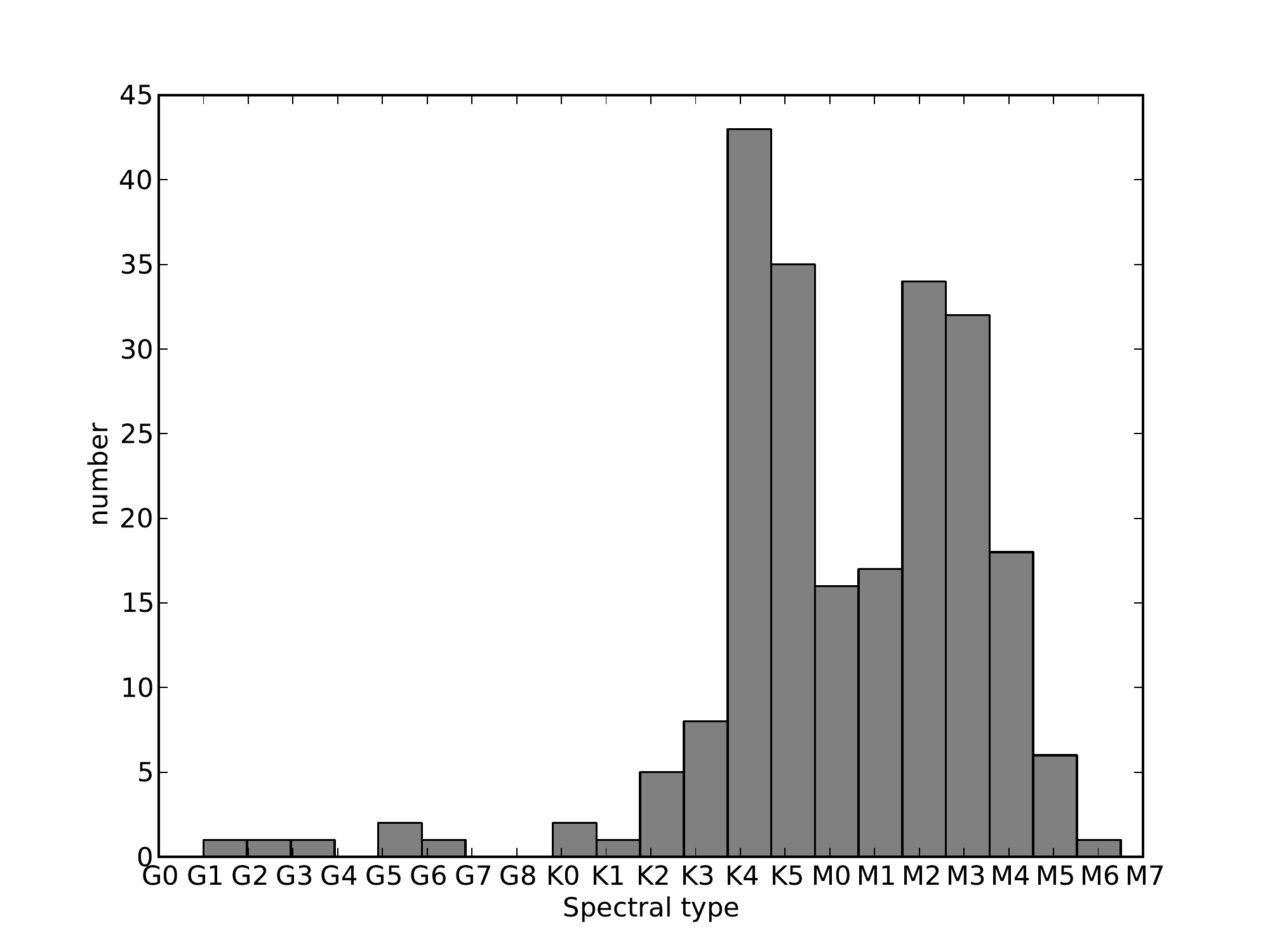}
   \caption{{\bf Left (\ref{histo}a):} Spectral type distribution for SMC CSGs  {\bf Right (\ref{histo}b):} Spectral type distribution for LMC CSGs.}
   \label{histo}
\end{figure*}

With the aim of checking the coherence of our spectral classification with respect to previous works, we have plotted the SpT distribution for each Cloud (see Fig.~\ref{histo}). For these histograms we have only used the 2012 SMC and 2013 LMC data: they contain all the stars from 2010 and 2011 too, and this way we avoid repetition. According to the literature, we expected distributions centred on K5\,--\,K7 for the SMC and on M2 for the LMC \citep{lev2013}. Our distributions show marked differences from this.

\cite{lev2012} noted that their SpT distributions from low metallicity galaxies, such as the SMC, might be incomplete because they only observed supergiants of K and M SpTs, while the CSG distribution seemed to continue into G~subtypes. We have observed a significant number of late-G~SGs in the SMC (84), confirming this suspicion (see Fig.~\ref{histo}a). The mean SpT for our SMC sample is K1, much earlier than in previous works: K5\,--\,K7 in \cite{mas2003b} and \cite{lev2013}, and M0 in \cite{eli1985}. This difference is partly due to the inclusion of G~CSGs in the SpT distribution moving the mean to earlier subtypes, but also because of the luminosity-SpT trend: in previous works the number of CSGs observed was smaller, and they were the brightest targets available. Therefore, these samples are biased towards the most luminous CSGs, which are also those with the latest SpTs.

The asymmetric SMC distribution toward later subtypes (K and M) is probably caused by two factors: an over-representation of RSGs with late types and a lack of G supergiants, both caused by our selection criteria. As we showed in Fig.~9 from Paper~I, our selection efficiency is significantly higher for late-K and M subtypes than for earlier ones, although it grows again for mid~G stars. Indeed, 49 G~SGs were found among the targets selected by our criteria. In spite of this, we can be certain that there is a large number of early or mid-G CSGs that we did not observe. The work by \cite{neu2010} identifies 176 YSG candidates, and none of the 49 G~SGs selected by our criteria are in common with their list. In addition to our own targtes, we also observed stars from this list of 176 YSG candidates as low priority targets to fill spare fibres. Of 43 stars in this list that turn out to have a late SpT (a significant fraction have B, A or F SpT), we have classified 35 as G~SGs, with the other eight being later. In view of this incompleteness in the sample of G-type SGs, the mean SpT for SMC CSGs could be even earlier than K1, which is already significantly earlier than in previous works.

Unexpectedly, the LMC distribution does not seem to have a clear central peak, but rather to be bimodal, with a maximum centred at K4\,--\,K5 and a secondary peak at M2\,--\,M3. A slightly similar behaviour has been found for the LMC before, by \cite{lev2012}, but the bimodality was not so clear. Apart from the maximum at M2, their data hint at a second maximum at mid-K, but the number of K-type stars in their LMC sample is too low (less than 10 CSGs with K~subtypes) to consider it statistically significant. This bimodality has not been commented, analysed or explained before in any work. The Galactic sample in \citet{lev2012} shows a more clear secondary maximum at mid\:K, but in the MW most of the K supergiants are low-luminosity objects, and many are found in clusters too old to host real RSGs. Therefore, a significant fraction of K\,Ib stars should not be real supergiants but high-luminosity red giants \citep{neg2012b}. Thus, the apparent bimodality in the MW arises from the inclusion of a number of objects that are physically red giants in spite of their morphological classification. In contrast, our LMC sample hardly contains Ib stars with $M_{\textrm{bol}}<-6$~mag, and thus it is impossible that bimodality would be caused by a similar effect.

\begin{figure}[h!]
   \centering
   \includegraphics[trim=1cm 0.5cm 1cm 1.5cm,clip,width=9cm]{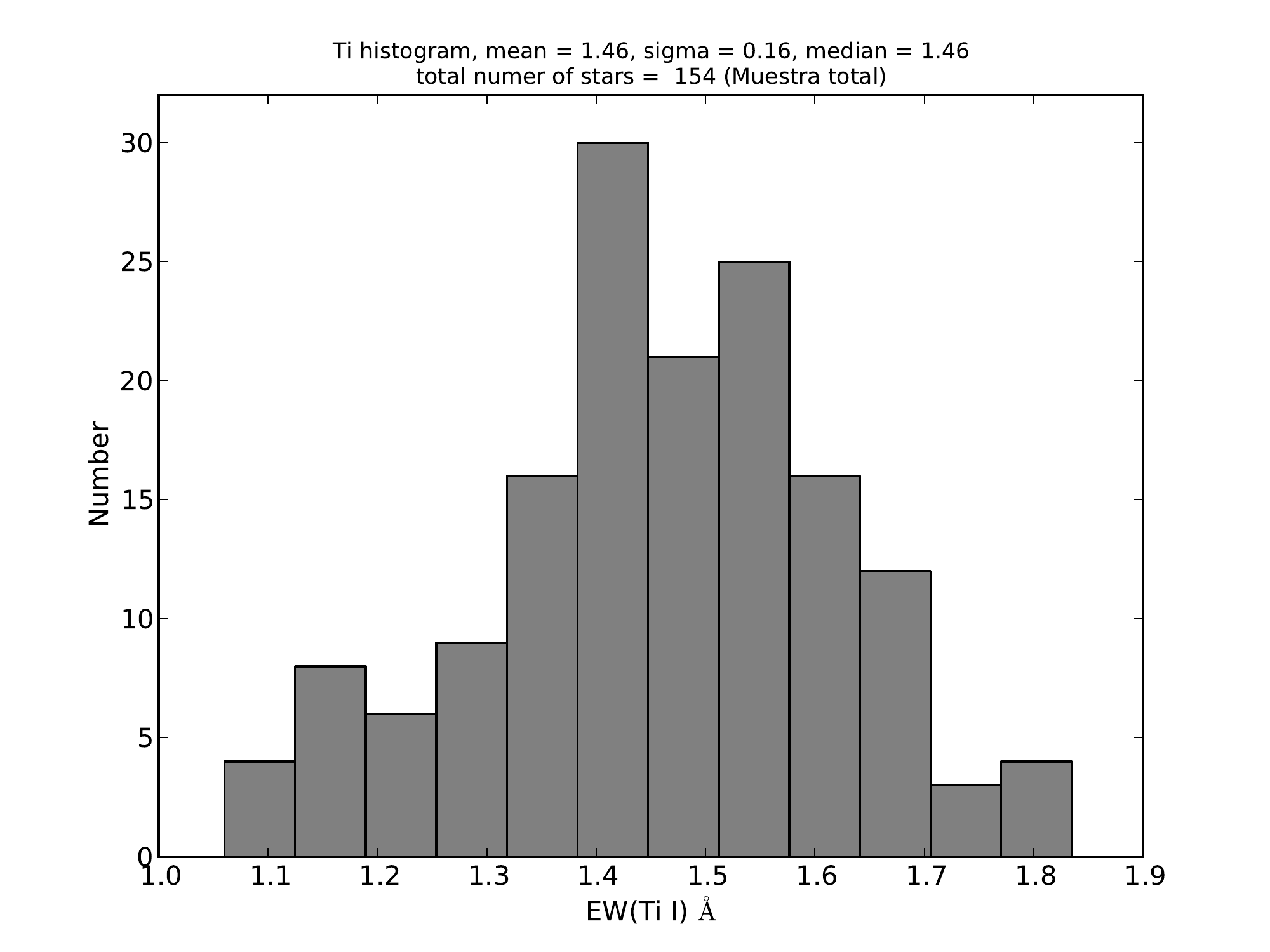}
   \caption{Histogram of the sums of the equivalent widths of Ti\,{\sc{i}}, for the LMC 2013 sample. The CSGs later than M3 have been not included here, because their EW(Ti\,{\sc{i}}) are affected by TiO bands (see Sect.~\ref{EW}). The few and scatter CSGs with values of EW(Ti\,{\sc{i}}) lower than $1$\:\AA{} have been considered outliers for this histogram, and have been not included.  The width of the bins have been calculated by the Freedman-Diaconis rule, multiplied by a factor of $0.8$.}
   \label{hist_Ti}
\end{figure}

Initially, we considered the possibility that this bimodality might be caused by a systematic error in the spectral classification, but this is easily ruled out. We have verified that SpT is a qualitative measurement of the temperature, and we have another variable related to temperature, EW(Ti\,{\sc{i}}), whose measurement is independent of the SpT\footnote{Ti\,{\sc{i}} lines were measured only in the CaT spectral region, while the SpTs were assigned by inspection of the optical range, and are thus totally independent.}, and therefore not affected by any systematic error in SpT classification. In Fig.~\ref{hist_Ti}, we represent the distribution of EW(Ti\,{\sc{i}}) in a histogram, where the size of the bin has been determined by the Freedman-Diaconis rule (with a scale factor 0.8). This histogram presents a minimum at EW(Ti\,{\sc{i}})$\sim1.5$\:\AA{}. This value roughly corresponds to SpTs M0\,--\,M1 (see Fig.~\ref{SpT_Ti}b) as we should expect if the minimum in the SpT distribution is not a product of a systematic error in the classification. However, we have to note that the minimum in the distribution of EW(Ti\,{\sc{i}}) does not seem so low as that in Fig.~\ref{histo}b. This is because of three reasons. Firstly, the correspondence between EW(Ti\,{\sc{i}}) and SpT is broad. At a given subtype, we have a dispersion as large as $0.4$\:\AA{} in EW(Ti\,{\sc{i}}). Secondly, the linear behaviour exhibited by EW(Ti\,{\sc{i}}) starts to change (becoming less  sloped) at M1\,--\,M2. Thus, we have many M2 and M3 stars with values of EW(Ti\,{\sc{i}}) around $1.5$\:\AA{}. Finally, because of the effect of the TiO bands over the Ti\,{\sc{i}} lines, we did not measure EW(Ti\,{\sc{i}}) in CSGs with SpTs later than M3. In consequence, the values of EW(Ti\,{\sc{i}}) larger than 1.5 are slightly underpopulated.

\begin{figure*}[th!]
   \centering
   \includegraphics[trim=1cm 0.5cm 1.5cm 1.2cm,clip,width=9cm]{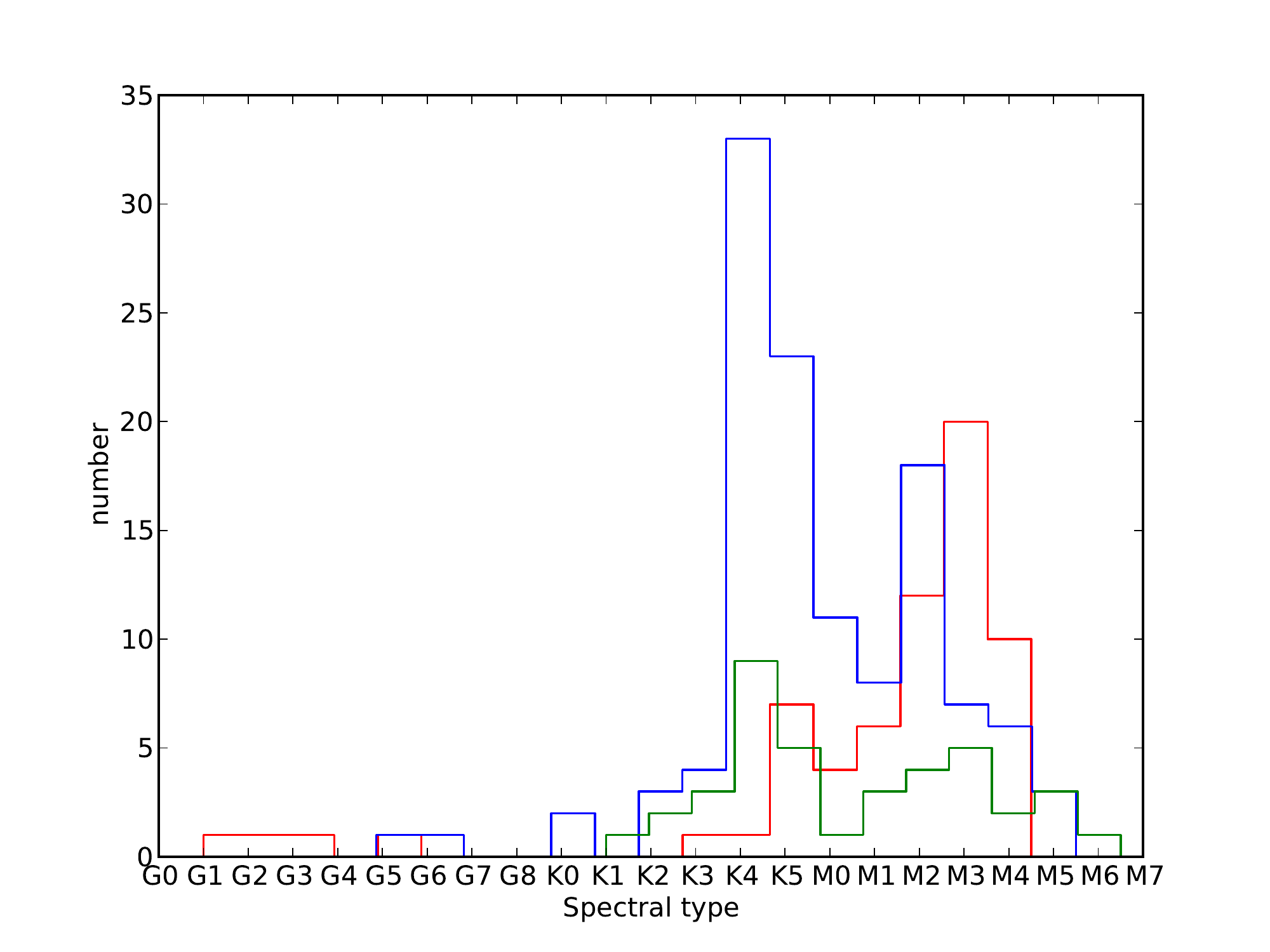}
   \includegraphics[trim=1cm 0.5cm 1.5cm 1.2cm,clip,width=9cm]{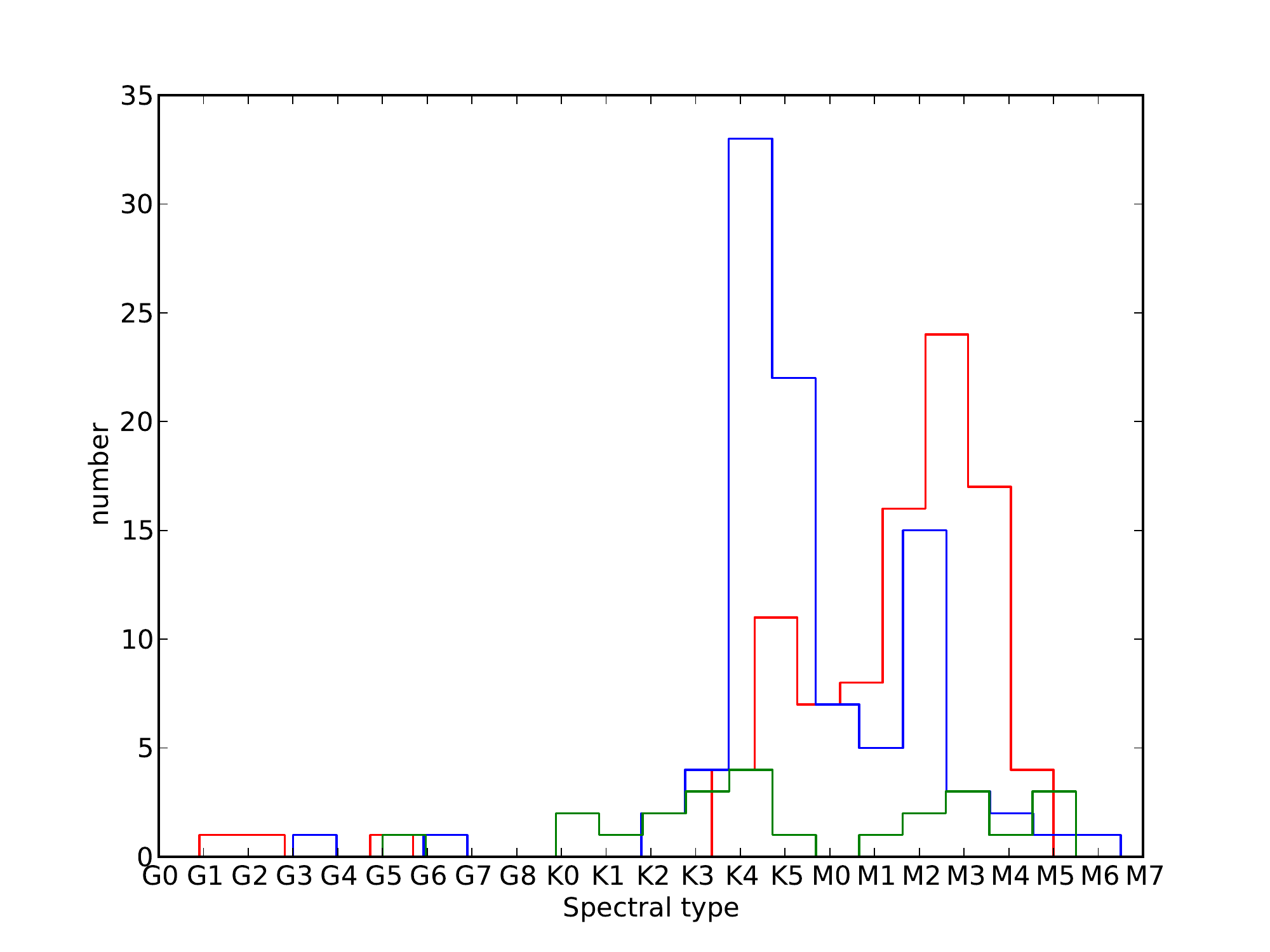}
   \caption{Spectral type distributions for stars in the LMC, segregated by luminosity. {\bf Left (\ref{histo_seg_LMC}a):}Segregation done by the assigned LC (Red: Ia and Ia-Iab; Blue: Iab and Iab-Ib; Green: Ib and Ib-II) {\bf Right (\ref{histo_seg_LMC}b):} Segregation done by $M_{\textrm{bol}}$ (Red: $M_{\textrm{bol}}<-6.92$; Blue: $-6>M_{\textrm{bol}}\leq-6.92$; Green: $M_{\textrm{bol}}\leq-6$).}
   \label{histo_seg_LMC}
\end{figure*}

A closer examination of the sample shows that most of the RSGs in the second maximum are high-luminosity objects, with luminosity classes Ia and Ia-Iab. To check the reality of this effect, we segregated the data by their LC (Fig.~\ref{histo_seg_LMC}a), and found that the bimodality is caused by the superposition of two different distributions, one dominated by high luminosity CSGs (Ia or Ia-Iab), and the other by mid-luminosity (Iab or Iab-Ib) CSGs. This is coherent with the SpT-luminosity relation. To make sure that there is no systematic effect in the assignation of luminosity classes leading to this bimodal distribution, we have used the $M_{\textrm{bol}}$ as a direct measurement of a star's actual luminosity. We calculated $M_{\textrm{bol}}$ in Sect.~\ref{mbols}, and its value is completely independent of the LC or SpT assigned. We split the LMC sample in three groups. The faint group, with $M_{\textrm{bol}}>-6$~mag, contains the lowest luminosity CSGs; as we have shown before (Sect.~\ref{SpT_lum}), these stars present a different behaviour to mid- and high-luminosity CSGs. We will not use this faint group, as our sampling at these luminosities is very sparse and certainly biased, and contamination by luminous AGB stars cannot be discarded. For stars with $M_{\textrm{bol}}<-6$~mag, we have used the median value ($M_{\textrm{bol}}=-6.92$) to split the sample, with the mid-luminosity group comprising stars with $-6.92 < M_{\textrm{bol}}<-6$~mag, and the bright group formed by stars with $M_{\textrm{bol}}<-6.92$~mag. In Fig.~\ref{histo_seg_LMC}b, we plot the result of this segregation finding a distribution very similar to that obtained using our LCs. If anything, the separation between the two populations becomes more obvious.

To evaluate whether the bimodality is statistically significant, we tested if both luminosity groups can be considered part of the same population (the bimodality is not statistically significant), or if they can be treated as different statistical populations. For this, we used the Kolmogorov-Smirnov test (KST), because we do not know {\em a priori} if these distributions are normal or not. The KST cannot be used for discrete data, as is the case of our SpT classification. Therefore we transformed our sample to a continuous distribution by following these steps: firstly, we split the sample in three groups by their LCs, as explained above. Secondly, we counted how many stars belong to each spectral subtype, separately for the high- and mid-luminosity samples. Then, for each subtype, we generated a random normal distribution of subtypes with the same number of objects that the subtype originally has, centred on the corresponding subtype and with a sigma of 1 subtype (the mean uncertainty). To avoid random fluctuations, we generated $10\,000$ continuous SpT distributions, calculating then the KST for each one. The fraction of randomly generated SpT distributions that pass the KST also gives us an idea of the likelihood that another observational sample similar to ours would give a similar bimodal behaviour in its SpT distribution. The same process has been also done for the groups segregated by $M_{\textrm{bol}}$ values. 

In the KST, the null hypothesis is that both groups (high- and mid-luminosity) come from the same statistical population. We obtained that all the continuous randomly generated SpT distributions have probabilities that the null hypothesis is correct lower than the 3-$\sigma$ significance level, with medians of $2\cdot10^{-6}$ and $10^{-8}$ for LC and $M_{\textrm{bol}}$ segregations respectively. Therefore, all of them reject the null hypothesis, meaning that the presence of two populations with different SpT distributions is confirmed as the cause of the bimodality.

\begin{figure*}[th!]
   \centering
   \includegraphics[trim=1cm 0.5cm 1.5cm 1.2cm,clip,width=9cm]{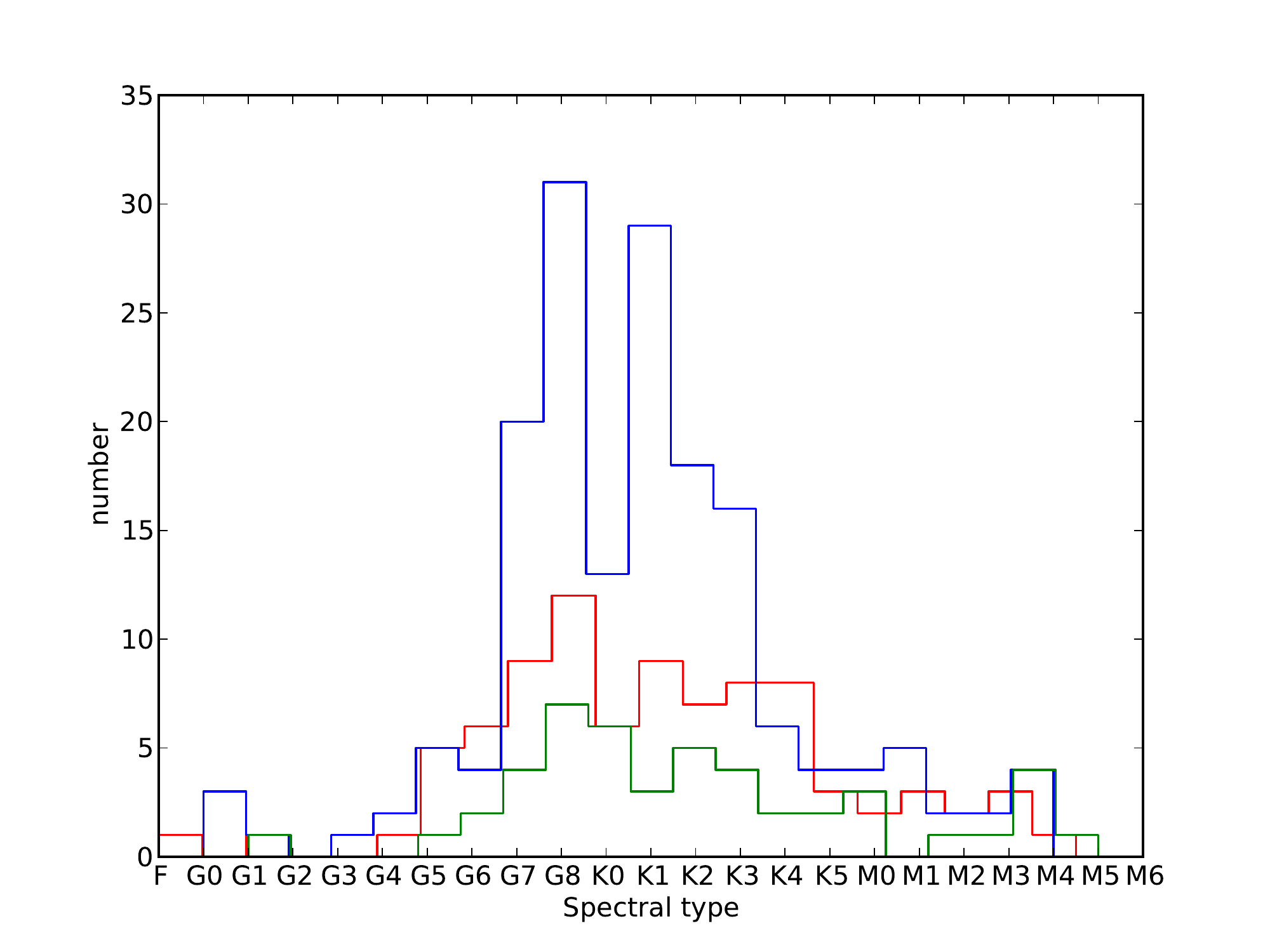}
   \includegraphics[trim=1cm 0.5cm 1.5cm 1.2cm,clip,width=9cm]{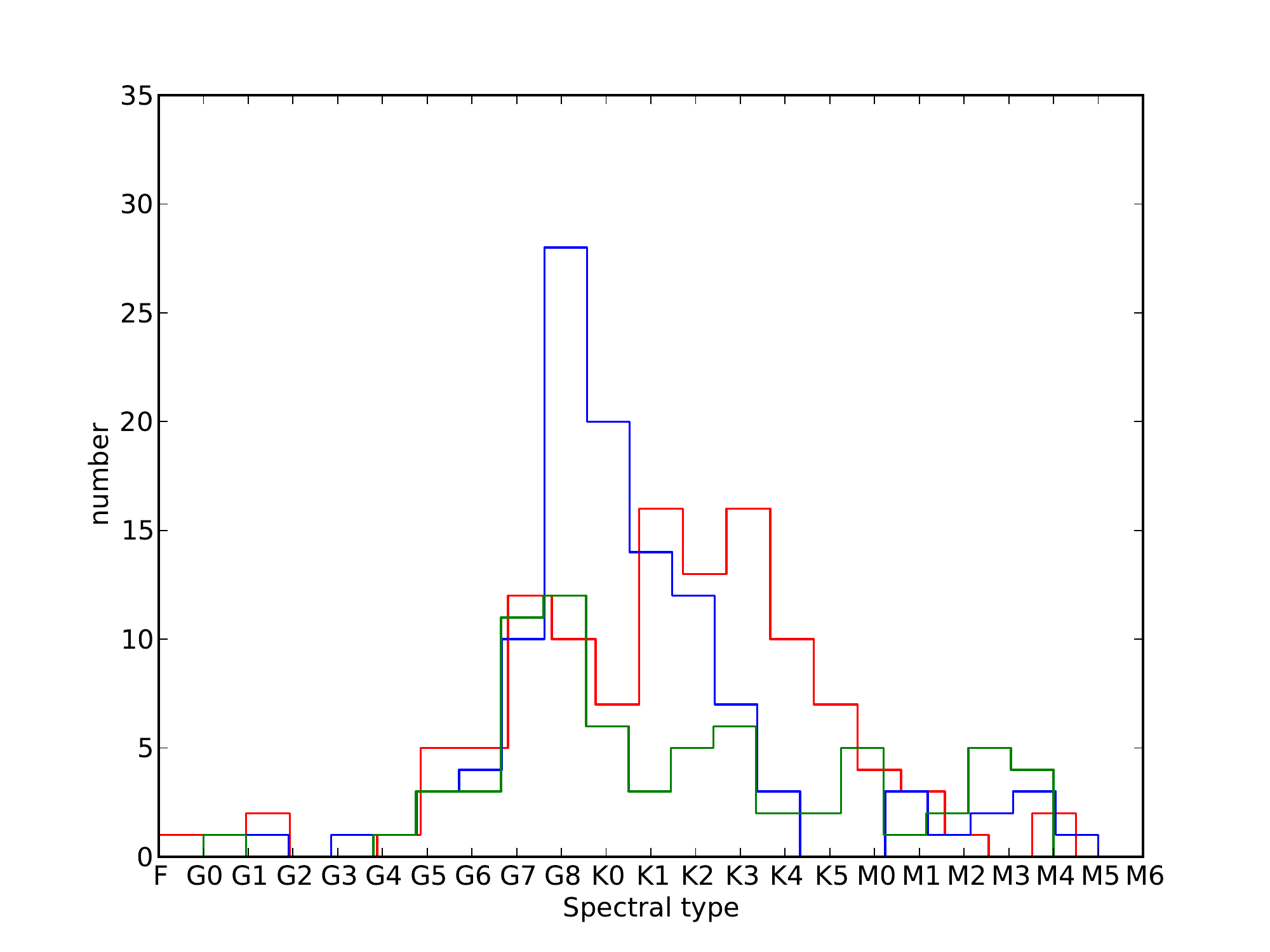}
   \caption{Spectral type distributions for stars in the SMC, segregated luminosity.{\bf Left (\ref{histo_seg_SMC}a):}Segregation done by the assigned LC (Red: Ia and Ia-Iab; Blue: Iab and Iab-Ib; Green: Ib and Ib-II) {\bf Right (\ref{histo_seg_SMC}b):} Segregation done by $M_{\textrm{bol}}$ (Red: $M_{\textrm{bol}}<-6.88$; Blue: $-6>M_{\textrm{bol}}\leq-6.88$; Green: $M_{\textrm{bol}}\leq-6$)}
   \label{histo_seg_SMC}
\end{figure*}

For the SMC, both segregation options, by $M_{\textrm{bol}}$ (with median $M_{\textrm{bol}}=-6.88$~mag) and by LC (see Fig.~\ref{histo_seg_SMC}), do not show the same bimodality we find in the LMC. Nevertheless, when we segregate the stars by $M_{\textrm{bol}}$, we find that high-luminosity CSGs have slightly later types than mid-luminosity ones, which is a logic consequence of the SpT-luminosity relation. The lack of bimodality for the SMC SpT distribution could be a consequence of the previously commented lack of mid-G CSGs in our sample, {\em if} all of them would be mid-luminosity CSGs. Even if this hypothesis is right, still the high-luminosity CSGs from the SMC span a wide variety of SpTs (from mid~G down to M4). Therefore, a hidden bimodality among SMC CSGs does not seem likely.

We have to take into account as well the possibility that the bimodality in the LMC is caused by a selection effect on our sample. As this effect would need to filter out a large number of CSGs with high or mid luminosities, it should affect the extremes of our luminosity-SpT distribution (stars too faint or too bright), not its centre. Moreover, as can be deduced from Fig.~9 in Paper~I, the $Q_{\textrm{IR}}$ pseudo-colour criterion that we used would leave out preferentially stars with early-K or late-M-types, but should be very effective for SpTs close to M0, where the separation between the two peaks lies. For these reasons, the possibility of a selection effect causing the bimodality is very unlikely. Finally, we have evaluated the spatial distribution of high- and mid-luminosity CSGs in the LMC, and we have not found any significant inhomogeneity that may explain the bimodality as a consequence of our spatial sampling.

The alternative to a selection effect is that the SMC and the LMC CSGs show a different physical behaviour. For the SMC, all (or almost all) of the observed CSGs are in the same evolutionary state, characterized by low dust production and probably low mass-loss (see Sect.~\ref{SpT_lum}) at all luminosities. Instead, the CSGs from the LMC seem to have two differentiated states, one (named state~I onward) composed by most of the mid-luminosity CSGs and a few of the high-luminosity ones (which are indeed the most luminous and the latest within this state), and the other one (state~II onward) formed by the majority of high-luminosity CSGs, with some mid-luminosity CSGs among the less luminous and earlier stars within it this state. The state~I CSGs are characterized by earlier SpTs (mean SpT of K5) than state~II (mean SpT of M2.5), and lower mass-loss rates and dust production. The CSGs from state~II have mostly M subtypes, significant dust production and, likely, higher mass-losses (see Fig.~\ref{SpT_Mbol_ML}).

Using the relation $\log{(\dot M_{d}/\dot M_{\sun}a^{-1})}=0.57(K-[12])-9.95$ from \cite{jos2000}, we can estimate the mass-loss rate in form of dust\footnote{Note that for LMC RSGs the gas-to-dust ratio in the mass-loss is usually took around 500 \citep{mau2011}.}, $\dot M_{d}$, in our samples. We use here the equivalence $[12]=W3-0.435$ calculated by Fraser et al. (in prep.). We caution that this is just a rough estimation, because \textit{we use the $(K_{\textrm{S}}-[W3])$ colour, which is similar to $(K-[12])$ but not exactly the same colour, and also because} the relation was obtained from a MW sample, while our samples have significantly different metallicities. The typical values of $(K_{\textrm{S}}-[W3])$ among CSGs from state~I are around $0\:$mag, which corresponds to $\log{(\dot M_{d}/\dot M_{\sun}a^{-1})}\sim-10.2$. For the state~II, the typical values of $(K_{\textrm{S}}-[W3])$ are around $\sim2$, which corresponds to $\log{(\dot M_{d}/\dot M_{\sun}a^{-1})}\sim-9.0$. Thus, the mass-loss rate of the CSGs in the state~II is higher than those of CSGs in the state~I by more than one order of magnitude. In the SMC, instead, almost all the CSGs have $(K_{\textrm{S}}-[W3])$ values between 0 and 1, which correspond to values of $\log{(\dot M_{d}/\dot M_{\sun}a^{-1})}$ between $-10.2$ and $-9.6$.

The question then is, why is there not a gradient from one state to the other? In the SMC we see a smooth variation of $(K_{\textrm{S}}-[W3])$ with SpT, but in the LMC, the state~II group is concentrated at late SpTs (see Fig.~\ref{SpT_Mbol_ML}), with its own SpT distribution. The shift in SpT and the lack of a gradient suggest that, when the stars reach some physical conditions, namely sufficiently high luminosity and sufficiently low temperature\footnote{In the LMC these conditions correspond to ($M_{\textrm{bol}}<-7$) and SpT M0\,--\,M1}, they become unstable and evolve quickly toward later SpTs and higher luminosities, while at the same time increasing significantly their mass-loss rate.

It is important to note that the evidence found does not necessarily imply that these state~II CSGs are more evolved than state~I ones, as \cite{dav2013} seem to suggest when they state that RSGs with high luminosity, late SpT and high mass-loss rate are more evolved. In view of standard evolutionary tracks (such as those in Fig.~\ref{mod}), the state~II may be formed by a mixture of RSGs sufficiently massive to have reached these conditions evolving from blue supergiants at roughly constant luminosity and some with lower masses that have reached a luminosity high enough to be in this state because of their evolution in the RSG phase (see the discussion in Sect.~\ref{variab}). Given the shape of the IMF, there should be more lower-mass RSGs reaching this state. However, the CSGs with lower masses should be in a more evolved stage to have reached this high luminosity, and so the time they spend in this state has to be short. Since these effects cancel out, there is no {\em a priori} reason to think that they should dominate the population of state~II. Moreover, a significant number of state~II RSGs have $M_{\textrm{bol}}\sim-8$, and the evolutionary tracks predict that only those RSGs with $M>15\:M_{\sun}$ may reach such luminosities. Therefore there must be a significant fraction of massive RSGs in state~II, corresponding to the massive RSGs with late SpTs in the MW discussed in Sect.~\ref{SpT_lum}.

Another interesting question is whether there is a population equivalent to state~II in the SMC. As pointed out by \cite{lev2007}, there is a number of RSGs in the SMC with luminosities and mass-loss rates that may be equivalent to state~II. Since these stars are extreme in their conditions with respect to other CSGs from the SMC, we do not expect a large number of them. In consequence, they are too scarce to cause a noticeable bimodality in the SpT distribution. There are four stars in our sample that, according to their position in the SpT-luminosity diagram and their mass-loss, could be in state~II: [M2002]~18592, the most luminous star in our SMC sample with $M_{\textrm{bol}}\sim-9$, [M2002]~55188 \citep[which was already studied in][]{lev2007} and SMC400, both with types later than M2 and with ($M_{\textrm{bol}}<-7$), and SkKM13, which has $M_{\textrm{bol}}\sim-8$ and was classified as M1.5 in our 2012 observation, but as M3 in 2010. Moreover, two of them, [M2002]~18592 and SkKM13, fall along variability trend~I (dominated by LMC luminous stars) in Fig.~\ref{max_var_ml}. The existence of these few CSGs in the SMC seems to support the scenario proposed, as their SpTs are extreme for the SMC SpT distribution. If so, these stars would have reached such late SpTs (really uncommon among SMC CSGs) through the feedback mechanism proposed earlier.

If we attribute the very different fraction of stars reaching the state~II in the SMC and the LMC to their difference in metallicity, we should then expect RSGs with properties similar to those in state~II to be significantly present in the MW. \cite{eli1985} show their sample of Ia and Iab RSGs for three galaxies in their Fig.~1. While the mean SpT for the combined population of Ia and Iab RSGs from the MW is around M2, for the Ia objects alone, it is around M3\,--\,M4. Thus, similarly to the MCs, the SpT distribution in the MW also presents a shift to later subtypes between mid- and high-luminosity RSGs. Moreover, many of the MW Ia RSGs are well-known to display signs of very heavy mass-loss \citep{hum1974b}. On the other hand, in the sample of \cite{eli1985} there is no evidence of a miminum in the SpT distribution between both states. This might have a physical reason if the instability conditions are fulfilled only at the early M subtypes (M0\,--\,M1), because the typical subtypes of Iab supergiants in the MW are around M2, and so the minimum would not be noticeable. However, this lack of a minimum may also be due to sampling effects, because our sample has a higher fraction of high-luminosity CSGs than the typical galactic samples. For example, some of the MW CSGs with the latest spectral types, highest luminosities and largest mass-loss rates \citep[e.g. VX~Sgr, S~Per or VY~CMa; see][]{sch2006,wit2012} are not present in the list of \cite{eli1985}.

Very luminous RSGs with SpTs later than the average have been observed in some Galactic open clusters, for example Stephenson~2 \citep[see the discussion about it in Sect.~\ref{SpT_lum} and the work of][]{neg2013}. This SpT bimodality among the galactic RSGs in open clusters also seems to support our hypothesis of two separate states. These clusters have a given age, but some of them contain one or a few RSGs whose SpTs are later and whose luminosities are quite higher than those of the rest of the RSGs in the cluster. Thus, if the difference between the two states is due to evolution along the RSG track, the pass from one state to the other has to be a relatively quick process, because the lifetime of these stars in the RSG phase is short. Unfortunately, clusters in the MW typically have a small number of RSGs, and so contain only a few, if any, late and bright RSGs with mass-loss. For example, NGC~7419, the cluster with the largest number of RSGs in the solar vicinity, only contains five RSGs, four Iab with SpTs around M1 and one with a late SpT (MY~Cep, M7.5\,I), which is much brighter \citep{mar2013}. Thus, the distribution of SpTs among RSGs in galactic open clusters also seems to point to a bimodal distribution, but the clusters do not provide the numbers needed to have statistical significance.

This separation in two evolutionary states has not been described before, even though many works have studied RSGs in the MCs. There are good reasons for this, as the separation becomes evident only once a large enough sample of CSGs is available. In addition, previous works that have studied large numbers of CSGs from the LMC lacked enough mid-luminosity stars, as they focused their efforts on the bright end of the distribution ($M_{\textrm{bol}}<-7$; \citealt{mas2003b,lev2006}). Conversely, \cite{neu2012} present a large number of CSGs and reach low enough luminosities, but they do not perform spectral classification, or evaluate their mass-loss. However, we have to note that some works have already highlighted the presence of some very late, high-luminosity and high-mass-loss CSGs in the MCs \citep[e.g.][]{hum1974a,eli1986,sch2006,lev2007}.

\section{Conclusions}
In this work we have analysed the behaviour of the main spectral features in the CaT range ($8400$--$8900$\:\AA{}) with SpT and luminosity ($M_{\textrm{bol}}$) in a large sample of CSGs from both MCs.

\begin{enumerate}

\item We have found a strong correlation between SpT and EW(Ti\,{\sc{i}}), a weaker correlation between SpT and EW(Fe\,{\sc{i}}) and an almost flat trend for EW(CaT) against SpT, in both MCs. Our data also show that $M_{\textrm{bol}}$ has a strong correlation with EW(Fe\,{\sc{i}}), and significant ones with EW(CaT) and EW(Ti\,{\sc{i}}), with EW(Ti\,{\sc{i}}) having the weakest correlation with $M_{\textrm{bol}}$ among the three indices measured. Finally, we have studied the changes in the spectral indices when a given star experiments changes in SpT (naturally, restricting our study to spectrally variable CSGs), finding that there is a significant correlation between changes in the EW(Ti\,{\sc{i}}) index and changes in SpT. On the other hand, there is no correlation at all (coefficients $\sim0$) between changes in SpT and variations in the indices EW(Fe\,{\sc{i}}) and EW(CaT), none of which seems to change significantly, even when the star experiments changes larger than half a spectral type. Our synthetic spectra (as well as classical spectral classification criteria) suggest that EW(Ti\,{\sc{i}}) should be the index with the strongest dependence on $T_{\textrm{eff}}$ and the weakest dependence on luminosity, while EW(Fe\,{\sc{i}}) and EW(CaT) should be more sensitive to luminosity and much less to $T_{\textrm{eff}}$. These correlations are hardly compatible with the hypothesis that all RSGs have the same $T_{\textrm{eff}}$ with their SpTs determined by luminosity, except if the behaviour predicted by {\em all} 1D models for atomic lines is completely wrong. In consequence, we consider much more reasonable that, at least in the range from G0 to early-M types, spectral type is determined by temperature much more directly than by any other physical properties. Therefore it should be possible to define a $T_{\textrm{eff}}$ scale for CSGs. The scale itself will likely depend to some degree on the atmosphere models used as references, and may even be slightly different if calculated with different spectral features, because not all features are produced at the same depth in the extended atmospheres of CGSs.

\item Despite this, we have confirmed through robust statistical tests previous suggestions that the CSGs from each galaxy show a clear trend towards later SpTs as the luminosity increases.  This correlation is much stronger in the LMC, because substantial spectral variability somewhat hides it in the SMC. None of the current families of evolutionary models predicts this sort of behaviour. We consider two possible interpretations: either more massive (and hence more luminous) stars tend to display later spectral types, or all stars evolve to higher luminosity and lower $T_{\textrm{eff}}$ after some time as RSGs. These two scenarios are not exclusive. In any case, since changes in SpT seem to reflect real variations in $T_{\textrm{eff}}$, the dependence of SpT on luminosity is indirect.

\item We find that the fraction of stars with observed spectral variability and the amplitude of variations are significantly larger for the sample from the SMC than for that from the LMC. Comparison to the work of \cite{whi1978} shows that RSGs from the MW display even less spectral variability and smaller amplitudes than the LMC sample. Thus, a relation between metallicity and variability appears to be present. Using data on photometric variability from the literature, we have studied the relationship between photometric and spectroscopic variability, finding that the number of observed spectral variables is significantly lower in the case of the sample from the LMC even among photometric variables. This may simply be a consequence of the later SpTs in the LMC. For a given change in colour, spectral variations are less noticeable for late SpTs, as was already suggested by \cite{eli1985}.

\item Using the ($K_{\textrm{S}}-[W3]$) colour as a proxy for mass loss, we have confirmed that, at a given luminosity, CSGs from the LMC have higher mass-loss rates than those from the the SMC \citep{bon2010}. We have found strong hints of correlations  between ($K_{\textrm{S}}-[W3]$) and the observed spectral variation when different sub-populations are considered. However, only a fraction of the CSGs in our samples follow these trends, making their interpretation unclear. 

\item We confirm the trend to earlier spectral types with decreasing metallicity found by previous works. As SpT \textit{is indicating effective temperature} seems to indicate $T_{\textrm{eff}}$, at a given luminosity, the CSGs from the LMC should be cooler than those from the SMC, and thus larger. 

\item The distribution of SpTs in the SMC is centred at K1, a subtype much earlier than found in previous works. This is partly due to the large number of G~supergiants that seem to complete the distribution toward early types, as \cite{lev2012} had already suspected. All previous works had only used K and M SGs. A second likely explanation is the extension of our sample to lower luminosities, as the luminosity/SpT correlation implies that many of the low-luminosity RSGs will be of early type. Since our sample is very incomplete for G~SGs, further displacement to earlier types of the peak is possible. In any case, we have to caution that spectral variability is high (with changes up to a whole spectral type) in the SMC, including some of the brightest RSGs. If this variability continues till the end of their life, the use of non-coeval observations to identify SN progenitors in low-metallicity environments may give rise to some inconsistencies. 

\item The distribution of SpTs in the LMC does not present a clear peak. There are two peaks centred at K4\,--\,K5 and M2\,--\,M3, with a statistically significant decrease in between. This gap is not caused by systematics in the spectral classification or selection effects, neither by the physical size of the spectral-type bins, because it is also seen in the distribution of EW(Ti), a directly observed magnitude. The peaks are a consequence of the bimodality of the distribution itself, which can be divided in two groups: one formed by mid-luminosity (Iab) CSGs, which have earlier types, and the other formed by the high-luminosity (Ia) CSGs, with predominantly M-types. The existence of a minimum between the two groups, around types M0\,--\,M1 strongly suggests that this difference is not simply a consequence of the relation between spectral type and luminosity. In addition, these groups display significantly different mass-loss rates (the difference is much larger than the difference between Ia and Iab CSGs in the SMC). Given the analogy with the distribution of spectral types for RSGs in Milky Way open clusters, we suggest that this difference is a consequence of two separate evolutionary states. 

\item If this interpretation is correct, stars that reach a sufficiently high luminosity and a sufficiently low temperature (roughly corresponding to M0\,Ia for the LMC) experiment some kind of instability that displaces them to even lower temperatures. This change results in an increase of the mass-loss rate by about one order of magnitude. Stars in the SMC, even those with high luminosities, do not seem to reach these conditions. There is indirect evidence of a similar behaviour in the MW, though the jump in the Galaxy may occur around spectral type M3. If this jump is, as suggested above, an evolutionary effect, some (or most) RSGs at typical LMC metallicity and higher will pass through a phase of enhanced mass loss and higher bolometric luminosity close to the end of their lives as RSGs. This may have some consequences for the detectability of type~II SN progenitors, since they could be more obscured than simple evolutionary models predict.   

\end{enumerate}

\acknowledgements
We thank the referee, Dr. Rolf Kudritzki, for constructive criticism, leading to substantial improvement of the paper. The observations have been partially supported by the OPTICON project (observing proposals 2010B/01, 2011A/014 and 2012A/015), which is funded by the European Commission under the Seventh Framework Programme (FP7). Part of the observations have been taken under service mode, (service proposal AO171) and the authors gratefully acknowledge the help of the AAO support astronomers. This research is partially supported by the Spanish Ministerio de Econom\'{\i}a y Competitividad (Mineco) under grants AYA2012-39364-C02-02, and FPI BES-2011-049345. The work reported on in this publication has been partially supported by the European Science Foundation (ESF), in the framework of the GREAT Research Networking Programme. This research made use of the Simbad, Vizier, and Aladin services developed at the Centre de Donn\'ees Astronomiques de Strasbourg, France. This publication makes use of data products from the Two Micron All Sky Survey, which is a joint project of the University of Massachusetts and the Infrared Processing and Analysis Center/California Institute of Technology, funded by the National Aeronautics and Space Administration and the National Science Foundation.
\endacknowledgements

\bibliographystyle{aa}
\bibliography{general}

\begin{thebibliography}{72}
\expandafter\ifx\csname natexlab\endcsname\relax\def\natexlab#1{#1}\fi

\bibitem[{{Arroyo-Torres} {et~al.}(2013){Arroyo-Torres}, {Wittkowski},
  {Marcaide}, \& {Hauschildt}}]{arr2013}
{Arroyo-Torres}, B., {Wittkowski}, M., {Marcaide}, J.~M., \& {Hauschildt},
  P.~H. 2013, \aap, 554, A76

\bibitem[{{Barklem} {et~al.}(2000){Barklem}, {Piskunov}, \& {O'Mara}}]{bar2000}
{Barklem}, P.~S., {Piskunov}, N., \& {O'Mara}, B.~J. 2000, \aaps, 142, 467

\bibitem[{{Bessell} \& {Wood}(1984)}]{bes1984}
{Bessell}, M.~S. \& {Wood}, P.~R. 1984, \pasp, 96, 247

\bibitem[{{Bonanos} {et~al.}(2010){Bonanos}, {Lennon}, {K{\"o}hlinger}, {van
  Loon}, {Massa}, {Sewilo}, {Evans}, {Panagia}, {Babler}, {Block}, {Bracker},
  {Engelbracht}, {Gordon}, {Hora}, {Indebetouw}, {Meade}, {Meixner}, {Misselt},
  {Robitaille}, {Shiao}, \& {Whitney}}]{bon2010}
{Bonanos}, A.~Z., {Lennon}, D.~J., {K{\"o}hlinger}, F., {et~al.} 2010, \aj,
  140, 416

\bibitem[{{Brott} {et~al.}(2011){Brott}, {de Mink}, {Cantiello}, {Langer}, {de
  Koter}, {Evans}, {Hunter}, {Trundle}, \& {Vink}}]{bro2011}
{Brott}, I., {de Mink}, S.~E., {Cantiello}, M., {et~al.} 2011, \aap, 530, A115

\bibitem[{{Cardelli} {et~al.}(1989){Cardelli}, {Clayton}, \&
  {Mathis}}]{car1989}
{Cardelli}, J.~A., {Clayton}, G.~C., \& {Mathis}, J.~S. 1989, \apj, 345, 245

\bibitem[{{Carquillat} {et~al.}(1997){Carquillat}, {Jaschek}, {Jaschek}, \&
  {Ginestet}}]{car1997}
{Carquillat}, M.~J., {Jaschek}, C., {Jaschek}, M., \& {Ginestet}, N. 1997,
  \aaps, 123, 5

\bibitem[{{Cohen} \& {Gaustad}(1973)}]{coh1973}
{Cohen}, M. \& {Gaustad}, J.~E. 1973, \apjl, 186, L131

\bibitem[{{Davies} {et~al.}(2007){Davies}, {Figer}, {Kudritzki}, {MacKenty},
  {Najarro}, \& {Herrero}}]{dav2007}
{Davies}, B., {Figer}, D.~F., {Kudritzki}, R.-P., {et~al.} 2007, \apj, 671, 781

\bibitem[{{Davies} {et~al.}(2010){Davies}, {Kudritzki}, \& {Figer}}]{dav2010}
{Davies}, B., {Kudritzki}, R.-P., \& {Figer}, D.~F. 2010, \mnras, 407, 1203

\bibitem[{{Davies} {et~al.}(2015){Davies}, {Kudritzki}, {Gazak}, {Plez},
  {Bergemann}, {Evans}, \& {Patrick}}]{dav2015}
{Davies}, B., {Kudritzki}, R.-P., {Gazak}, Z., {et~al.} 2015, \apj, 806, 21

\bibitem[{{Davies} {et~al.}(2013){Davies}, {Kudritzki}, {Plez}, {Trager},
  {Lan{\c c}on}, {Gazak}, {Bergemann}, {Evans}, \& {Chiavassa}}]{dav2013}
{Davies}, B., {Kudritzki}, R.-P., {Plez}, B., {et~al.} 2013, \apj, 767, 3

\bibitem[{{Diaz} {et~al.}(1989){Diaz}, {Terlevich}, \& {Terlevich}}]{dia1989}
{Diaz}, A.~I., {Terlevich}, E., \& {Terlevich}, R. 1989, \mnras, 239, 325

\bibitem[{{Drout} {et~al.}(2012){Drout}, {Massey}, \& {Meynet}}]{dro2012}
{Drout}, M.~R., {Massey}, P., \& {Meynet}, G. 2012, \apj, 750, 97

\bibitem[{{Ekstr{\"o}m} {et~al.}(2012){Ekstr{\"o}m}, {Georgy}, {Eggenberger},
  {Meynet}, {Mowlavi}, {Wyttenbach}, {Granada}, {Decressin}, {Hirschi},
  {Frischknecht}, {Charbonnel}, \& {Maeder}}]{eks2012}
{Ekstr{\"o}m}, S., {Georgy}, C., {Eggenberger}, P., {et~al.} 2012, \aap, 537,
  A146

\bibitem[{{Ekstr{\"o}m} {et~al.}(2013){Ekstr{\"o}m}, {Georgy}, {Meynet},
  {Groh}, \& {Granada}}]{eks2013}
{Ekstr{\"o}m}, S., {Georgy}, C., {Meynet}, G., {Groh}, J., \& {Granada}, A.
  2013, in EAS Publications Series, Vol.~60, EAS Publications Series, ed.
  P.~{Kervella}, T.~{Le Bertre}, \& G.~{Perrin}, 31--41

\bibitem[{{Elias} {et~al.}(1985){Elias}, {Frogel}, \& {Humphreys}}]{eli1985}
{Elias}, J.~H., {Frogel}, J.~A., \& {Humphreys}, R.~M. 1985, \apjs, 57, 91

\bibitem[{{Elias} {et~al.}(1986){Elias}, {Frogel}, \& {Schwering}}]{eli1986}
{Elias}, J.~H., {Frogel}, J.~A., \& {Schwering}, P.~B.~W. 1986, \apj, 302, 675

\bibitem[{{Gazak} {et~al.}(2014){Gazak}, {Davies}, {Kudritzki}, {Bergemann}, \&
  {Plez}}]{gaz2014}
{Gazak}, J.~Z., {Davies}, B., {Kudritzki}, R., {Bergemann}, M., \& {Plez}, B.
  2014, \apj, 788, 58

\bibitem[{{Gazak} {et~al.}(2015){Gazak}, {Kudritzki}, {Evans}, {Patrick},
  {Davies}, {Bergemann}, {Plez}, {Bresolin}, {Bender}, {Wegner}, {Bonanos}, \&
  {Williams}}]{gaz2015}
{Gazak}, J.~Z., {Kudritzki}, R., {Evans}, C., {et~al.} 2015, \apj, 805, 182

\bibitem[{{Georgy} {et~al.}(2013){Georgy}, {Ekstr{\"o}m}, {Eggenberger},
  {Meynet}, {Haemmerl{\'e}}, {Maeder}, {Granada}, {Groh}, {Hirschi}, {Mowlavi},
  {Yusof}, {Charbonnel}, {Decressin}, \& {Barblan}}]{geo2013}
{Georgy}, C., {Ekstr{\"o}m}, S., {Eggenberger}, P., {et~al.} 2013, \aap, 558,
  A103

\bibitem[{{Ginestet} {et~al.}(1994){Ginestet}, {Carquillat}, {Jaschek}, \&
  {Jaschek}}]{gin1994}
{Ginestet}, N., {Carquillat}, J.~M., {Jaschek}, M., \& {Jaschek}, C. 1994,
  \aaps, 108, 359

\bibitem[{{Gonz{\'a}lez-Fern{\'a}ndez}
  {et~al.}(2015){Gonz{\'a}lez-Fern{\'a}ndez}, {Dorda}, {Negueruela}, \&
  {Marco}}]{gon2015}
{Gonz{\'a}lez-Fern{\'a}ndez}, C., {Dorda}, R., {Negueruela}, I., \& {Marco}, A.
  2015, \aap, 578, A3

\bibitem[{{Graczyk} {et~al.}(2014){Graczyk}, {Pietrzy{\'n}ski}, {Thompson},
  {Gieren}, {Pilecki}, {Konorski}, {Udalski}, {Soszy{\'n}ski}, {Villanova},
  {G{\'o}rski}, {Suchomska}, {Karczmarek}, {Kudritzki}, {Bresolin}, \&
  {Gallenne}}]{gra2014}
{Graczyk}, D., {Pietrzy{\'n}ski}, G., {Thompson}, I.~B., {et~al.} 2014, \apj,
  780, 59

\bibitem[{{Gray} \& {Corbally}(1994)}]{gra1994}
{Gray}, R.~O. \& {Corbally}, C.~J. 1994, \aj, 107, 742

\bibitem[{{Gustafsson} {et~al.}(2008){Gustafsson}, {Edvardsson}, {Eriksson},
  {J{\o}rgensen}, {Nordlund}, \& {Plez}}]{gus08}
{Gustafsson}, B., {Edvardsson}, B., {Eriksson}, K., {et~al.} 2008, \aap, 486,
  951

\bibitem[{{Heiter} \& {Eriksson}(2006)}]{hei2006}
{Heiter}, U. \& {Eriksson}, K. 2006, \aap, 452, 1039

\bibitem[{{Humphreys}(1974)}]{hum1974a}
{Humphreys}, R.~M. 1974, \apj, 188, 75

\bibitem[{{Humphreys}(1979)}]{hum1979a}
{Humphreys}, R.~M. 1979, \apj, 231, 384

\bibitem[{{Humphreys} \& {Davidson}(1979)}]{hum1979}
{Humphreys}, R.~M. \& {Davidson}, K. 1979, \apj, 232, 409

\bibitem[{{Humphreys} \& {McElroy}(1984)}]{hum1984}
{Humphreys}, R.~M. \& {McElroy}, D.~B. 1984, \apj, 284, 565

\bibitem[{{Humphreys} \& {Ney}(1974)}]{hum1974b}
{Humphreys}, R.~M. \& {Ney}, E.~P. 1974, \apj, 194, 623

\bibitem[{{Humphreys} {et~al.}(1972){Humphreys}, {Strecker}, \&
  {Ney}}]{hum1972}
{Humphreys}, R.~M., {Strecker}, D.~W., \& {Ney}, E.~P. 1972, \apj, 172, 75

\bibitem[{{Josselin} {et~al.}(2000){Josselin}, {Blommaert}, {Groenewegen},
  {Omont}, \& {Li}}]{jos2000}
{Josselin}, E., {Blommaert}, J.~A.~D.~L., {Groenewegen}, M.~A.~T., {Omont}, A.,
  \& {Li}, F.~L. 2000, \aap, 357, 225

\bibitem[{{Keenan}(1945)}]{kee1945}
{Keenan}, P.~C. \&~{Hynek}, J.~A. 1945, \apj, 101, 265

\bibitem[{{Keller} \& {Wood}(2006)}]{kel2006}
{Keller}, S.~C. \& {Wood}, P.~R. 2006, \apj, 642, 834

\bibitem[{{Kirkpatrick} {et~al.}(1991){Kirkpatrick}, {Henry}, \&
  {McCarthy}}]{kir1991}
{Kirkpatrick}, J.~D., {Henry}, T.~J., \& {McCarthy}, Jr., D.~W. 1991, \apjs,
  77, 417

\bibitem[{{Kupka} {et~al.}(2000){Kupka}, {Ryabchikova}, {Piskunov}, {Stempels},
  \& {Weiss}}]{kup2000}
{Kupka}, F.~G., {Ryabchikova}, T.~A., {Piskunov}, N.~E., {Stempels}, H.~C., \&
  {Weiss}, W.~W. 2000, Baltic Astronomy, 9, 590

\bibitem[{{Lee}(1970)}]{lee1970}
{Lee}, T.~A. 1970, \apj, 162, 217

\bibitem[{{Levesque}(2013)}]{lev2013}
{Levesque}, E.~M. 2013, in EAS Publications Series, Vol.~60, EAS Publications
  Series, ed. P.~{Kervella}, T.~{Le Bertre}, \& G.~{Perrin}, 269--277

\bibitem[{{Levesque} \& {Massey}(2012)}]{lev2012}
{Levesque}, E.~M. \& {Massey}, P. 2012, \aj, 144, 2

\bibitem[{{Levesque} {et~al.}(2007){Levesque}, {Massey}, {Olsen}, \&
  {Plez}}]{lev2007}
{Levesque}, E.~M., {Massey}, P., {Olsen}, K.~A.~G., \& {Plez}, B. 2007, \apj,
  667, 202

\bibitem[{{Levesque} {et~al.}(2005){Levesque}, {Massey}, {Olsen}, {Plez},
  {Josselin}, {Maeder}, \& {Meynet}}]{lev2005}
{Levesque}, E.~M., {Massey}, P., {Olsen}, K.~A.~G., {et~al.} 2005, \apj, 628,
  973

\bibitem[{{Levesque} {et~al.}(2006){Levesque}, {Massey}, {Olsen}, {Plez},
  {Meynet}, \& {Maeder}}]{lev2006}
{Levesque}, E.~M., {Massey}, P., {Olsen}, K.~A.~G., {et~al.} 2006, \apj, 645,
  1102

\bibitem[{{Marco} \& {Negueruela}(2013)}]{mar2013}
{Marco}, A. \& {Negueruela}, I. 2013, \aap, 552, A92

\bibitem[{{Massey}(2003)}]{mas2003}
{Massey}, P. 2003, \araa, 41, 15

\bibitem[{{Massey} \& {Olsen}(2003)}]{mas2003b}
{Massey}, P. \& {Olsen}, K.~A.~G. 2003, \aj, 126, 2867

\bibitem[{{Mauron} \& {Josselin}(2011)}]{mau2011}
{Mauron}, N. \& {Josselin}, E. 2011, \aap, 526, A156

\bibitem[{{M{\'e}sz{\'a}ros} {et~al.}(2012){M{\'e}sz{\'a}ros}, {Allende
  Prieto}, {Edvardsson}, {Castelli}, {Garc{\'{\i}}a P{\'e}rez}, {Gustafsson},
  {Majewski}, {Plez}, {Schiavon}, {Shetrone}, \& {de Vicente}}]{mes12}
{M{\'e}sz{\'a}ros}, S., {Allende Prieto}, C., {Edvardsson}, B., {et~al.} 2012,
  \aj, 144, 120

\bibitem[{{Meynet} \& {Maeder}(2000)}]{mey2000}
{Meynet}, G. \& {Maeder}, A. 2000, \aap, 361, 101

\bibitem[{{Munari} \& {Tomasella}(1999)}]{mun1999}
{Munari}, U. \& {Tomasella}, L. 1999, \aaps, 137, 521

\bibitem[{{Negueruela} {et~al.}(2013){Negueruela},
  {Gonz{\'a}lez-Fern{\'a}ndez}, {Dorda}, {Marco}, \& {Clark}}]{neg2013}
{Negueruela}, I., {Gonz{\'a}lez-Fern{\'a}ndez}, C., {Dorda}, R., {Marco}, A.,
  \& {Clark}, J.~S. 2013, in EAS Publications Series, Vol.~60, EAS Publications
  Series, ed. P.~{Kervella}, T.~{Le Bertre}, \& G.~{Perrin}, 279--285

\bibitem[{{Negueruela} \& {Marco}(2012)}]{neg2012b}
{Negueruela}, I. \& {Marco}, A. 2012, \aj, 143, 46

\bibitem[{{Neugent} {et~al.}(2010){Neugent}, {Massey}, {Skiff}, {Drout},
  {Meynet}, \& {Olsen}}]{neu2010}
{Neugent}, K.~F., {Massey}, P., {Skiff}, B., {et~al.} 2010, \apj, 719, 1784

\bibitem[{{Neugent} {et~al.}(2012){Neugent}, {Massey}, {Skiff}, \&
  {Meynet}}]{neu2012}
{Neugent}, K.~F., {Massey}, P., {Skiff}, B., \& {Meynet}, G. 2012, \apj, 749,
  177

\bibitem[{{Patrick} {et~al.}(2015){Patrick}, {Evans}, {Davies}, {Kudritzki},
  {Gazak}, {Bergemann}, {Plez}, \& {Ferguson}}]{pat2015}
{Patrick}, L.~R., {Evans}, C.~J., {Davies}, B., {et~al.} 2015, \apj, 803, 14

\bibitem[{{Piskunov} {et~al.}(1995){Piskunov}, {Kupka}, {Ryabchikova}, {Weiss},
  \& {Jeffery}}]{pis1995}
{Piskunov}, N.~E., {Kupka}, F., {Ryabchikova}, T.~A., {Weiss}, W.~W., \&
  {Jeffery}, C.~S. 1995, \aaps, 112, 525

\bibitem[{{Robitaille} {et~al.}(2008){Robitaille}, {Meade}, {Babler},
  {Whitney}, {Johnston}, {Indebetouw}, {Cohen}, {Povich}, {Sewilo}, {Benjamin},
  \& {Churchwell}}]{rob2008}
{Robitaille}, T.~P., {Meade}, M.~R., {Babler}, B.~L., {et~al.} 2008, \aj, 136,
  2413

\bibitem[{{Schuster} {et~al.}(2006){Schuster}, {Humphreys}, \&
  {Marengo}}]{sch2006}
{Schuster}, M.~T., {Humphreys}, R.~M., \& {Marengo}, M. 2006, \aj, 131, 603

\bibitem[{{Skrutskie} {et~al.}(2006){Skrutskie}, {Cutri}, {Stiening},
  {Weinberg}, {Schneider}, {Carpenter}, {Beichman}, {Capps}, {Chester},
  {Elias}, {Huchra}, {Liebert}, {Lonsdale}, {Monet}, {Price}, {Seitzer},
  {Jarrett}, {Kirkpatrick}, {Gizis}, {Howard}, {Evans}, {Fowler}, {Fullmer},
  {Hurt}, {Light}, {Kopan}, {Marsh}, {McCallon}, {Tam}, {Van Dyk}, \&
  {Wheelock}}]{skr2006}
{Skrutskie}, M.~F., {Cutri}, R.~M., {Stiening}, R., {et~al.} 2006, \aj, 131,
  1163

\bibitem[{{Solf}(1978)}]{sol1978}
{Solf}, J. 1978, \aaps, 34, 409

\bibitem[{{Soszynski} {et~al.}(2002){Soszynski}, {Udalski}, {Szymanski},
  {Kubiak}, {Pietrzynski}, {Wozniak}, {Zebrun}, {Szewczyk}, \&
  {Wyrzykowski}}]{sos2002}
{Soszynski}, I., {Udalski}, A., {Szymanski}, M., {et~al.} 2002, \actaa, 52, 369

\bibitem[{{Subramanian} \& {Subramaniam}(2012)}]{sub2012}
{Subramanian}, S. \& {Subramaniam}, A. 2012, \apj, 744, 128

\bibitem[{{van Loon} {et~al.}(2005){van Loon}, {Cioni}, {Zijlstra}, \&
  {Loup}}]{loo2005}
{van Loon}, J.~T., {Cioni}, M.-R.~L., {Zijlstra}, A.~A., \& {Loup}, C. 2005,
  \aap, 438, 273

\bibitem[{{Vollmann} \& {Eversberg}(2006)}]{vol2006}
{Vollmann}, K. \& {Eversberg}, T. 2006, Astronomische Nachrichten, 327, 862

\bibitem[{{Walker}(2012)}]{wal2012}
{Walker}, A.~R. 2012, \apss, 341, 43

\bibitem[{{White} \& {Wing}(1978)}]{whi1978}
{White}, N.~M. \& {Wing}, R.~F. 1978, \apj, 222, 209

\bibitem[{{Wittkowski} {et~al.}(2012){Wittkowski}, {Hauschildt},
  {Arroyo-Torres}, \& {Marcaide}}]{wit2012}
{Wittkowski}, M., {Hauschildt}, P.~H., {Arroyo-Torres}, B., \& {Marcaide},
  J.~M. 2012, \aap, 540, L12

\bibitem[{{Wood} {et~al.}(1983){Wood}, {Bessell}, \& {Fox}}]{woo1983}
{Wood}, P.~R., {Bessell}, M.~S., \& {Fox}, M.~W. 1983, \apj, 272, 99

\bibitem[{{Wright} {et~al.}(2010){Wright}, {Eisenhardt}, {Mainzer}, {Ressler},
  {Cutri}, {Jarrett}, {Kirkpatrick}, {Padgett}, {McMillan}, {Skrutskie},
  {Stanford}, {Cohen}, {Walker}, {Mather}, {Leisawitz}, {Gautier}, {McLean},
  {Benford}, {Lonsdale}, {Blain}, {Mendez}, {Irace}, {Duval}, {Liu}, {Royer},
  {Heinrichsen}, {Howard}, {Shannon}, {Kendall}, {Walsh}, {Larsen}, {Cardon},
  {Schick}, {Schwalm}, {Abid}, {Fabinsky}, {Naes}, \& {Tsai}}]{wri2010}
{Wright}, E.~L., {Eisenhardt}, P.~R.~M., {Mainzer}, A.~K., {et~al.} 2010, \aj,
  140, 1868

\bibitem[{{Yang} \& {Jiang}(2011)}]{yan2011}
{Yang}, M. \& {Jiang}, B.~W. 2011, \apj, 727, 53

\bibitem[{{Yang} \& {Jiang}(2012)}]{yan2012}
{Yang}, M. \& {Jiang}, B.~W. 2012, \apj, 754, 35

\end{thebibliography}

\appendix
\onecolumn
\section{List of lines measured}

\begin{table*}[th!]
\caption{Atomic lines measured. The EWs are calculated over the measurement range indicated, using the continua calculated through the linear regression of the data from the continuum ranges. The last column indicates the references where the chemical species are identified.}
\label{lines}
\centering
\begin{tabular}{c c | c c | c c | c}
\hline\hline
\noalign{\smallskip}
\multicolumn{2}{c|}{Atomic Line}&\multicolumn{2}{c|}{Range of EW Measurement (\AA{})}&\multicolumn{2}{c|}{Continuum Ranges (\AA{})}&\\
$\lambda$\:(\AA{})&Chem. species&Lower Limit&Upper Limit&At Blue&At Red&Reference\\
\noalign{\smallskip}
\hline
\noalign{\smallskip}
8498.0&Ca\,{\sc{ii}}&8492.5&8503.5&8489.2-8490.4&8507.6-8509.6&\cite{sol1978}\\
8514.1&Fe\,{\sc{i}}&8512.5&8516.3&8507.6-8509.6&8557.5-8559.0&\cite{car1997}\\
8518.1&Ti\,{\sc{i}}&8516.8&8519.8&8507.6-8509.6&8557.5-8559.0&\cite{car1997}\\
8542.0&Ca\,{\sc{ii}}&8532.0&8553.0&8507.6-8509.6&8557.5-8559.0&\cite{sol1978}\\
8582.0&Fe\,{\sc{i}}&8581.0&8583.7&8579.5-8580.8&8600.0-8602.0&\cite{car1997}\\
8611.0&Fe\,{\sc{i}}&8610.9&8612.7&8600.0-8602.0&8619.6-8620.6&\cite{car1997}\\
8662.0&Ca\,{\sc{ii}}&8651.0&8673.0&8634.5-8640.4&8684.4-8686.0&\cite{sol1978}\\
8679.4&Fe\,{\sc{i}}&8676.9&8681.1&8634.5-8640.4&8684.4-8686.0&\cite{gin1994}\\
8683.0&Ti\,{\sc{i}}&8681.6&8684.2&8634.5-8640.4&8684.4-8686.0&\cite{gin1994}\\
8688.5&Fe\,{\sc{i}}&8687.3&8690.6&8684.4-8686.0&8695.5-8698.0&\cite{car1997}\\
8692.0&Ti\,{\sc{i}}&8691.0&8693.0&8684.4-8686.0&8695.5-8698.0&\cite{mun1999}\\
8710.2&Fe\,{\sc{i}}&8708.5&8711.3&8704.2-8706.3&8714.5-8715.5&\cite{kir1991}\\
8712.8&Fe\,{\sc{i}}&8711.3&8714.5&8704.2-8706.3&8714.5-8715.5&\cite{kir1991}\\
8730.5&Ti\,{\sc{i}}&8729.8&8731.7&8721.7-8723.4&8731.7-8733.8&\cite{kup2000}\\
8734.5&Ti\,{\sc{i}}&8733.5&8735.5&8731.7-8733.8&8753.5-8755.6&\cite{gin1994}\\
8757.0&Fe\,{\sc{i}}&8755.6&8758.8&8753.5-8755.6&8758.8-8761.0&\cite{kir1991}\\
8793.2\tablefootmark{a}&Fe\,{\sc{i}}&8791.5&8794.2&8786.0-8788.5&8810.0-8812.0&\cite{kir1991}\\
8805.0\tablefootmark{a}&Fe\,{\sc{i}}&8803.3&8805.6&8786.0-8788.5&8810.0-8812.0&\cite{kir1991}\\
8824.0\tablefootmark{a}&Fe\,{\sc{i}}&8823.2&8825.5&8810.0-8812.0&8828.5-8830.5&\cite{kir1991}\\
8838.0\tablefootmark{a}&Fe\,{\sc{i}}&8837.5&8840.0&8828.5-8830.5&8850.0-8854.0&\cite{kir1991}\\
\noalign{\smallskip}
\hline
\end{tabular}
\tablefoot{
\tablefoottext{a}{Outside the 2010 spectral range}
}
\end{table*}

\end{document}